\documentclass[a4paper,11pt]{article}

\usepackage{jheppub} 
\pdfoutput=1 
\usepackage[T1]{fontenc}
\usepackage{slashed}
\usepackage{subcaption}
\usepackage{bbm}
\usepackage{tikz}
\usepackage{tabularx}
\usepackage{physics}
\usepackage{multirow}
\usepackage{array} 
\usepackage{diagbox} 

\usepackage{float}

\tikzstyle{startstop1} = [rectangle, rounded corners, 
minimum width=2cm, 
minimum height=1cm,
text centered,
draw=black, 
fill=white!20]

\tikzstyle{none} = [rectangle, rounded corners, 
minimum width=2cm, 
minimum height=1cm,
text centered,
draw=none, 
fill=none]

\tikzstyle{arrow} = [thick,->,>=stealth]

\definecolor{mycolor}{rgb}{0.8, 0.4, 0.4}

\title{\boldmath
	Topology in Holographic Mean-Field Theory at  Zero and Finite Temperature
}

\author{Moongul Byun, Taewon Yuk, Young-Kwon Han, Debabrata Ghorai, and Sang-Jin Sin}

\affiliation{Department of Physics, Hanyang University, Seoul 04763, South Korea}

\emailAdd{moongulbyun@hanyang.ac.kr, tae1yuk@gmail.com,  youngkwonhan346@gmail.com, dghorai123@gmail.com, sangjin.sin@gmail.com}

\abstract{
	We investigate topological invariants in strongly interacting many-body systems within holographic mean-field theory (H-MFT) framework.
	Analytic expressions for retarded Green’s functions are obtained for all possible fermionic bilinear interactions in the limit of probe background limit $\mathrm{AdS}_4$, from which we construct topological Hamiltonians. 
	Integrating  Berry curvature over the momentum domain for the gapped spectra yields well-defined and quantized Chern numbers, enabling a systematic classification of them across interaction types.
	These topological invariants remain robust under deformation parameters like  interaction  and temperature, indicating that H-MFT encodes effective single-particle-state topology near a quantum critical point in strongly correlated systems. We point out why  topological number is defined in the holographic theories while it is not in the perturbative field theory. 
}


\begin{document} 
	\maketitle
	\flushbottom
	
	\section{Introduction and summary} 
	The AdS/CFT correspondence has provided a powerful framework for investigating strongly coupled quantum systems through their dual classical gravity descriptions \cite{Maldacena:1997re, Witten:1998qj}.
	Early developments in holography established the foundation for computing Green’s functions, thereby illuminating key aspects of boundary field theory dynamics.
	Since then, the correspondence has found widespread applications in condensed matter physics, such as in the study of holographic superconductors \cite{Hartnoll:2008kx, Hartnoll:2008vx, Gubser:2008px, Faulkner:2009am, Horowitz_2009, Horowitz_2011, Kim:2013oba} and fermionic dynamics \cite{Iqbal:2009fd, Laia:2011zn, Faulkner:2013bna}.
	
	Holographic mean-field theory is defined as the coupled system of asymptotic AdS gravity with `order parameter tensor fields' together with a fermion field in the bulk. In case there is a globally conserved charge, we must introduce local U(1) gauge field as is well known. 
	Since the fermion back reaction to other fields is subtle\footnote{If there is a fermion background, its interpretation is unclear: e.g., what is the meaning of the fermion condensation?}, we first decouple it from other fields and 
	once bosonic field configurations are obtained, we then couple the fermion field as a fluctuation and calculate the Green functions and spectral function, from which we can read off the boundary fermion dynamics according to the dictionary.

	The bulk theory is formulated as a continuum field theory without  microscopic lattice details. 
	This means that we are zooming in the Brillouin zone's small region containing the interesting special feature created by the order parameter field.  In reality such features are due to the lattice structures and chemical compositions. However, the surprising feature of the holographic mean field theory  \cite{Oh:2021, Sukrakarn:2023ncp} is that 
	simple order parameters can  reproduce almost  all  interesting  features of band theory including the gap, Dirac band,  flat bands of various dimension, nodal line, nodal circle and  nodal sphere. 
	This strongly indicates that the effect of the crystal structure is    encoded in the symmetry breaking pattern and this is partially because, in the low  momentum and energy scale  of meV $\sim$ eV scale where the usual experiments are done for the material property, the electron can not see the   lattice structure, because it is keV scale which is thousand to million times smaller than its wave length. 
	
	Therefore, in this H-MFT approach   it is advantageous to systematically examine the fermion coupling to various types of  order parameters  and classify their spectral behavior accordingly, assuming the dominance of interaction with  minimal scaling dimensions.  
	Along  this direction, we introduced bulk order parameter fields $B_{I}$, which gives certain symmetry breaking effect via fermionic bilinear couplings of the form \cite{Oh:2021} 
	\begin{equation}
		\mathcal{L}_{\mathrm{int}} = B_{I}\bar{\psi}\Gamma^{I}\psi,
	\end{equation}
	representing various types of condensations in the holographic setup analogous to fermion bilinear coupled to the Hubbard-Stratonovich field in the conventional mean field theory. 
	Within this framework, the holographic mean-field theory (H-MFT) has been systematically developed as a  bottom up approach to encode essential features of fermion dynamics in strongly interacting system, including spectral features observed in ARPES experiments on real materials \cite{Oh:2021, Sukrakarn:2023ncp}.
	The H-MFT   approach has been successfully applied to diverse systems, including holographic superconductors \cite{Yuk:2022lof, Ghorai:2023wpu, PhysRevD.109.066004}, Lieb lattice \cite{Han2023}, ABC stacked multilayer graphene \cite{Seo2022}, and Kondo lattice models \cite{Im:2023ffg, Han:2024rbr}.
	
	One should not forget that in the AdS space, the presence of the mass scale in the bulk does not necessarily mean that the boundary theory is gapped. It turns out only particular tensor type, the pseudoscalar order parameter can create boundary mass gap \cite{Oh:2021, Sukrakarn:2023ncp}.  
	If there is no mass scale at the boundary, topological insulator is hard to think about. In holography, there is no boundary of the boundary and therefore there can not be real bulk-boundary correspondence is possible.  	
	Therefore the pseudoscalar coupling has a special role to generate topological holographic matter. 
	
	Despite significant progress in H-MFT, its application to topological classification has remained largely unexplored.
	Identifying a physically meaningful topological invariant within a holographic framework would have significant implications:
	it would provide a new method for defining topology in strongly correlated many-body systems, such as topological Mott \cite{PhysRevLett.100.156401} and Kondo insulator \cite{Dzero_2016}.
	In such systems, conventional Berry  curvature formalism \cite{doi:10.1098/rspa.1984.0023, PhysRevLett.62.2747} should break down due to the absence of well-defined single-particle states.
	While the Uhlmann phase~\cite{UHLMANN1986229, UHLMANN1993253, Uhlmann1991} provides a many-body generalization of the geometric phase, it fails in strongly interacting regimes and typically does not yield integer-valued topological invariants.
	As a result, the task of defining robust topological invariants in strongly interacting systems remains as an open problem and even its possibility   is unclear.  
	
	In this work, we advance H-MFT as a new framework to address this problem.
	Building on previous H-MFT constructions of fermionic Green’s functions~\cite{Oh:2021, Sukrakarn:2023ncp}, where various fermionic bilinear interactions in AdS were studied, we extend these results by incorporating the prescription of~\cite{PhysRevB.86.165116} to compute effective Hamiltonians from Green’s functions and extract the associated topological invariants.
	In particular, we classify and analyze all 16 bilinear fermion interaction types in $\mathrm{AdS}_{4}$: scalar, pseudoscalar, vector, axial vector, and antisymmetric 2-tensor couplings.
	
	Our main finding is that \textit{H-MFT   yields well-defined and robust topological invariant, the Chern number, not only for 0-temperature but also at finite temperature.}  It is derived  using  the Berry  potential formalism for gapped phases in strongly interacting many-body systems.
	In H-MFT, this number can be systematically classified according to the number of fermion flavors and the structure of the interaction terms, and it remains invariant under deformation parameters, especially under temperature change. 
	
	In the one-flavor case, we find a universal Chern number $C = \frac{1}{2}\mathrm{sgn}(\mathcal{B}_{5})$ for gapped spectra where $\mathcal{B}_{5}$ is the order parameter of pseudoscalar coupling.
	This fractional value, which is different from the lattice models,  originates from the continuum limit in our approach, where absence of the  lattice 
	induces that of the Brillouin zone which enforce integer Chern number.
	In the two-flavor case, the Chern number becomes integer-valued, $C \in \{0, \pm 1\}$, and depends on both the structure of the interaction terms and the choice of quantization scheme.
	The resulting topological phases can be classified into four distinct cases:
	\begin{itemize}
		\item \textbf{case 1.} Topology is determined by single gapping parameter.
		\item \textbf{case 2.} Topology is determined by two gapping parameters.
		\item \textbf{case 3.} There exist two  competing gapping orders.
		\item \textbf{case 4.} Chern number is Zero, otherwise.
	\end{itemize}
	In the presence of the symmetry breaking scale, the topological property is protected by the gap, therefore it is natural to expect that no gap means no topology.  Indeed, the topology of gapped fermionic spectra at the boundary of $\mathrm{AdS}_{4}$ is classified according to the table~\ref{table1}.
	\begin{table}[tbp]
		\centering
		\resizebox{\textwidth}{!}{
			\begin{tabular}{|c| p{0.7\textwidth} | p{0.25\textwidth} |}
				\hline
				\textbf{Flavor} & \centering\textbf{Gapping term structure in Chern number} & \begin{minipage}{0.25\textwidth}
					\centering
					\textbf{Chern number}
				\end{minipage} \\
				\hline
				1 & \centering Single gapping parameter & \begin{minipage}{0.25\textwidth}
					\centering
					$C \in \{-1/2, +1/2\}$
				\end{minipage} \\
				\hline
				\multirow{4}{*}{2} & \centering Single gapping parameter & \multirow{2}{*}{\begin{minipage}{0.25\textwidth}
						\centering
						$C \in \{-1, +1\}$
				\end{minipage}} \\
				\cline{2-2}
				& \centering Sum of two independent gapping parameters &  \\
				\cline{2-3}
				& \centering Competing two independent gapping parameters & \begin{minipage}{0.25\textwidth}
					\centering
					$C \in \{-1, 0, +1\}$
				\end{minipage} \\
				\cline{2-3}
				& \centering Vanishing net topology & \begin{minipage}{0.25\textwidth}
					\centering
					$0$
				\end{minipage} \\
				\hline
			\end{tabular}
		}
		\caption{\label{table1}Classification of Chern numbers for gapped fermionic spectra via holographic mean-field theory in $\mathrm{AdS}_{4}$, organized by flavor and structure of gapping terms in analytic Chern number expression.}
	\end{table}

\begin{figure}[tbp]
		\centering\sffamily
		\resizebox{1.0\textwidth}{!}{
			\begin{tikzpicture}[node distance=5cm, scale=0.7, transform shape]
				\node (temp1) [startstop1] {
					\begin{minipage}{7cm}
						\centering{\footnotesize Weakly-interacting system}\\
						\includegraphics[width=1.0\textwidth]{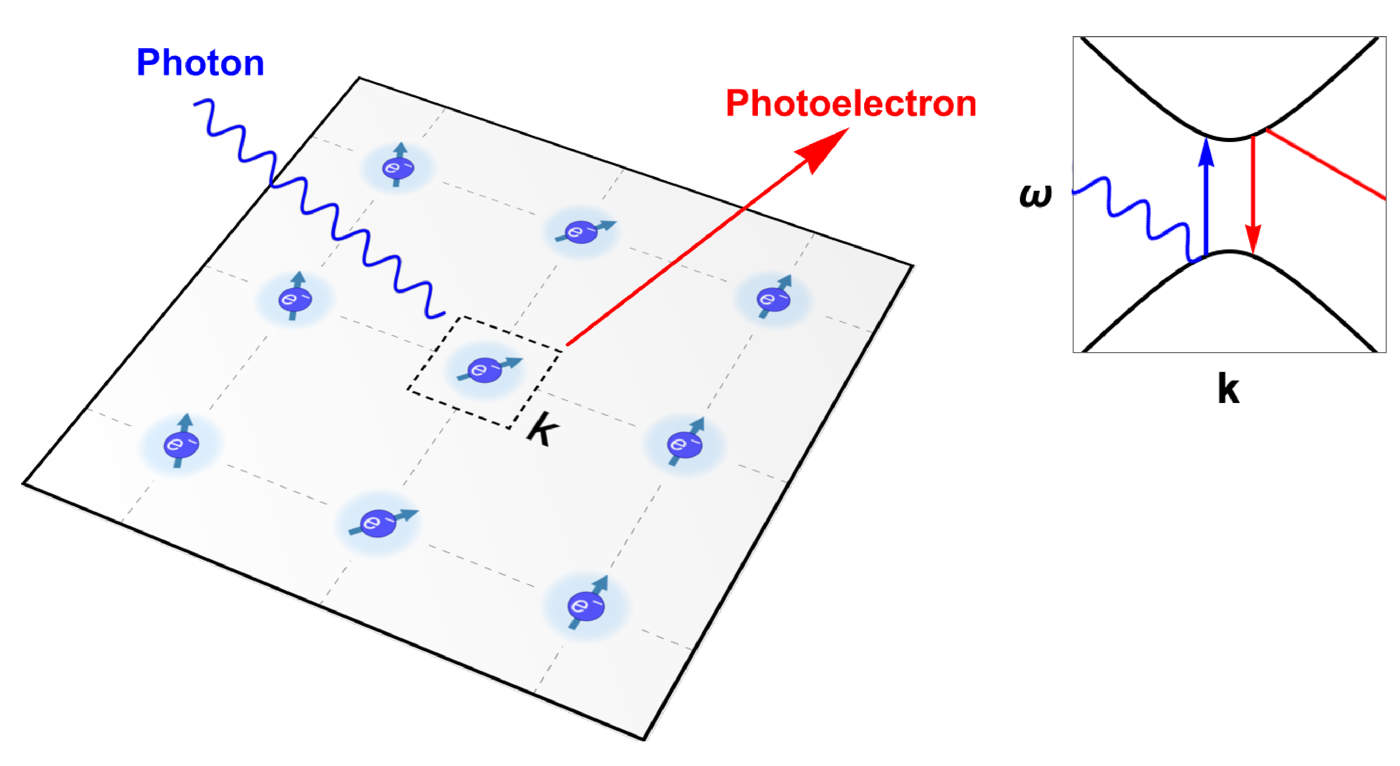}
					\end{minipage}
				};
				\node (temp2) [startstop1, below of=temp1, yshift=1.0cm] {
					\begin{minipage}{2.5cm}
						\centering\footnotesize
						Uhlmann phase
					\end{minipage}
				};
				\node (temp3) [startstop1, right of=temp1, xshift=3cm] {
					\begin{minipage}{7cm}
						\centering{\footnotesize Strongly-interacting system}\\
						\includegraphics[width=1.0\textwidth]{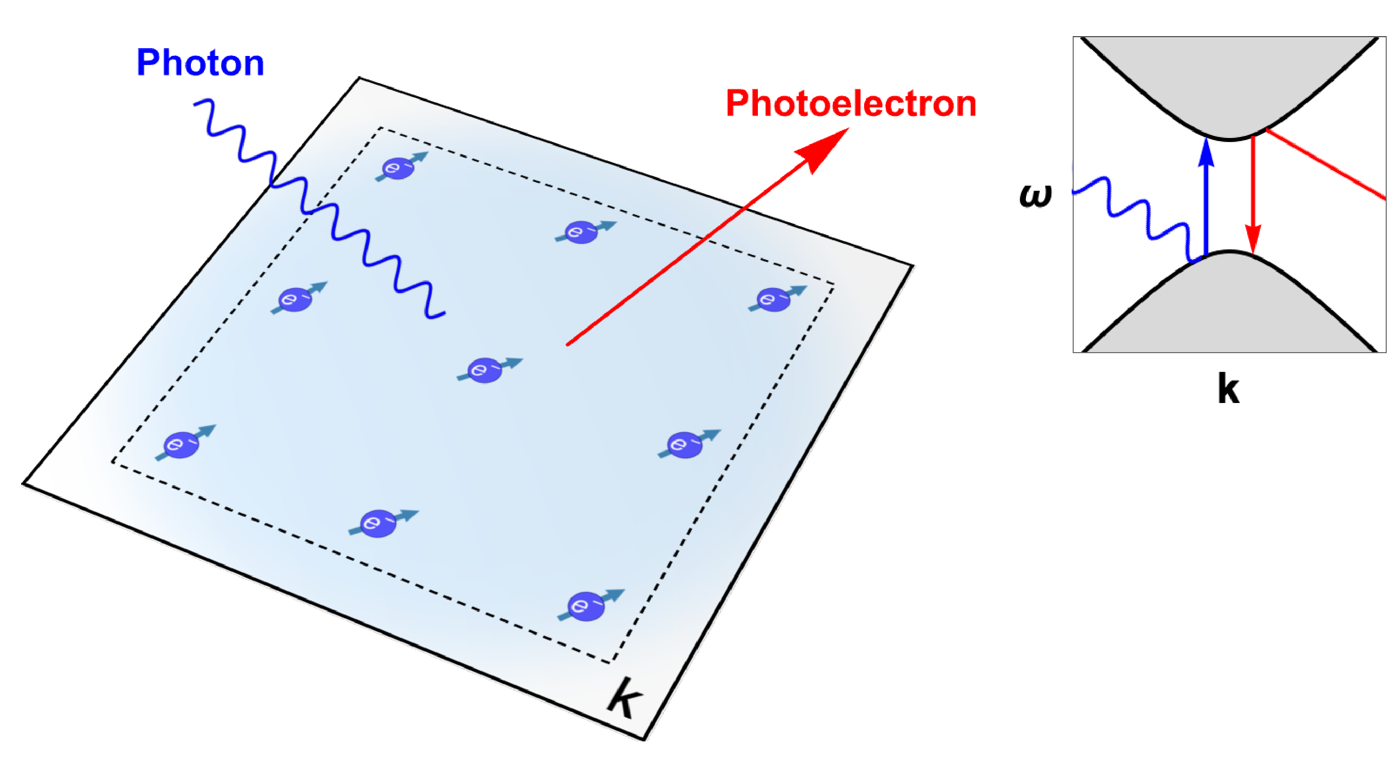}
					\end{minipage}
				};
				\node (temp4) [startstop1, above of=temp3, yshift=-1.0cm] {
					\begin{minipage}{4.0cm}
						\centering\footnotesize
						Fixed AdS background
					\end{minipage}
				};
				\node (temp5) [startstop1, below of=temp3, yshift=1.0cm] {
					\begin{minipage}{2.0cm}
						\centering\footnotesize
						Berry phase
					\end{minipage}
				};
				\node (temp6) [none, right of=temp3, xshift=1.2cm] {
					\begin{minipage}{5cm}
						\includegraphics[width=1.0\textwidth]{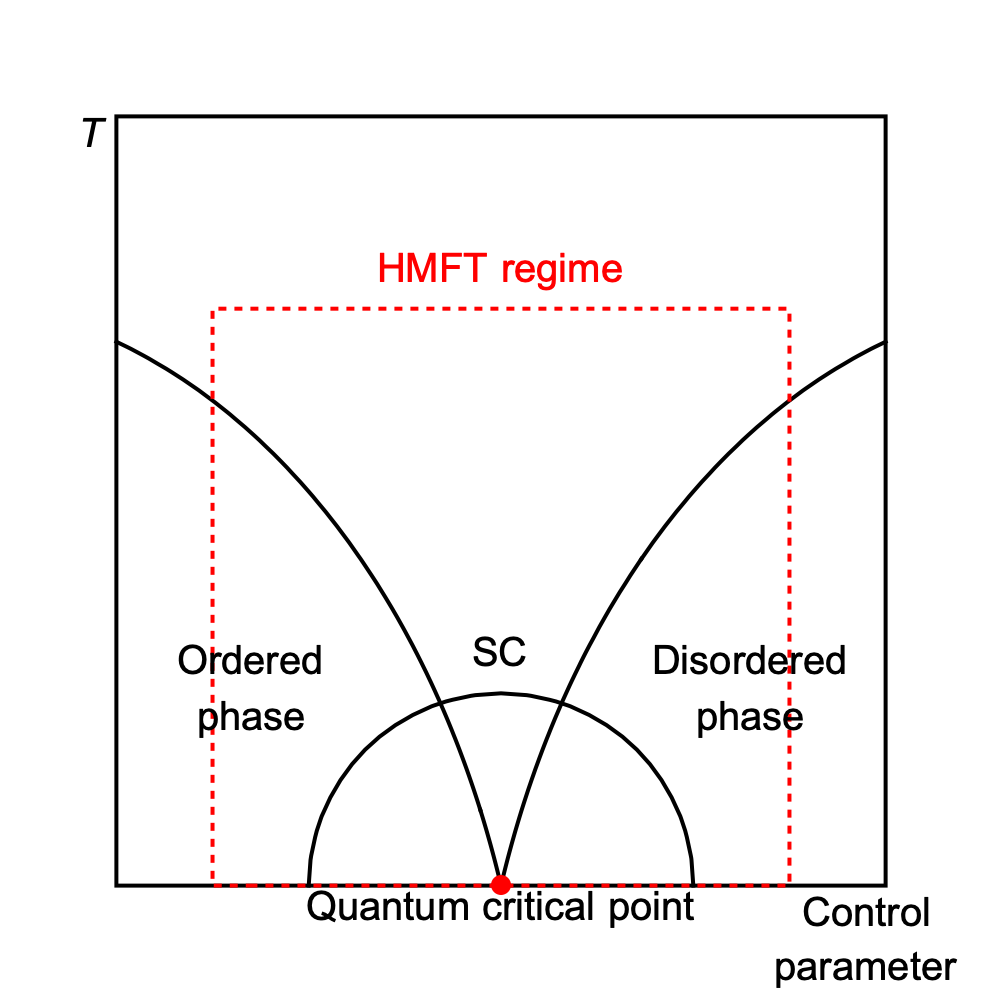}
					\end{minipage}
				};
				
				\draw[arrow] (temp1) -- node[anchor=center, fill=white, inner sep=3pt] {
					\footnotesize Many-body system
				} (temp2);
				\draw[<->, thick] (temp4) -- node[anchor=center, fill=white, inner sep=3pt] {
					\footnotesize Dual gravity: H-MFT  
				} (temp3);
				\draw[arrow] (temp3) -- node[anchor=center, fill=white, inner sep=3pt] {
					\footnotesize Effective single-particle state
				} (temp5);
				
				\draw[dotted, very thick] ([xshift=0.4cm, yshift=-4.5cm]temp1.east) -- ([xshift=0.4cm, yshift=4.4cm]temp1.east);
			\end{tikzpicture}
		}
		\caption{\label{Figure1}
			Regime in which holographic mean-field theory (H-MFT) yields a well-defined and robust Chern number.
			In weakly interacting systems (left), Uhlmann phase applies to sharp band structures of single-particle states labeled by momentum $\mathbf{k}$, as extracted from ARPES data.
			By contrast, strongly interacting systems in the H-MFT regime (right), with a fixed dual gravitational background, are described by effective single-particle states with $\mathbf{k}$, allowing Berry phase to be applied to broadened and filled spectra.
		}
	\end{figure}
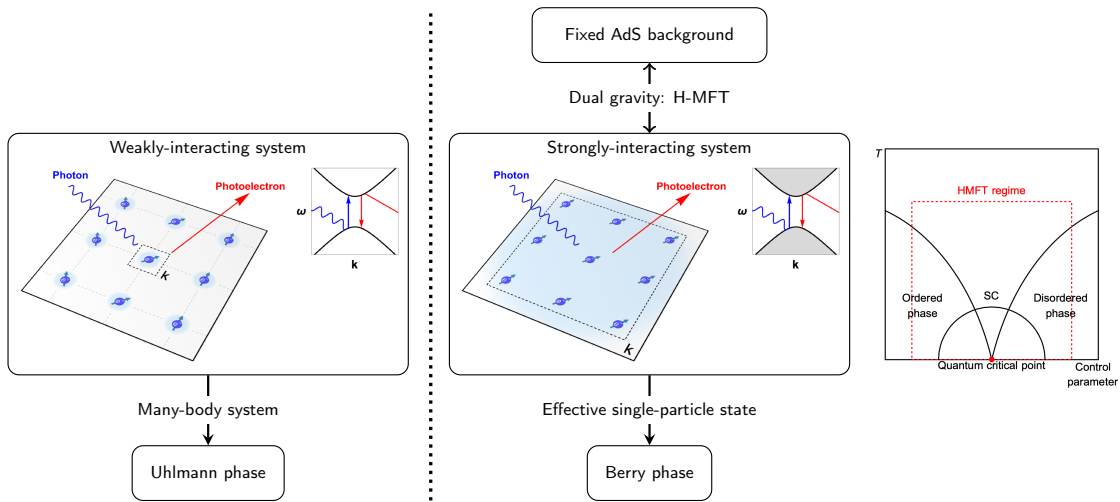

	Our Chern numbers are specifically defined within the strongly interacting many-body regime near a quantum critical point, where H-MFT   is applicable \cite{Sukrakarn:2023ncp}. 
	In this regime, the entire many-body system behaves as a macroscopic fluid dual to a fixed AdS background. 
	For instance, holographic model to describe hydrodynamic feature of Dirac fluid was reported in \cite{Seo:2016vks}.
	Therefore, the  Berry potential formalism for single-particle states becomes applicable, yielding a well-defined and robust topological invariant for gapped spectra.
	Experimentally, this distinction is reflected in ARPES measurements: while weakly interacting systems yield sharp quasiparticle peaks corresponding to single-particle excitations, strongly correlated systems exhibit broadened spectra and incoherent features \cite{RevModPhys.75.473, RevModPhys.93.025006}.
	In summary, within the H-MFT   regime, the topological invariant for strongly interacting many-body system is defined via  Berry potential applied to a single-particle state obtained from the dual gravitational description.
	This stands in contrast to the Uhlmann parallel transport, which is typically employed in weakly interacting many-body systems.
	We illustrate this summary in the figure~\ref{Figure1}.

	For our purpose, the analytic expressions and overall feature of the spectral functions and spectrum are more important than the precise backreacted configuration of each field components, and 	
	in the absence of the availability of the analytic expressions of  gravitational dual for a given strongly correlated system, we need to  adopt an effective description based on the assumption that key features of the system can be captured   from the  exact  AdS backgrounds of 0 and non-zero temperature solutions  without  backreaction of  the order parameter  to the gravity. 
	
	The rest of this paper is organized as follows.
	In the section~\ref{section2}, we introduce the holographic setup for fermions in $\text{AdS}_4$ and defines the retarded Green’s function and spectral densities.
	In the section~\ref{section3}, we formulate the topological framework and apply it to the one-flavor case at zero temperature.
	We then extend the analysis to the two-flavor case in the section~\ref{section4} by employing the non-Abelian Berry phase to address degeneracies in the eigensystem.
	In the section~\ref{section5}, we examine the Chern numbers at finite temperature and delineate the regime in which our holographic approach is applicable.
	We summarize our findings and outline future directions in the section~\ref{sec:conclusion}.
	Technical derivations and supplementary calculations are presented in the appendices.

	\section{\label{section2}Holographic fermion  and its Green's function}
	We set up the notations for  holographic fermions. 
	Throughout this paper, $M, N = t, x, y, z$ represent bulk indices on a four-dimensional manifold $\mathcal{M}$, while $a, b, \cdots$ or $\underline{t}, \underline{x}, \cdots$ denote the local tangent space indices in the vielbein formalism.
	Additionally, the indices $\mu, \nu, \cdots$ represent   boundary   indices on a four-dimensional   $\partial\mathcal{M}$, while $i, j, \cdots$  denotes  the spatial indices at the   boundary.
	Now we define the action by 
	\begin{equation}
		\label{eq2.1}
		S_{\mathrm{tot}} = S_{\mathrm{bulk}} + S_{\mathrm{bdy}} + S_{g} + S_{B} + S_{\mathrm{int}},
	\end{equation}
	\begin{equation}
		\label{eq2.2}
		S_{\mathrm{bulk}} = i\int_{\mathcal{M}} d^{4}x\sqrt{-g}\,\bar{\psi}(\slashed{D} - m)\psi,
	\end{equation}
	\begin{equation}
		\label{eq2.3}
		S_{\mathrm{bdy}} = -i\int_{\partial\mathcal{M}}d^{3}x\sqrt{-gg^{zz}} \, \bar{\psi}\psi,
	\end{equation}
	\begin{equation}
		\label{eq2.4}
		S_{g} = \int_{\mathcal{M}}d^{4}x\sqrt{-g}\left(R - 2\Lambda\right),
	\end{equation}
	\begin{equation}
		\label{eq2.5}
		S_{B} = \int_{\mathcal{M}}d^{4}x\sqrt{-g}\left(-\dfrac{1}{2}\left(\nabla_{M}B_{I}\right)^{2} - \dfrac{1}{2}m_{B}^{2}B_{I}^{2}\right)
	\end{equation}
	\begin{equation}
		\label{eq2.6}
		S_{\mathrm{int}} = i\int_{\mathcal{M}}d^{4}x\sqrt{-g}\,\bar{\psi}B_{I}\Gamma^{I}\psi.
	\end{equation}
	Here, $\slashed{D} = \Gamma^{M}D_{M} = \Gamma^{M}\left(\partial_{M} + \frac{1}{4}\omega_{abM}\Gamma^{ab}\right)$ and $\omega_{abM}$ denotes the spin connection. 
	The field $g_{MN}$ is the background metric tensor associated with the cosmological constant $\Lambda$.
	In \eqref{eq2.3}, $g^{zz}$ denotes the radial-radial component of the inverse metric.
	
	\noindent
	\textbf{Classification of interaction types.}
	In \eqref{eq2.6}, we use $B_{I}\Gamma^{I}$ as an abstract notation for various types of order parameter fields $B_{I}$, coupled to bulk fermions, and Gamma matrices $\Gamma^{I}$.
	Throughout this paper, we use $I, J, \cdots$ as schematic tensor indices labeling an order parameter and its conjugate fermion bilinear.
	In $\text{AdS}_{4}$, we have 16 types of interaction in \eqref{eq2.6} which are given by \cite{Oh:2021, Sukrakarn:2023ncp}
	\begin{equation}
		\label{eq2.7}
		B_{I} = \left\{B, B_{5}, B_{M}, B_{5M}, B_{MN}\right\} \quad\text{and}\quad \Gamma^{I} = \left\{-i\mathbbm{1}_{4\times4}, \Gamma^{5}, \Gamma^{M}, \Gamma^{M}\Gamma^{5}, \Gamma^{MN}\right\}.
	\end{equation}
	We classify the interaction types $B_{I}$ from the bulk perspective into scalar, pseudoscalar, vector, axial vector, and antisymmetric 2-tensor couplings.
	The interaction Lagrangian for such interaction type is given by
	\begin{equation}
		\label{eq2.8}
		\mathcal{L}_{\mathrm{int}} = i\bar{\psi}B_{I}\Gamma^{I}\psi = \begin{cases}
			-i\bar{\psi}B\psi & \text{(scalar)},\\
			\bar{\psi}B_{5}\Gamma^{5}\psi & \text{(pseudoscalar)},\\
			\bar{\psi}B_{M}\Gamma^{M}\psi & \text{(vector)},\\
			\bar{\psi}B_{5M}\Gamma^{M}\Gamma^{5}\psi & \text{(axial vector)},\\
			\dfrac{1}{2}\bar{\psi}B_{MN}\Gamma^{MN}\psi & \text{(antisymmetric 2-tensor)},
		\end{cases}
	\end{equation}
	We set the ansatz for the fields $B_{I}$ in \eqref{eq2.8} as solvable solutions of equations of motion from \eqref{eq2.5} by choosing appropriate $m_{B}$ value, and only choose the source terms as non-zero parameters in probe limit such that \cite{Oh:2021, Sukrakarn:2023ncp}
	\begin{equation}
		\label{eq2.9}
		B_{I} = \mathcal{B}_{I}z^{p_{I}} \quad \text{where} \quad \mathcal{B}_{I} = \left\{\mathcal{B}, \mathcal{B}_{5}, \mathcal{B}_{M}, \mathcal{B}_{5M}, \mathcal{B}_{MN}\right\} \quad\text{and}\quad p_{I} = \left\{1, 1, 0, 0, -1\right\}.
	\end{equation}
	Here, each component of $\mathcal{B}_{I}$ is a constant order parameter.
	
	\subsection{Dirac equation and source identification}
	\label{section3.1}
	Throughout this paper, we adopt the gamma matrix representation as
	\begin{equation}
		\label{eq2.10}
		\begin{gathered}
			\Gamma^{\underline{t}} = -i\sigma_{1}\otimes\mathbbm{1}_{2\times2},\quad\Gamma^{\underline{x}} = \sigma_{2}\otimes\sigma_{1},\quad\Gamma^{\underline{y}} = \sigma_{2}\otimes\sigma_{2},\quad\Gamma^{\underline{z}} = \sigma_{3}\otimes\mathbbm{1}_{2\times2},\\
			\Gamma^{5} = \sigma_{2}\otimes\sigma_{3},\quad\Gamma^{ab} = \dfrac{1}{2}[\Gamma^{a}, \Gamma^{b}].
		\end{gathered}
	\end{equation}
	The gamma matrices satisfy the Clifford algebra $\{\Gamma^{M}, \Gamma^{N}\} = 2g^{MN} \mathbbm{1}_{4\times4}$ in four dimension so that $\Gamma^{MN}$ contains some metric information. 
	We define the metric of the bulk in the inverse radius coordinate as
	\begin{equation}
		\label{eq2.11}
		ds^{2} = -\dfrac{f(z)}{z^{2}}dt^{2} + \dfrac{dx^{2} + dy^{2}}{z^{2}} + \dfrac{dz^{2}}{z^{2}f(z)}
	\end{equation}
	which is asymptotically $\text{AdS}_{4}$. 
	Also, in \eqref{eq2.4}, $\Lambda = -3$ and $f(z)$ is given by
	\begin{equation}
		\label{eq2.12}
		f(z) = 1 - \left(\dfrac{z}{z_{\mathrm{H}}}\right)^{3} \quad\text{where}\quad z_{\mathrm{H}} = \dfrac{3}{4\pi T},
	\end{equation}
	where $z_{\mathrm{H}}$ is the horizon and $T$ is the temperature.
	
	From the total action, the bulk equation of motion for $\psi$ is given by
	\begin{equation}
		\label{eq2.13}
		\left(\slashed{D} - m + B_{I}\Gamma^{I}\right)\psi = 0
	\end{equation}
	We put an ansatz of Dirac field into Dirac equation as
	\begin{equation}
		\label{eq2.14}
		\psi = (-gg^{zz})^{-1/4}e^{-i\omega t + ik_{x}x + ik_{y}y}\zeta(z),
	\end{equation}
	where $\zeta(z)$ is a four-component spinor field. 
	The term $(-gg^{zz})^{-1/4}$ was introduced to remove spin-connection term in Dirac equation, which greatly simplifies the system. 
	
	To define the retarded Green’s function, we first identify the source and condensation. 
	For this, we decompose the spinor field in \eqref{eq2.14} as
	\begin{equation}
		\label{eq2.15}
		\psi = \begin{pmatrix}
			\psi_{+}\\
			\psi_{-}
		\end{pmatrix}\quad\text{and}\quad\zeta = \begin{pmatrix}
			\zeta_{+}\\
			\zeta_{-}
		\end{pmatrix}.
	\end{equation}
	Then, by applying the ansatz \eqref{eq2.14}, the boundary action in \eqref{eq2.3} can be rewritten as
	\begin{equation}
		\label{eq2.16}
		\begin{gathered}
			S_{\mathrm{bdy}} = -i\int_{\partial\mathcal{M}}d^{3}x \, \bar{\zeta}\zeta = -\int_{\partial\mathcal{M}}d^{3}x \, \zeta^{\dagger}(\sigma_{1}\otimes\sigma_{0})\zeta =  -\int_{\partial\mathcal{M}}d^{3}x \, \zeta^{\dagger}_{+}\zeta_{-} + \text{h.c.}
		\end{gathered}
	\end{equation}
	By varying the bulk action with respect to $\psi$ and add the variation of \eqref{eq2.16}, it can be shown that the total action variation can be represented only in terms of $\zeta_{+}$ if the equation of motion is satisfied \cite{Iqbal:2009fd}.
	Therefore, we can identify the boundary quantities of $\zeta_{+}$ and $\zeta_{-}$ as the two-component source and condensation, respectively.
	For notational clarity, we redefine 
	\begin{equation}
		\label{eq2.17}
		\xi^{(\mathrm{S})} \equiv \zeta_{+}\quad\text{and}\quad\xi^{(\mathrm{C})} \equiv \zeta_{-},
	\end{equation}
	corresponding to the source and condensation at the boundary, respectively.
	To explicitly extract the source and condensation terms from these bulk quantities, we examine the boundary behavior of $\xi^{(\mathrm{S})}$ and $\xi^{(\mathrm{C})}$ by solving the Dirac equation.
	The leading $z$-dependent terms of the spinors $\xi^{(\mathrm{S})}$ and $\xi^{(\mathrm{C})}$ near the boundary are extracted as
	\begin{equation}
		\begin{aligned}
			\label{eq2.18}
			\xi^{(\mathrm{S})} \approx z^{m}\mathcal{J}\quad&\text{and}\quad\xi^{(\mathrm{C})} \approx z^{-m}\mathcal{C}.
		\end{aligned}
	\end{equation}
	Here, $\mathcal{J}$ and $\mathcal{C}$ are two-component spinors, representing the source and condensation, respectively.
	This identification is used to define the Green’s function throughout our analysis.
	
	\subsection{\label{section2.2}Green's function and spectral densities}
	Based on the chosen source and condensation, the Green’s function is obtained from the corresponding extended bulk quantities, with explicit calculations provided in Appendix~\ref{appendix:A}.
	From these results, the bulk spinors $\xi^{(\mathrm{S})}$ and $\xi^{(\mathrm{C})}$ can be written in matrix form, separating their solution basis from their coefficients, which can be generally expressed as
	\begin{equation}
		\label{eq2.19}
		\xi^{(\mathrm{S})} = \mathbb{S}(z)\mathbf{c} \quad \text{and} \quad \xi^{(\mathrm{C})} = \mathbb{C}(z)\mathbf{c},
	\end{equation}
	where $\mathbb{S}(z)$ and $\mathbb{C}(z)$ are $2\times2$ matrices, and $\mathbf{c}$ is a constant two-component vector.
	For a given coupling with $B_{I}\Gamma^{I}$, we obtain the functions $\mathbb{S}(z)$ and $\mathbb{C}(z)$ by solving the Dirac equation \eqref{eq2.13} under the infalling boundary condition for spinors.
	The retarded Green’s function is then given by
	\begin{equation}
		\label{eq2.20}
		G_{R} = \lim_{z \to 0} z^{2m} \mathbb{C}(z) \mathbb{S}^{-1}(z),
	\end{equation}
	which is a $2\times2$ matrix-valued function of $(\omega, \mathbf{k})$ where $\mathbf{k} \equiv (k_{x}, k_{y})$.
	
	The spectral density is then defined in terms of the retarded Green’s function as
	\begin{equation}
		\label{eq2.21}
		A(\omega, \mathbf{k}) = \text{Im}\text{Tr}G_{R}(\omega, \mathbf{k}).
	\end{equation}
	We get the spectral density by taking the imaginary part of $\mathrm{Tr}G_{R}(\omega, \mathbf{k})$ \textit{after taking} $\omega \to \omega + i\epsilon$ ($\epsilon > 0$) to regularize pole structures.
	This spectral density provides a physically meaningful criterion for identifying topological phases—such as gapless or gapped phases—which can be directly compared with ARPES data.
	
	\section{\label{section3}One-flavor spinors at zero temperature}
	In this section, we apply the holographic framework developed in the section~\ref{section2} to derive analytic expressions for the retarded Green’s function and the associated topological quantities for a one-flavor spinor case.
	Throughout this section, we consider the pure $\text{AdS}_4$ background ($T = 0$) by setting $f(z) = 1$ in \eqref{eq2.11}, and focus on massless spinors ($m = 0$).
	
	\subsection{\label{sec:topology}Topology from holographic fermions}
	Following the prescription of \cite{PhysRevB.86.165116}, the effective topological Hamiltonian is defined as the inverse of zero-frequency retarded Green’s function:
	\begin{equation}
		\label{eq3.1}
		H_{\mathrm{topo}}(\mathbf{k}) \equiv -G_{R}^{-1}(0, \mathbf{k}).
	\end{equation}
	Interesting point is that whatever method one uses, once we get Green functions, we can define the the topological Hamiltonian and we use the holography to generate a non-trivial Green function for a strongly interacting system at the boundary. 
	
	The associated Berry connection is computed from the eigenvector $|\mathbf{k}\rangle$ which is the lowest-energy eigenstate of $H_{\mathrm{topo}}(\mathbf{k})$ as
	\begin{equation}
		\label{eq3.2}
		\mathcal{A}(\mathbf{k}) = i\langle\mathbf{k}|\nabla_{\mathbf{k}}|\mathbf{k}\rangle.
	\end{equation}
	The corresponding Berry curvature is then defined as
	\begin{equation}
		\label{eq3.3}
		\mathcal{F}(\mathbf{k}) = 
		\partial_{k_{x}}\mathcal{A}_{k_{y}}(\mathbf{k}) - \partial_{k_{y}}\mathcal{A}_{k_{x}}(\mathbf{k}),
	\end{equation}
	which is a scalar-valued function.
	The Chern number is defined as the integral of the Berry curvature over the entire momentum plane in our continuum system:
	\begin{equation}
		\label{eq3.4}
		C = \dfrac{1}{2\pi}\int_{\mathbb{R}^{2}}dk_{x}dk_{y}\,\mathcal{F}(\mathbf{k}).
	\end{equation}
	
	For analytic evaluation, it is convenient to decompose the topological Hamiltonian into Pauli matrices:
	\begin{equation}
		\label{eq3.5}
		H_{\mathrm{topo}}(\mathbf{k}) = \sum_{i = 0}^{3}h_{i}(\mathbf{k})\sigma_{i} = h_{0}(\mathbf{k})\sigma_{0} + \mathbf{h}(\mathbf{k})\cdot\boldsymbol{\sigma},
	\end{equation}
	where each $h_{i}(\mathbf{k}) \in \mathbb{R}$, and we define $\mathbf{h}(\mathbf{k}) \equiv \left(h_{1}(\mathbf{k}), h_{2}(\mathbf{k}), h_{3}(\mathbf{k})\right)$.
	Then, the Berry curvature simplifies to
	\begin{equation}
		\label{eq3.6}
		\mathcal{F}(\mathbf{k}) = \dfrac{1}{2}\hat{\mathbf{h}}(\mathbf{k})\cdot\left[\partial_{k_{x}}\hat{\mathbf{h}}(\mathbf{k}) \times \partial_{k_{y}}\hat{\mathbf{h}}(\mathbf{k})\right],
	\end{equation}
	where $\hat{\mathbf{h}}(\mathbf{k})$ indicates the normalized vector of $\mathbf{h}(\mathbf{k})$.
	This allows one to compute the Berry curvature directly from the topological Hamiltonian.
	
\begin{figure}[tbp]
		\centering
		\includegraphics[width=0.35\linewidth]{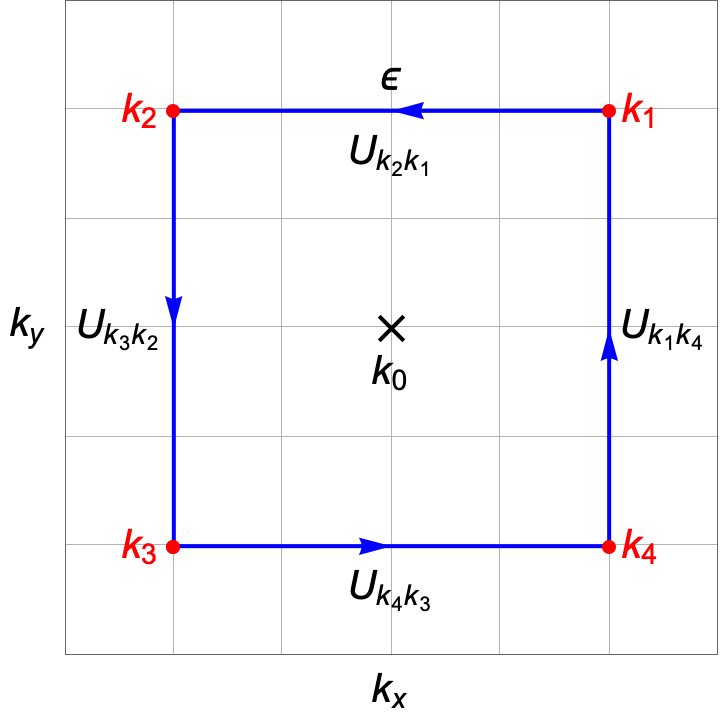}
		\caption{\label{Figure2}
			Rectangular Wilson loop used to discretize the Berry phase on an infinitesimal ($\epsilon \ll 1$) plaquette, defined to surround the reference point $\mathbf{k}_{0}$. Each edge carries an overlap function $U_{\mathbf{k}\mathbf{k}'} = \bra{\mathbf{k}}\ket{\mathbf{k}^{\prime}}$; their counter-clockwise ordered product gives the plaquette phase $\phi(\mathbf{k}_{0})$ defined in \eqref{eq3.7}.
		}
	\end{figure}

	For numerical evaluations, we discretize the Berry phase using the Wilson-loop method on an infinitesimal rectangular plaquette centered at $\mathbf{k}_{0}$ (see the figure~\ref{Figure2}).
	Given four corners $\mathbf{k}_{1}, \dots, \mathbf{k}_{4}$ ordered counterclockwise, the phase accumulation over the path is \cite{Vanderbilt_2018}
	\begin{equation}
		\label{eq3.7}
		\phi(\mathbf{k}_{0}) = \text{Im}\log(U_{\mathbf{k}_{1}\mathbf{k}_{4}}U_{\mathbf{k}_{4}\mathbf{k}_{3}}U_{\mathbf{k}_{3}\mathbf{k}_{2}}U_{\mathbf{k}_{2}\mathbf{k}_{1}}),
	\end{equation}
	where $U_{\mathbf{k}\mathbf{k}^{\prime}} \equiv \bra{\mathbf{k}}\ket{\mathbf{k}^{\prime}}$ denotes the overlap function between neighboring eigenstates.
	The Chern number is then computed by summing these local phases over all plaquettes:
	\begin{equation}
		\label{eq3.8}
		C = \dfrac{1}{2\pi}\sum_{\mathbf{k}\in\mathbb{R}^{2}}\phi(\mathbf{k}).
	\end{equation}
	Even though the integration in \eqref{eq3.4} or summation in \eqref{eq3.8} extends over all of $\mathbb{R}^2$, the resulting Chern number remains finite due to the rapid decay of the Berry curvature at large $|\mathbf{k}|$.
	
	\subsection{\label{section3.2}Critical phase}
	We first apply the topological framework to the critical case without any bulk coupling.
	In this case, the retarded Green’s function is given by
	\begin{equation}
		\label{eq3.9}
		G_{R}(\omega, \mathbf{k}) = \dfrac{\omega\sigma_{0} + \boldsymbol{\sigma}\cdot\mathbf{k}}{\sqrt{\abs{\mathbf{k}}^{2} - \omega^{2}}}.
	\end{equation}
	Using \eqref{eq3.1}, the corresponding topological Hamiltonian is
	\begin{equation}
		\label{eq3.10}
		H_{\mathrm{topo}}(\mathbf{k}) = -\dfrac{\boldsymbol{\sigma}\cdot\mathbf{k}}{\abs{\mathbf{k}}}.
	\end{equation}
	The associated Berry connection takes the form
	\begin{equation}
		\label{eq3.11}
		\mathcal{A}(\mathbf{k}) = \dfrac{1}{2\abs{\mathbf{k}}^{2}}\left(-k_{y}, k_{x}\right).
	\end{equation}
	
	Meanwhile, the Berry connection diverges at $\mathbf{k} = (0, 0)$, and so does the Berry curvature.
	This divergence arises because there is degeneracy at the gap-closing point, a feature that persists in interacting cases in the absence of a gapping parameter.
	We provide a detailed discussion of this singular behavior in Appendix~\ref{appendix:B}, along with a proposed method for defining topological invariants in gapless phases.
	
	To avoid such ambiguities, we compute the Chern numbers \textit{only when a spectral gap is present}.
	Introducing the gap resolves the degeneracies and regularizes the singularities, making the Berry curvature a smooth and finite function.
	To this end, we identify a gapping parameter and introduce this into the Dirac equation across all interaction types, as discussed in the following section.
	
	\begin{table}[t]
		\centering
		\resizebox{1.0\textwidth}{!}
		{
			\begin{tabular}[c]{| c | c | c | c |}
				\hline
				\textbf{Interaction} & \textbf{Spectral densities} & \textbf{Berry curvature ($\mathcal{B}_{5} > 0$)} & \textbf{Berry curvature ($\mathcal{B}_{5} < 0$)} \\
				\hline
				\begin{minipage}{0.20\textwidth}
					\centering
					\textbf{Pseudoscalar}\\
					$\mathcal{L}_{\mathrm{int}} = \bar{\psi}B_{5}\Gamma^{5}\psi$
				\end{minipage} & \begin{minipage}{0.5\textwidth}
					\vspace{10pt}
					\centering
					$\mathcal{B}_{5} = 1$\\
					\vspace{5pt}
					\includegraphics[width =1.0\textwidth]{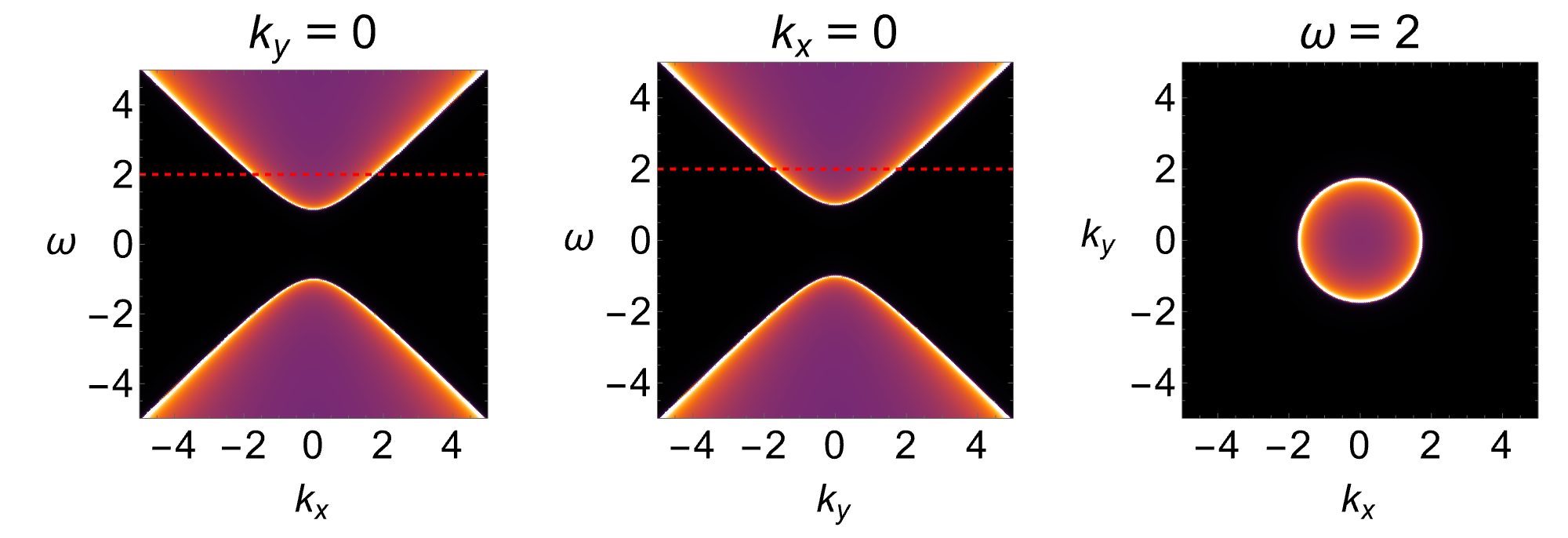}
					\vspace{0pt}
				\end{minipage} & \begin{minipage}{0.3\textwidth}
					\centering
					$\mathcal{B}_{5} = 1$\\
					\vspace{5pt}
					\includegraphics[width =1.0\textwidth]{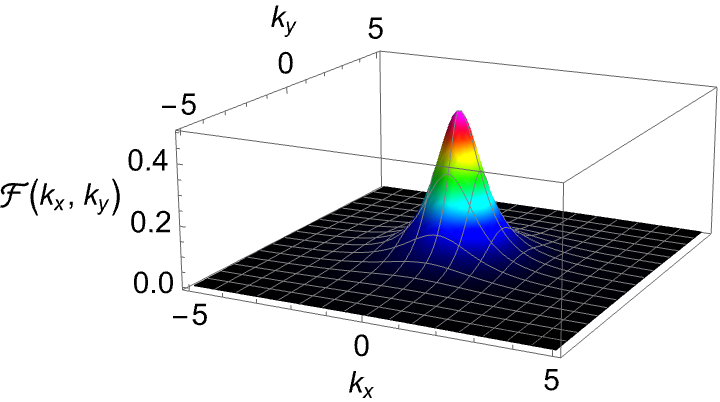}
				\end{minipage} & \begin{minipage}{0.3\textwidth}\vspace{5pt}
					\centering
					$\mathcal{B}_{5} = -1$\\
					\vspace{5pt}
					\includegraphics[width =1.0\textwidth]{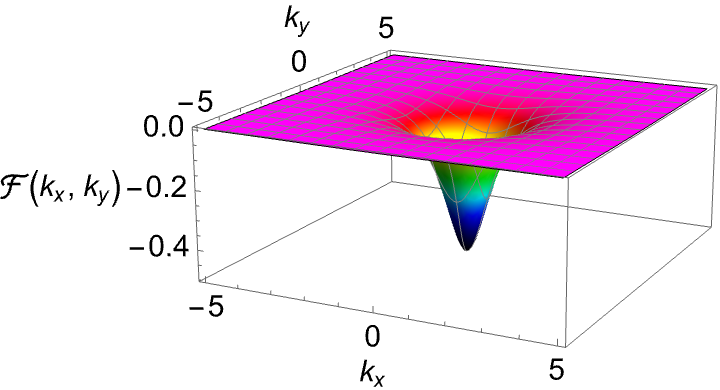}
				\end{minipage}\\
				\hline
			\end{tabular}
		}
		\caption{\label{table2}Constant-momentum and constant-frequency slices (red dotted lines) of the spectral density within the range $A(\omega, \mathbf{k}) \in [0, 10]$ are shown for the pseudoscalar coupling $\mathcal{B}_{5} = 1$. The corresponding Berry curvatures are illustrated for both $\mathcal{B}_{5} = 1$ and $\mathcal{B}_{5} = -1$.}
	\end{table}

	\begin{table}[t]\small
		\centering
		\resizebox{1.0\textwidth}{!}
		{
			\begin{tabular}[c]{| c | c | c |}
				\hline
				\begin{minipage}{0.25\textwidth}
					\vspace{5pt}
					\centering
					\large\textbf{Interaction}
					\vspace{5pt}
				\end{minipage} & \multicolumn{2}{c|}{
					\begin{minipage}{1.0\textwidth}
						\vspace{5pt}
						\centering
						\large\textbf{Retarded Green's function}
						\vspace{5pt}
					\end{minipage}
				} \\
				\hline
				\begin{minipage}{0.25\textwidth}\vspace{5pt}
					\centering
					{\large\textbf{Scalar}}\\
					$\mathcal{L}_{\mathrm{int}} = -i\bar{\psi}B\psi$
					\vspace{0pt}
				\end{minipage} & \multicolumn{2}{c|}{\begin{minipage}{1.0\textwidth}\vspace{5pt}
						\begin{equation*}
							G_{R}(\omega, \mathbf{k}) = -\dfrac{\omega\sigma_{0} + \boldsymbol{\sigma}\cdot\mathbf{k} - \mathcal{B}_{5}\sigma_{3}}{\mathcal{B} - \sqrt{\abs{\mathbf{k}}^{2} + \mathcal{B}^{2} + \mathcal{B}_{5}^{2} - \omega^{2}}}
						\end{equation*}
						\vspace{0pt}
				\end{minipage}}  \\
				\cline{1-3}
				\multirow{3}{*}{\begin{minipage}{0.25\textwidth}\vspace{5pt}
						\centering
						{\large\textbf{Vector}}\\
						$\mathcal{L}_{\mathrm{int}} = \bar{\psi}B_{M}\Gamma^{M}\psi$
						\vspace{0pt}
				\end{minipage}} & $B_{t}$ & \begin{minipage}{0.5\textwidth}\vspace{5pt}
					\begin{equation*}
						G_{R}(\omega, \mathbf{k}) = \dfrac{(\omega + \mathcal{B}_{t})\sigma_{0} + \boldsymbol{\sigma}\cdot\mathbf{k} - \mathcal{B}_{5}\sigma_{3}}{\sqrt{\abs{\mathbf{k}}^{2} + \mathcal{B}_{5}^{2} - (\omega + \mathcal{B}_{t})^{2}}}
					\end{equation*}
					\vspace{0pt}
				\end{minipage} \\
				\cline{2-3}
				& $B_{i}$ & \begin{minipage}{1.0\textwidth}\vspace{5pt}
					\begin{equation*}
						G_{R}(\omega, \mathbf{k}) = \dfrac{\omega\sigma_{0} + (k_{i} - \mathcal{B}_{i})\sigma_{i} + k_{\perp}\sigma_{\perp} - \mathcal{B}_{5}\sigma_{3}}{\sqrt{(k_{i} - \mathcal{B}_{i})^{2} + k_{\perp}^{2} + \mathcal{B}_{5}^{2} - \omega^{2}}}
					\end{equation*}\vspace{0pt}
				\end{minipage}  \\
				\cline{2-3}
				& $B_{z}$ & \begin{minipage}{1.0\textwidth}\vspace{5pt}
					\centering
					\large Identical to pseudoscalar case.\vspace{5pt}
				\end{minipage} \\
				\cline{1-3}
				\multirow{3}{*}{\begin{minipage}{0.25\textwidth}\vspace{5pt}
						\centering
						{\large\textbf{Axial vector}}\\
						$\mathcal{L}_{\mathrm{int}} = \bar{\psi}B_{5M}\Gamma^{M}\Gamma^{5}\psi$
						\vspace{0pt}
				\end{minipage}}  & $B_{5t}\Gamma^{5}$ & \begin{minipage}{1.3\textwidth}\vspace{5pt}
					\begin{equation*}
						\begin{gathered}
							G_{R}(\omega, \mathbf{k}) = \dfrac{4\abs{\mathbf{k}}\left(\omega\sigma_{0} - \mathcal{B}_{5}\sigma_{3}\right)}{(s_{+} + s_{-})\sqrt{4\abs{\mathbf{k}}^{2} - (s_{+} - s_{-})^{2}}} - \left(\dfrac{s_{+} - s_{-}}{s_{+} + s_{-}}\right)\left(\dfrac{k_{y}\sigma_{1} - k_{x}\sigma_{2}}{\abs{\mathbf{k}}}\right) + \left(\dfrac{\sqrt{4\abs{\mathbf{k}}^{2} - (s_{+} - s_{-})^{2}}}{s_{+} + s_{-}}\right)\dfrac{\boldsymbol{\sigma}\cdot\mathbf{k}}{\abs{\mathbf{k}}}\\
							\text{where}\quad s_{\pm} = \sqrt{(\abs{\mathbf{k}}\pm\mathcal{B}_{5t})^{2} + \mathcal{B}_{5}^{2} - \omega^{2}}
						\end{gathered}
					\end{equation*}\vspace{0pt}
				\end{minipage} \\
				\cline{2-3}
				& $B_{5i}\Gamma^{5}$ & \begin{minipage}{1.0\textwidth}\vspace{5pt}
					\begin{equation*}
						\begin{gathered}
							G_{R}(\omega, \mathbf{k}) = \dfrac{4[2(k_{i}^{2} + \mathcal{B}_{5}^{2})(\omega\sigma_{0} + k_{\perp}\sigma_{\perp}) - \sqrt{k_{i}^{2} + \mathcal{B}_{5}^{2}}(s_{+} - s_{-})(k_{\perp}\sigma_{0} + \omega\sigma_{\perp})]}{(s_{+} + s_{-})\left((s_{+} - s_{-})^{2} - 4\sqrt{k_{i}^{2} + \mathcal{B}_{5}^{2}}\right)} + \dfrac{2(k_{i}\sigma_{i} - \mathcal{B}_{5}\sigma_{3})}{s_{+} + s_{-}}\\
							\text{where}\quad s_{\pm} = \sqrt{\left(\mathcal{B}_{5i} - \sqrt{k_{i}^{2} + \mathcal{B}_{5}^{2}}\right)^{2} + k_{\perp}^{2} - \omega^{2}}
						\end{gathered}
					\end{equation*}\vspace{0pt}
				\end{minipage}   \\
				\cline{2-3}
				& $B_{5z}\Gamma^{5}$ & \begin{minipage}{1.3\textwidth}\vspace{5pt}
					\begin{equation*}
						\begin{gathered}
							G_{R}(\omega, \mathbf{k}) = \dfrac{\omega\sigma_{0} + \boldsymbol{\sigma}\cdot\mathbf{k}}{2}\left(\dfrac{1}{s_{+}} + \dfrac{1}{s_{-}} - \dfrac{i(s_{+} - s_{-})(\mathcal{B}_{5z} - \mathcal{B}_{5})}{s_{+}s_{-}\sqrt{\abs{\mathbf{k}}^{2} - \omega^{2}}}\right) + \dfrac{1}{2}\left[\left(\dfrac{1}{s_{+}} + \dfrac{1}{s_{-}}\right)(\mathcal{B}_{5z} - \mathcal{B}_{5}) - i\left(\dfrac{1}{s_{+}} - \dfrac{1}{s_{-}}\right)\sqrt{\abs{\mathbf{k}}^{2} - \omega^{2}}\right]\\
							\text{where}\quad
							s_{\pm} = \sqrt{\left(\sqrt{\abs{\mathbf{k}}^{2} - \omega^{2}} \pm i\mathcal{B}_{5z}\right)^{2} + \mathcal{B}_{5}^{2}}
						\end{gathered}
					\end{equation*}\vspace{0pt}
				\end{minipage}  \\
				\cline{1-3}
				\multirow{4}{*}{\begin{minipage}{0.25\textwidth}\vspace{5pt}
						\centering
						{\large\textbf{Antisymmetric 2-tensor}}\\
						\vspace{0.2cm}
						$\mathcal{L}_{\mathrm{int}} = \dfrac{1}{2}\bar{\psi}B_{MN}\Gamma^{MN}\psi$
						\vspace{0pt}
				\end{minipage}}  & $B_{ti}$ &  \begin{minipage}{1.2\textwidth}\vspace{5pt}
					\begin{equation*}
						\begin{gathered}
							G_{R}(\omega, \mathbf{k}) = -\dfrac{8(k_{i}^{2} - \omega^{2})(k_{\perp}\sigma_{\perp} - \mathcal{B}_{5}\sigma_{3}) - 4i(s_{+} - s_{-})\sqrt{k_{i}^{2} - \omega^{2}}(\mathcal{B}_{5}\sigma_{\perp} + k_{\perp}\sigma_{3})}{(s_{+} + s_{-})\left[(s_{+} - s_{-})^{2} - 4(k_{i}^{2} - \omega^{2})\right]} + \left(\dfrac{2}{s_{+} + s_{-}}\right)(\omega\sigma_{0} + k_{i}\sigma_{i})\\
							\text{where}\quad s_{\pm} = \sqrt{\left(i\mathcal{B}_{ti} \pm \sqrt{k_{i}^{2} - \omega^{2}}\right)^{2} + k_{\perp}^{2} + \mathcal{B}_{5}^{2}}
						\end{gathered}
					\end{equation*}\vspace{0pt}
				\end{minipage} \\
				\cline{2-3}
				& $B_{tz}$ &  \begin{minipage}{1.3\textwidth}\vspace{5pt}
					\begin{equation*}
						\begin{gathered}
							G_{R}(\omega, \mathbf{k}) = \left(\dfrac{(\omega + \mathcal{B}_{tz})(s_{+} + s_{-}) - (s_{+} - s_{-})\sqrt{\abs{\mathbf{k}}^{2} + \mathcal{B}_{5}^{2}}}{2s_{+}s_{-}}\right)\sigma_{0} + \dfrac{1}{2}\left(\dfrac{1}{s_{+}} + \dfrac{1}{s_{-}} - \dfrac{(s_{+} - s_{-})(\omega + \mathcal{B}_{tz})}{s_{+}s_{-}\sqrt{\abs{\mathbf{k}}^{2} + \mathcal{B}_{5}^{2}}}\right)(\boldsymbol{\sigma}\cdot\mathbf{k} - \mathcal{B}_{5}\sigma_{3})\\
							\text{where}\quad s_{\pm} = \sqrt{\left(\mathcal{B}_{tz} \pm \sqrt{\abs{\mathbf{k}}^{2} + \mathcal{B}_{5}^{2}}\right)^{2} - \mathcal{B}_{5}^{2}}
						\end{gathered}
					\end{equation*}\vspace{0pt}
				\end{minipage} \\
				\cline{2-3}
				& $B_{ij}$ &  \begin{minipage}{1.0\textwidth}\vspace{5pt}
					\begin{equation*}
						\begin{gathered}
							G_{R}(\omega, \mathbf{k}) = \dfrac{4\abs{\mathbf{k}}^{2} - (s_{+} + s_{-})^{2}}{2(\omega - \mathcal{B}_{5})(s_{+} + s_{-})}\left[\sigma_{0} - \left(\dfrac{s_{+} - s_{-}}{\abs{\mathbf{k}}}\right)\sigma_{3}\right] - \dfrac{4\mathcal{B}_{5}\abs{\mathbf{k}}}{s_{+} - s_{-} + 2\abs{\mathbf{k}}}\left(\dfrac{\sigma_{0} + \sigma_{3}}{s_{+} + s_{-}}\right) + \dfrac{2\boldsymbol{\sigma}\cdot\mathbf{k}}{s_{+} + s_{-}}\\
							\text{where}\quad s_{\pm} = \sqrt{\left(\mathcal{B}_{ij}\pm\abs{\mathbf{k}}\right)^{2} + \mathcal{B}_{5}^{2} - \omega^{2}}
						\end{gathered}
					\end{equation*}\vspace{0pt}
				\end{minipage}  \\
				\cline{2-3}
				& $B_{iz}$ & \begin{minipage}{1.3\textwidth}\vspace{5pt}
					\begin{equation*}
						G_{R}(\omega, \mathbf{k}) = \dfrac{i}{2\abs{\mathbf{k}}}\left(\dfrac{\sqrt{k_{i}^{2} - (\mathcal{B}_{iz} + i\abs{\mathbf{k}})^{2}}}{\mathcal{B}_{iz} + i\abs{\mathbf{k}} + k_{i}}\right)\left(\omega\sigma_{0} + \boldsymbol{\sigma}\cdot\mathbf{k} - \mathcal{B}_{5}\sigma_{3}\right) - \dfrac{i}{2\abs{\mathbf{k}}}\left(\dfrac{\sqrt{k_{i}^{2} - (\mathcal{B}_{iz} - i\abs{\mathbf{k}})^{2}}}{\mathcal{B}_{iz} - i\abs{\mathbf{k}} + k_{i}}\right)\left(\omega\sigma_{0} - k_{i}\sigma_{i} + k_{\perp}\sigma_{\perp} - \mathcal{B}_{5}\sigma_{3}\right)
					\end{equation*}\vspace{0pt}
				\end{minipage}  \\
				\hline
			\end{tabular}
		}
		\caption{\label{table3}
			Analytic expressions for retarded Green's function of one-flavor spinors under each interaction type in $\text{AdS}_{4}$. 
			All results are obtained by introducing additional pseudoscalar coupling $B_{5}$.
		}
	\end{table}
	\begin{table}[t]\small
		\centering
		\resizebox{1.0\textwidth}{!}
		{
			\begin{tabular}[c]{| c | c | c |}
				\hline
				\begin{minipage}{0.25\textwidth}
					\vspace{5pt}\centering
					\large\textbf{Interaction}
					\vspace{5pt}
				\end{minipage} & \multicolumn{2}{c|}{
					\begin{minipage}{1.0\textwidth}
						\vspace{5pt}\centering
						\large\textbf{Topological Hamiltonian}
						\vspace{5pt}
					\end{minipage}
				} \\
				\hline
				\begin{minipage}{0.25\textwidth}\vspace{5pt}
					\centering
					{\large\textbf{Scalar}}\\
					$\mathcal{L}_{\mathrm{int}} = -i\bar{\psi}B\psi$
					\vspace{0pt}
				\end{minipage} & \multicolumn{2}{c|}{\begin{minipage}{1.0\textwidth}\vspace{5pt}
						\begin{equation*}
							H_{\mathrm{topo}}(\mathbf{k}) = -\dfrac{\boldsymbol{\sigma}\cdot\mathbf{k} - \mathcal{B}_{5}\sigma_{3}}{\mathcal{B} + \sqrt{\abs{\mathbf{k}}^{2} + \mathcal{B}^{2} + \mathcal{B}_{5}^{2}}}
						\end{equation*}
						\vspace{0pt}
				\end{minipage}}  \\
				\cline{1-3}
				\multirow{3}{*}{\begin{minipage}{0.25\textwidth}\vspace{5pt}
						\centering
						{\large\textbf{Vector}}\\
						$\mathcal{L}_{\mathrm{int}} = \bar{\psi}B_{M}\Gamma^{M}\psi$
						\vspace{0pt}
				\end{minipage}} & $B_{t}$ & \begin{minipage}{0.5\textwidth}\vspace{5pt}
					\begin{equation*}
						H_{\mathrm{topo}}(\mathbf{k}) = \dfrac{\sigma_{0}\mathcal{B}_{t} - \boldsymbol{\sigma}\cdot\mathbf{k} + \mathcal{B}_{5}\sigma_{3}}{\sqrt{\abs{\mathbf{k}}^{2} + \mathcal{B}_{5}^{2} - \mathcal{B}_{t}^{2}}}
					\end{equation*}
					\vspace{0pt}
				\end{minipage} \\
				\cline{2-3}
				& $B_{i}$ & \begin{minipage}{1.0\textwidth}\vspace{5pt}
					\begin{equation*}
						H_{\mathrm{topo}}(\mathbf{k}) = -\dfrac{(k_{i} - \mathcal{B}_{i})\sigma_{i} + k_{\perp}\sigma_{\perp} - \mathcal{B}_{5}\sigma_{3}}{\sqrt{(k_{i} - \mathcal{B}_{i})^{2} + k_{\perp}^{2} + \mathcal{B}_{5}^{2}}}
					\end{equation*}\vspace{0pt}
				\end{minipage}  \\
				\cline{2-3}
				& $B_{z}$ & \begin{minipage}{1.0\textwidth}\vspace{5pt}
					\centering
					\large Identical to pseudoscalar case.
					\vspace{5pt}
				\end{minipage} \\
				\cline{1-3}
				\multirow{3}{*}{\begin{minipage}{0.25\textwidth}\vspace{5pt}
						\centering
						{\large\textbf{Axial vector}}\\
						$\mathcal{L}_{\mathrm{int}} = \bar{\psi}B_{5M}\Gamma^{M}\Gamma^{5}\psi$
						\vspace{0pt}
				\end{minipage}}  & $B_{5t}\Gamma^{5}$ & \begin{minipage}{1.3\textwidth}\vspace{5pt}
					\begin{equation*}
						\begin{gathered}
							H_{\mathrm{topo}}(\mathbf{k}) = \left(\dfrac{\sqrt{4\abs{\mathbf{k}}^{2} - (s_{+} - s_{-})^{2}}}{s_{+} + s_{-}}\right)\dfrac{\boldsymbol{\sigma}\cdot\mathbf{k}}{\abs{\mathbf{k}}} + \left(\dfrac{s_{+} - s_{-}}{s_{+} + s_{-}}\right)\left(\dfrac{k_{y}\sigma_{1} - k_{x}\sigma_{2}}{\abs{\mathbf{k}}}\right) + \left(\dfrac{4\mathcal{B}_{5}\abs{\mathbf{k}}}{(s_{+} + s_{-})\sqrt{4\abs{\mathbf{k}}^{2} - (s_{+} - s_{-})^{2}}}\right)\sigma_{3}\\
							\text{where}\quad s_{\pm} = \sqrt{(\abs{\mathbf{k}}\pm\mathcal{B}_{5t})^{2} + \mathcal{B}_{5}^{2}}
						\end{gathered}
					\end{equation*}\vspace{0pt}
				\end{minipage} \\
				\cline{2-3}
				& $B_{5i}\Gamma^{5}$ & \begin{minipage}{1.0\textwidth}\vspace{5pt}
					\begin{equation*}
						\begin{gathered}
							H_{\mathrm{topo}}(\mathbf{k}) = \dfrac{4k_{\perp}\sqrt{k_{i}^{2} + \mathcal{B}_{5}^{2}}\left[(s_{-} - s_{+})\sigma_{0} + 2\sqrt{k_{x}^{2} + \mathcal{B}_{5}^{2}}\sigma_{\perp}\right]}{(s_{+} + s_{-})\left[(s_{+} - s_{-})^{2} - 4(k_{i}^{2} + \mathcal{B}_{5}^{2})\right]} - \left(\dfrac{2}{s_{+} + s_{-}}\right)(k_{i}\sigma_{i} - \mathcal{B}_{5}\sigma_{3})\\
							\text{where}\quad s_{\pm} = \sqrt{\left(\mathcal{B}_{5i} \pm \sqrt{k_{i}^{2} + \mathcal{B}_{5}^{2}}\right)^{2} + k_{\perp}^{2}}
						\end{gathered}
					\end{equation*}\vspace{0pt}
				\end{minipage}   \\
				\cline{2-3}
				& $B_{5z}\Gamma^{5}$ & \begin{minipage}{1.0\textwidth}\vspace{5pt}
					\begin{equation*}
						\begin{gathered}
							H_{\mathrm{topo}}(\mathbf{k}) = \dfrac{\boldsymbol{\sigma}\cdot\mathbf{k}}{2\abs{\mathbf{k}}}\left(\dfrac{i\abs{\mathbf{k}} + \mathcal{B}_{5z} + \mathcal{B}_{5}}{s_{+}} + \dfrac{i\abs{\mathbf{k}} - \mathcal{B}_{5z} - \mathcal{B}_{5}}{s_{-}}\right) - \dfrac{i}{2}\left(\dfrac{i\abs{\mathbf{k}} + \mathcal{B}_{5z} + \mathcal{B}_{5}}{s_{+}} - \dfrac{i\abs{\mathbf{k}} - \mathcal{B}_{5z} - \mathcal{B}_{5}}{s_{-}}\right)\sigma_{3}\\
							\text{where}\quad
							s_{\pm} = \sqrt{(\mathcal{B}_{5z} \pm i\abs{\mathbf{k}})^{2} - \mathcal{B}_{5}^{2}}
						\end{gathered}
					\end{equation*}\vspace{0pt}
				\end{minipage}  \\
				\cline{1-3}
				\multirow{4}{*}{\begin{minipage}{0.25\textwidth}\vspace{5pt}
						\centering
						{\large\textbf{Antisymmetric 2-tensor}}\\
						\vspace{0.2cm}
						$\mathcal{L}_{\mathrm{int}} = \dfrac{1}{2}\bar{\psi}B_{MN}\Gamma^{MN}\psi$
						\vspace{0pt}
				\end{minipage}}  & $B_{ti}$ &  \begin{minipage}{1.0\textwidth}\vspace{5pt}
					\begin{equation*}
						\begin{gathered}
							H_{\mathrm{topo}}(\mathbf{k}) = \dfrac{8k_{i}^{2}(k_{\perp}\sigma_{\perp} - \mathcal{B}_{5}\sigma_{3}) - 4i(s_{+} - s_{-})k_{i}(\mathcal{B}_{5}\sigma_{\perp} + k_{\perp}\sigma_{3})}{(s_{+} + s_{-})\left[(s_{+} - s_{-})^{2} - 4k_{i}^{2}\right]} - \dfrac{2k_{i}\sigma_{i}}{s_{+} + s_{-}}\\
							\text{where}\quad s_{\pm} = \sqrt{(k_{i}\pm i\mathcal{B}_{ti})^{2} + k_{\perp}^{2} + \mathcal{B}_{5}^{2}}
						\end{gathered}
					\end{equation*}\vspace{0pt}
				\end{minipage} \\
				\cline{2-3}
				& $B_{tz}$ &  \begin{minipage}{1.3\textwidth}\vspace{5pt}
					\centering
					\large Identical to pseudoscalar case.
					\vspace{5pt}
				\end{minipage} \\
				\cline{2-3}
				& $B_{ij}$ &  \begin{minipage}{1.0\textwidth}\vspace{5pt}
					\begin{equation*}
						\begin{gathered}
							H_{\mathrm{topo}}(\mathbf{k}) = \left(\dfrac{2\mathcal{B}_{ij}\mathcal{B}_{5}}{\abs{\mathbf{k}}^{2} - \mathcal{B}_{ij}^{2} + \mathcal{B}_{5}^{2} + s_{+}s_{-}}\right)\sigma_{0} - \dfrac{2\boldsymbol{\sigma}\cdot\mathbf{k}}{s_{+} + s_{-}} + \dfrac{2\abs{\mathbf{k}}}{\mathcal{B}_{5}}\left(\dfrac{\mathcal{B}_{ij}}{s_{+} - s_{-}} + \dfrac{\abs{\mathbf{k}}}{s_{+} + s_{-}}\right)\sigma_{3}\\
							\text{where}\quad s_{\pm} = \sqrt{\left(\abs{\mathbf{k}}\pm\mathcal{B}_{ij}\right)^{2} + \mathcal{B}_{5}^{2}}
						\end{gathered}
					\end{equation*}\vspace{0pt}
				\end{minipage}  \\
				\cline{2-3}
				& $B_{iz}$ & \begin{minipage}{1.0\textwidth}\vspace{5pt}
					\begin{equation*}
						\begin{gathered}
							H_{\mathrm{topo}}(\mathbf{k}) = -\left(\dfrac{\mathrm{sgn}(k_{i})\sqrt{4k_{i}^{2} - (s_{+} - s_{-})^{2}}}{2s_{-}}\right)\sigma_{i} - \dfrac{2k_{i}\mathrm{sgn}(k_{i})}{s_{-}\sqrt{4k_{i}^{2} - (s_{+} - s_{-})^{2}}}(k_{\perp}\sigma_{\perp} - \mathcal{B}_{5}\sigma_{3})\\
							\text{where}\quad 
							s_{\pm} = \sqrt{(\mathcal{B}_{iz}\pm k_{i})^{2} + k_{\perp}^{2} + \mathcal{B}_{5}^{2}}
						\end{gathered}
					\end{equation*}\vspace{0pt}
				\end{minipage}  \\
				\hline
			\end{tabular}
		}
		\caption{\label{table4}
			Analytic expressions for topological Hamiltonian of one-flavor spinors under each interaction type in $\text{AdS}_{4}$. All results are obtained from gapped spectral densities by introducing a pseudoscalar coupling $B_{5}$.
		}
	\end{table}
	\begin{table}[t]\small
		\centering
		\resizebox{1.0\textwidth}{!}
		{
			\begin{tabular}[c]{| c | c | c | c |}
				\hline
				\begin{minipage}{0.25\textwidth}
					\vspace{5pt}\centering
					\large\textbf{Interaction}
					\vspace{5pt}
				\end{minipage} & \multicolumn{2}{c|}{
					\begin{minipage}{1.0\textwidth}
						\vspace{5pt}\centering
						\large\textbf{Berry curvature}
						\vspace{5pt}
					\end{minipage}
				} & \begin{minipage}{0.21\textwidth}
					\vspace{5pt}\centering
					\large\textbf{Chern number}
					\vspace{5pt}
				\end{minipage} \\
				\hline
				\begin{minipage}{0.25\textwidth}\vspace{5pt}
					\centering
					{\large\textbf{Scalar}}\\
					$\mathcal{L}_{\mathrm{int}} = -i\bar{\psi}B\psi$
					\vspace{5pt}
				\end{minipage} & \multicolumn{2}{c|}{\begin{minipage}{1.0\textwidth}\vspace{5pt}
						\centering
						\large Identical to pseudoscalar case.
						\vspace{5pt}
				\end{minipage}} & \multirow{35}{*}{\begin{minipage}{0.21\textwidth}\vspace{5pt}
						\begin{equation*}
							C = \dfrac{\mathrm{sgn}(\mathcal{B}_{5})}{2}
						\end{equation*}\vspace{0pt}
				\end{minipage}} \\
				\cline{1-3}
				\multirow{3}{*}{\begin{minipage}{0.25\textwidth}\vspace{5pt}
						\centering
						{\large\textbf{Vector}}\\
						$\mathcal{L}_{\mathrm{int}} = \bar{\psi}B_{M}\Gamma^{M}\psi$
						\vspace{0pt}
				\end{minipage}} & $B_{t}$ & \begin{minipage}{0.5\textwidth}\vspace{5pt}
					\centering
					\large Identical to pseudoscalar case.
					\vspace{5pt}
				\end{minipage} &  \\
				\cline{2-3}
				& $B_{i}$ & \begin{minipage}{0.9\textwidth}\vspace{5pt}
					\begin{equation*}
						\mathcal{F}(\mathbf{k}) = \dfrac{\mathcal{B}_{5}}{2\left((k_{i} - \mathcal{B}_{i})^{2} + k_{\perp}^{2} + {\mathcal{B}_{5}}^{2}\right)^{3/2}}
					\end{equation*}\vspace{0pt}
				\end{minipage} &  \\
				\cline{2-3}
				& $B_{z}$ & \begin{minipage}{0.9\textwidth}\vspace{5pt}
					\centering
					\large Identical to pseudoscalar case.
					\vspace{5pt}
				\end{minipage} &  \\
				\cline{1-3}
				\multirow{3}{*}{\begin{minipage}{0.25\textwidth}\vspace{5pt}
						\centering
						{\large\textbf{Axial vector}}\\
						$\mathcal{L}_{\mathrm{int}} = \bar{\psi}B_{5M}\Gamma^{M}\Gamma^{5}\psi$
						\vspace{0pt}
				\end{minipage}}  & $B_{5t}\Gamma^{5}$ & \begin{minipage}{1.0\textwidth}\vspace{5pt}
					\begin{equation*}
						\mathcal{F}(\mathbf{k}) = \dfrac{2\mathcal{B}_{5}\abs{\mathbf{k}}}{s_{+}s_{-}(s_{+} + s_{-})\sqrt{4\abs{\mathbf{k}}^{2} - (s_{+} - s_{-})^{2}}}\quad\text{where}\quad s_{\pm} = \sqrt{(\abs{\mathbf{k}} \pm \mathcal{B}_{5t})^{2} + \mathcal{B}_{5}^{2}}
					\end{equation*}\vspace{0pt}
				\end{minipage} &  \\
				\cline{2-3}
				& $B_{5i}\Gamma^{5}$ & \begin{minipage}{0.9\textwidth}\vspace{5pt}
					\begin{equation*}
						\begin{gathered}
							\mathcal{F}(\mathbf{k}) = -\dfrac{\mathcal{B}_{5}\sqrt{4(k_{i}^{2} + \mathcal{B}_{5}^{2}) - (s_{+} - s_{-})^{2}}\left[(s_{+}^{2} - s_{-}^{2})^{2} - 4(k_{i}^{2} + \mathcal{B}_{5}^{2})(s_{+}^{2} + s_{-}^{2})\right]}{{32(k_{i} + \mathcal{B}_{5})^{3}(s_{+}s_{-})^{5/2}}}\\
							\text{where}\quad s_{\pm} = \sqrt{\left(\mathcal{B}_{5i}\pm\sqrt{k_{i}^{2} + \mathcal{B}_{5}^{2}}\right)^{2} + k_{\perp}^{2}}
						\end{gathered}
					\end{equation*}\vspace{0pt}
				\end{minipage} &  \\
				\cline{2-3}
				& $B_{5z}\Gamma^{5}$ & \begin{minipage}{1.15\textwidth}\vspace{5pt}
					\begin{equation*}
						\begin{gathered}
							\mathcal{F}(\mathbf{k}) = \dfrac{\mathcal{B}_{5z}\mathcal{B}_{5}(\abs{\mathbf{k}}^{2} - \mathcal{B}_{5z}^{2} + \mathcal{B}_{5}^{2})\sqrt{\abs{\mathbf{k}}^{2} + (\mathcal{B}_{5z} - \mathcal{B}_{5})^{2}}}{(s_{+}s_{-})^{7/2}}\left(\dfrac{\mathcal{B}_{5z} + \mathcal{B}_{5}}{s_{+} + s_{-}} - \dfrac{i\abs{\mathbf{k}}}{s_{+} - s_{-}}\right)\\
							\text{where}\quad
							s_{\pm} = \sqrt{(\mathcal{B}_{5z} \pm i\abs{\mathbf{k}})^{2} + \mathcal{B}_{5}^{2}}
						\end{gathered}
					\end{equation*}\vspace{0pt}
				\end{minipage} &  \\
				\cline{1-3}
				\multirow{4}{*}{\begin{minipage}{0.25\textwidth}\vspace{5pt}
						\centering
						{\large\textbf{Antisymmetric 2-tensor}}\\
						\vspace{0.2cm}
						$\mathcal{L}_{\mathrm{int}} = \dfrac{1}{2}\bar{\psi}B_{MN}\Gamma^{MN}\psi$
						\vspace{0pt}
				\end{minipage}}  & $B_{ti}$ &  \begin{minipage}{0.9\textwidth}\vspace{5pt}
					\begin{equation*}
						\begin{gathered}
							\mathcal{F}(\mathbf{k}) = \dfrac{\mathcal{B}_{ti}\mathcal{B}_{5}(s_{+} + s_{-}) + i\mathcal{B}_{5}(s_{+} - s_{-})k_{i} - k_{\perp}(\mathcal{B}_{ti}^{2} - s_{+}s_{-} + \abs{\mathbf{k}}^{2}) - \mathcal{B}_{5}^{2}k_{\perp}}{4\mathcal{B}_{ti}s_{+}s_{-}(k_{\perp}^{2} + \mathcal{B}_{5})}\\
							\text{where}\quad
							s_{\pm} = \sqrt{(k_{i} \pm i\mathcal{B}_{ti})^{2} + k_{\perp}^{2} + \mathcal{B}_{5}^{2}}
						\end{gathered}
					\end{equation*}\vspace{0pt}
				\end{minipage} & \\
				\cline{2-3}
				& $B_{tz}$ &  \begin{minipage}{0.9\textwidth}\vspace{5pt}
					\centering
					\large Identical to pseudoscalar case.
					\vspace{5pt}
				\end{minipage} & \\
				\cline{2-3}
				& $B_{ij}$ &  \begin{minipage}{0.9\textwidth}\vspace{5pt}
					\begin{equation*}
						\begin{gathered}
							\mathcal{F}(\mathbf{k}) = \dfrac{4\sqrt{2}\mathcal{B}_{ij}^{2}\mathcal{B}_{5}(\mathcal{B}_{ij}^{2} + \mathcal{B}_{5}^{2} + s_{+}s_{-} - \abs{\mathbf{k}}^{2})(\abs{\mathbf{k}}^{2} + \mathcal{B}_{ij}^{2} + \mathcal{B}_{5}^{2})\abs{\mathcal{B}_{ij}}\abs{\mathcal{B}_{5}}}{s_{+}s_{-}(s_{+} + s_{-})^{3}\left[\mathcal{B}_{5}^{4} + \mathcal{B}_{5}^{2}(2\mathcal{B}_{ij}^{2} - s_{+}s_{-} + 2\abs{\mathbf{k}}^{2}) + (\mathcal{B}_{ij}^{2} - \mathcal{B}_{5}^{2})(\mathcal{B}_{ij}^{2} + s_{+}s_{-} - \abs{\mathbf{k}}^{2})\right]^{3/2}}
							\\
							\text{where}\quad s_{\pm} = \sqrt{(\abs{\mathbf{k}}\pm\mathcal{B}_{ij})^{2} + {\mathcal{B}_{5}}^{2}}
						\end{gathered}
					\end{equation*}\vspace{0pt}
				\end{minipage} &  \\
				\cline{2-3}
				& $B_{iz}$ & \begin{minipage}{0.9\textwidth}\vspace{5pt}
					\begin{equation*}
						\mathcal{F}(\mathbf{k}) = \dfrac{\mathcal{B}_{5}(s_{+}^{2} + s_{-}^{2})\abs{k_{i}}}{2(s_{+}s_{-})^{5/2}\sqrt{4k_{i}^{2} - (s_{+} - s_{-})^{2}}}\quad\text{where}\quad s_{\pm} = \sqrt{(k_{i}\pm\mathcal{B}_{iz})^{2} + k_{\perp}^{2} + \mathcal{B}_{5}^{2}}
					\end{equation*}\vspace{0pt}
				\end{minipage} &  \\
				\hline
			\end{tabular}
		}
		\caption{\label{table5}
			Analytic expressions for the Berry curvature and Chern number of one-flavor spinors under each interaction type in $\text{AdS}_{4}$. All results are obtained from gapped spectral densities by introducing a pseudoscalar coupling $B_{5}$.
		}
	\end{table}

	\subsection{\label{section3.3}Gapped phase}
	We begin by considering a pseudoscalar coupling of the form $B_{5}(z) = \mathcal{B}_{5} z$, the fundamental example representing the universal topology of the system in the presence of a spectral gap.
	The corresponding retarded Green’s function computed by holography  is
	\begin{equation}
		\label{eq3.12}
		G_{R}(\omega, \mathbf{k}) = \dfrac{\sigma_{0}\omega + \boldsymbol{\sigma}\cdot\mathbf{k} - \mathcal{B}_{5}\sigma_{3}}{\sqrt{\abs{\mathbf{k}}^{2} + \mathcal{B}_{5}^{2} - \omega^{2}}}.
	\end{equation}
	The resulting spectral density exhibits a direct gap as it is shown in the table~\ref{table2}.
	From \eqref{eq3.1}, the corresponding topological Hamiltonian is
	\begin{equation}
		\label{eq3.13}
		H_{\mathrm{topo}}(\mathbf{k}) = -\dfrac{\boldsymbol{\sigma}\cdot\mathbf{k} - \mathcal{B}_{5}\sigma_{3}}{\sqrt{\abs{\mathbf{k}}^{2} + \mathcal{B}_{5}^{2}}}.
	\end{equation}
	The associated Berry connection is given by
	\begin{equation}
		\label{eq3.14}
		\mathcal{A}(\mathbf{k}) = \dfrac{1}{2\abs{\mathbf{k}}^{2}}\left(1 - \dfrac{\mathcal{B}_{5}}{\sqrt{\abs{\mathbf{k}}^{2} + \mathcal{B}_{5}^{2}}}\right)\left(-k_{y}, k_{x}\right).
	\end{equation}
	This Berry connection yields finite and smooth Berry curvature, which is given by
	\begin{equation}
		\label{eq3.15}
		\mathcal{F}(\mathbf{k}) = \dfrac{\mathcal{B}_{5}}{2\left(\abs{\mathbf{k}}^{2} + \mathcal{B}_{5}^{2}\right)^{3/2}}.
	\end{equation}
	The curvature profiles are illustrated in the table~\ref{table2}.
	The corresponding Chern number is a quantized fractional number:
	\begin{equation}
		\label{eq3.16}
		C = \dfrac{1}{2}\mathrm{sgn}(\mathcal{B}_{5}).
	\end{equation}
	
	The scalar coupling with negative sign ($\mathcal{B} < 0$) also opens a spectral gap \cite{Oh:2021p}. 
	In this case, the Green’s function becomes
	\begin{equation}
		\label{eq3.17}
		G_{R}(\omega, \mathbf{k}) = -\dfrac{\omega\sigma_{0} + \boldsymbol{\sigma}\cdot\mathbf{k}}{\mathcal{B} - \sqrt{\abs{\mathbf{k}}^{2} + \mathcal{B}^{2} - \omega^{2}}},
	\end{equation}
	which has the same matrix structure as in the critical case, so that the associated Berry connection is identical to the critical case. 
	For this reason, we identify only the pseudoscalar coupling $\mathcal{B}_{5}$ as the genuine gapping parameter responsible for topological phases.
	
	Because the other interaction cases yield gapless spectra, we introduce additional pseudoscalar coupling into the interaction action, which is then modified to
	\begin{equation}
		\label{eq3.18}
		S_{\mathrm{int}} = i\int_{\mathcal{M}} d^{4}x \sqrt{-g} \left( \bar{\psi} B_{I} \Gamma^{I} \psi - i \bar{\psi} B_{5} \Gamma^{5} \psi \right),
	\end{equation}
	where $B_{I}$ denotes the remaining 15 couplings.
	Analytic expressions for the retarded Green’s function, the topological Hamiltonian, and the Berry curvature in the gapped phase for each interaction type are summarized in the table~\ref{table3}, \ref{table4}, and~\ref{table5}, respectively.
	Here, $k_{i}$ denotes the momentum component along the interaction direction, while $k_{\perp}$ denotes the component perpendicular to it.
	The corresponding plots of gapped spectral densities and Berry curvatures are listed in the table~\ref{table6}.
	\begin{table}[p]\small
		\centering
		\resizebox{1.0\textwidth}{!}
		{
			\begin{tabular}[c]{| c | c | c | c | c |}
				\hline
				\begin{minipage}{0.25\textwidth}
					\vspace{5pt}\centering
					\large\textbf{Interaction}
					\vspace{5pt}
				\end{minipage} & \multicolumn{2}{c|}{
					\begin{minipage}{0.5\textwidth}
						\vspace{5pt}\centering
						\large\textbf{Spectral densities}
						\vspace{5pt}
					\end{minipage}
				} & \begin{minipage}{0.35\textwidth}
					\vspace{5pt}\centering
					\large\textbf{Berry curvature ($\mathcal{B}_{5} > 0$)}
					\vspace{5pt}
				\end{minipage} & \begin{minipage}{0.35\textwidth}
					\vspace{5pt}\centering
					\large\textbf{Berry curvature ($\mathcal{B}_{5} < 0$)}
					\vspace{5pt}
				\end{minipage}\\
				\hline
				\begin{minipage}{0.25\textwidth}\vspace{5pt}
					\centering
					{\large\textbf{Scalar}}\\
					$\mathcal{L}_{\mathrm{int}} = -i\bar{\psi}B\psi$
					\vspace{0pt}
				\end{minipage} & \multicolumn{2}{c|}{
					\begin{minipage}{0.45\textwidth}
						\vspace{10pt}
						\centering
						\footnotesize
						$\mathcal{B} = 2$ and $\mathcal{B}_{5} = 1$\\
						\vspace{3pt}
						\includegraphics[width =1.0\textwidth]{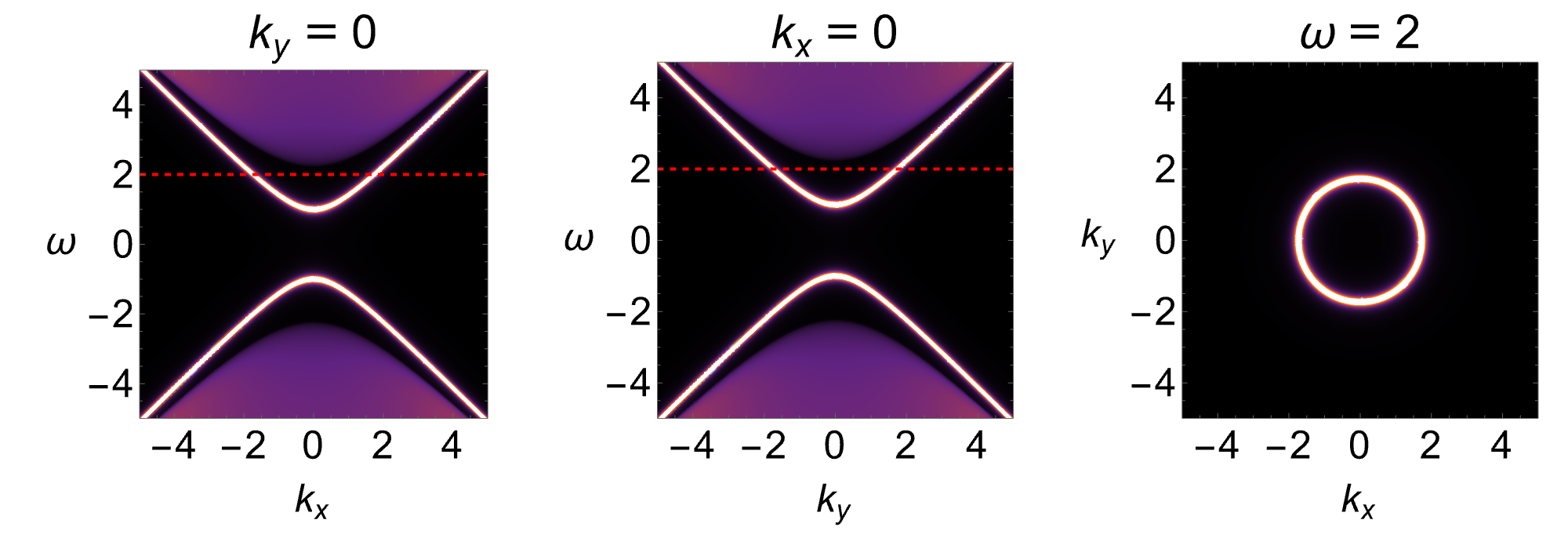}
					\end{minipage}
				} & \multicolumn{2}{c|}{\large Identical to pseudoscalar case.}\\
				\hline
				\multirow{3}{*}{\begin{minipage}{0.25\textwidth}\vspace{5pt}
						\centering
						{\large\textbf{Vector}}\\
						$\mathcal{L}_{\mathrm{int}} = \bar{\psi}B_{M}\Gamma^{M}\psi$
						\vspace{0pt}
				\end{minipage}} & $B_{t}$ & \begin{minipage}{0.45\textwidth}
					\vspace{10pt}
					\centering
					\footnotesize
					$\mathcal{B}_{t} = 1$ and $\mathcal{B}_{5} = 2$\\
					\vspace{3pt}
					\includegraphics[width =1.0\textwidth]{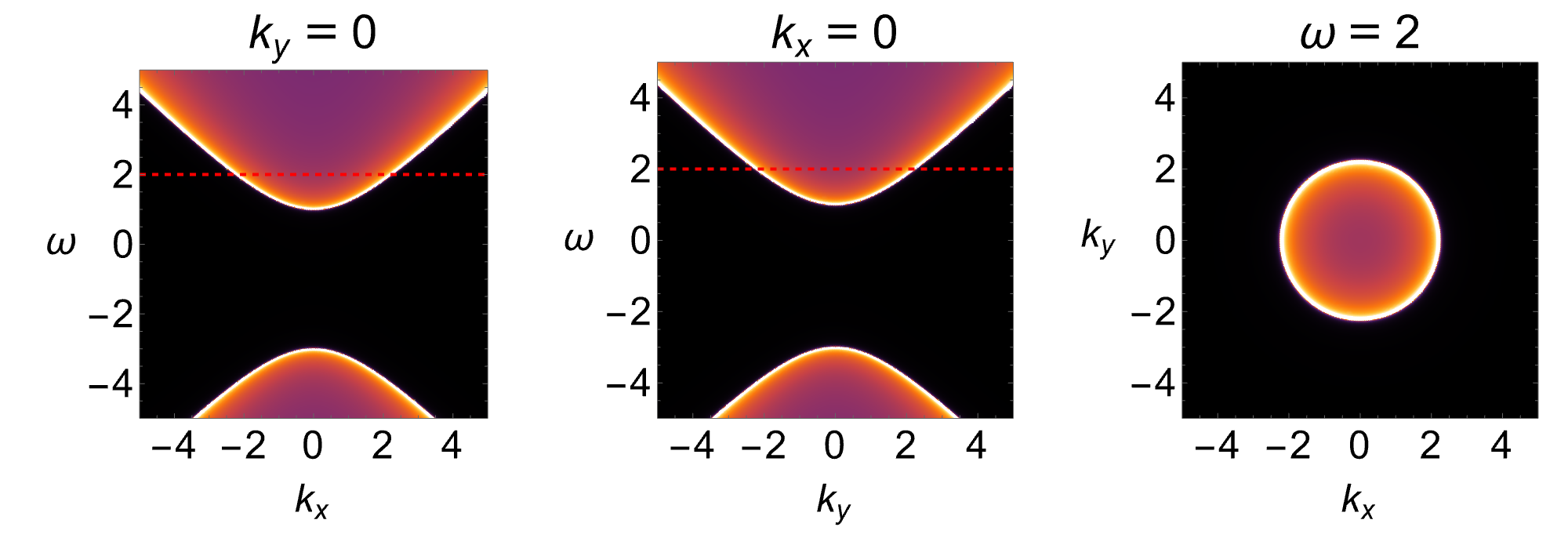}
				\end{minipage} & \multicolumn{2}{c|}{\begin{minipage}{0.5\textwidth}
						\centering
						\large Identical to pseudoscalar case.
				\end{minipage}}\\
				\cline{2-5}
				& $B_{i}$ & \begin{minipage}{0.45\textwidth}
					\vspace{10pt}
					\centering
					\footnotesize
					$\mathcal{B}_{x} = 2$ and $\mathcal{B}_{5} = 1$\\
					\vspace{3pt}
					\includegraphics[width =1.0\textwidth]{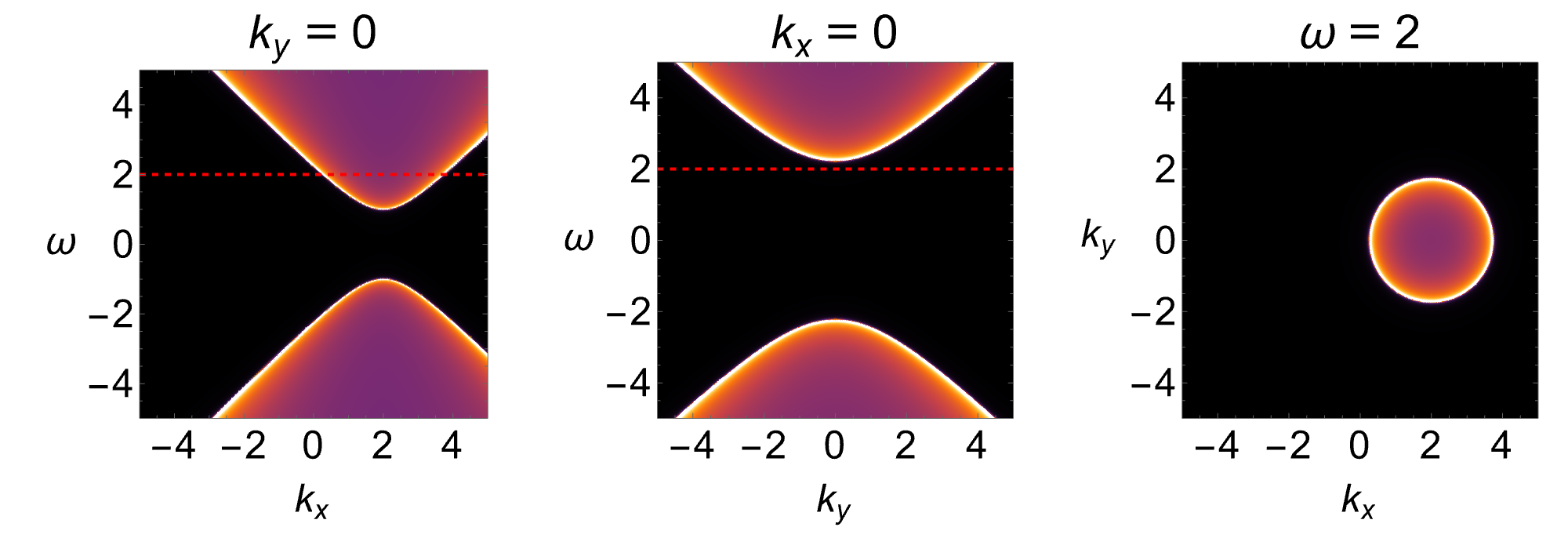}
				\end{minipage} & \begin{minipage}{0.25\textwidth}
					\centering
					\footnotesize
					$\mathcal{B}_{x} = 2$ and $\mathcal{B}_{5} = 1$\\
					\vspace{5pt}
					\includegraphics[width =1.0\textwidth]{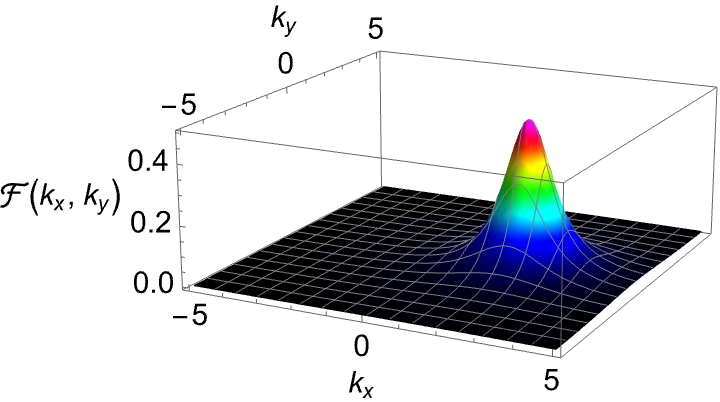}
				\end{minipage} & \begin{minipage}{0.25\textwidth}
					\centering
					\footnotesize
					$\mathcal{B}_{x} = 2$ and $\mathcal{B}_{5} = -1$\\
					\vspace{5pt}
					\includegraphics[width =1.0\textwidth]{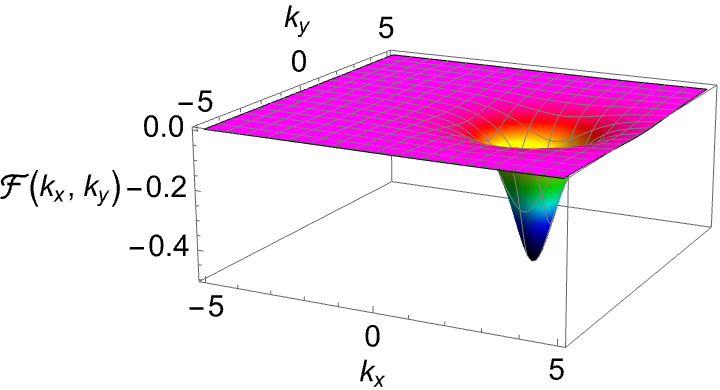}
				\end{minipage} \\
				\cline{2-5}
				& $B_{z}$ & \multicolumn{3}{c|}{\begin{minipage}{1.0\textwidth}\vspace{5pt}
						\centering
						\large Identical to pseudoscalar case.\vspace{5pt}
				\end{minipage}} \\
				\hline
				\multirow{3}{*}{\begin{minipage}{0.25\textwidth}\vspace{5pt}
						\centering
						{\large\textbf{Axial vector}}\\
						$\mathcal{L}_{\mathrm{int}} = \bar{\psi}B_{5M}\Gamma^{M}\Gamma^{5}\psi$
						\vspace{0pt}
				\end{minipage}} & $B_{5t}$ & \begin{minipage}{0.45\textwidth}
					\vspace{10pt}
					\centering
					\footnotesize
					$\mathcal{B}_{5t} = 2$ and $\mathcal{B}_{5} = 0.5$\\
					\vspace{3pt}
					\includegraphics[width =1.0\textwidth]{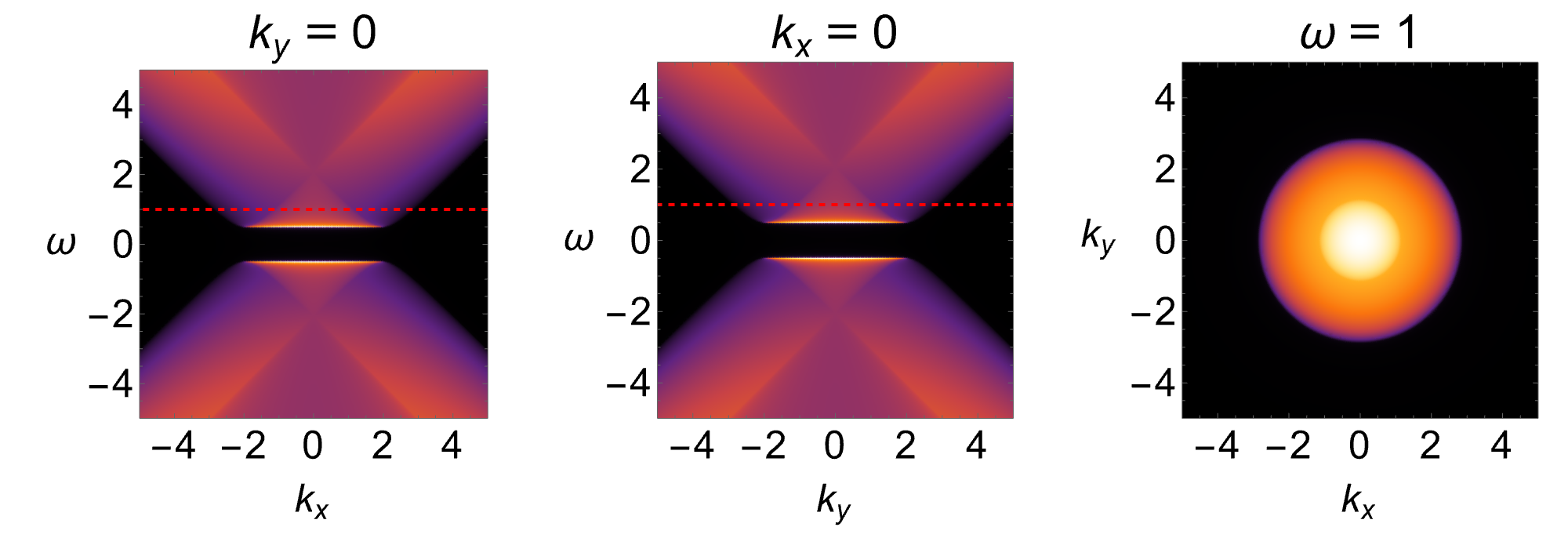}
				\end{minipage} & \begin{minipage}{0.25\textwidth}
					\centering
					\footnotesize
					$\mathcal{B}_{5t} = 2$ and $\mathcal{B}_{5} = 0.5$\\
					\vspace{5pt}
					\includegraphics[width =1.0\textwidth]{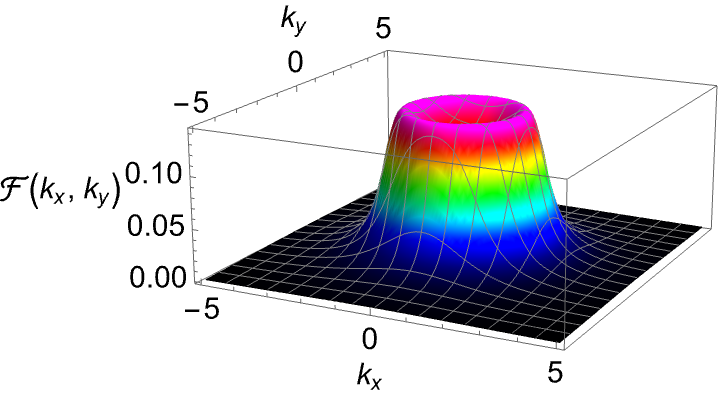}
				\end{minipage} & \begin{minipage}{0.25\textwidth}
					\centering
					\footnotesize
					$\mathcal{B}_{5t} = 2$ and $\mathcal{B}_{5} = -0.5$\\
					\vspace{5pt}
					\includegraphics[width =1.0\textwidth]{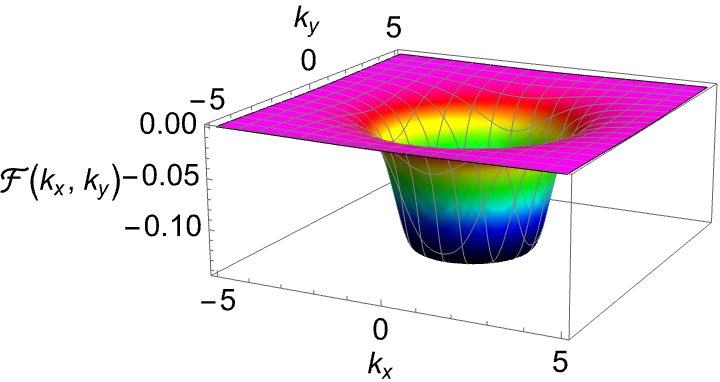}
				\end{minipage} \\
				\cline{2-5}
				& $B_{5i}$ & \begin{minipage}{0.45\textwidth}
					\vspace{10pt}
					\centering
					\footnotesize
					$\mathcal{B}_{5x} = 2$ and $\mathcal{B}_{5} = 2.5$\\
					\vspace{3pt}
					\includegraphics[width =1.0\textwidth]{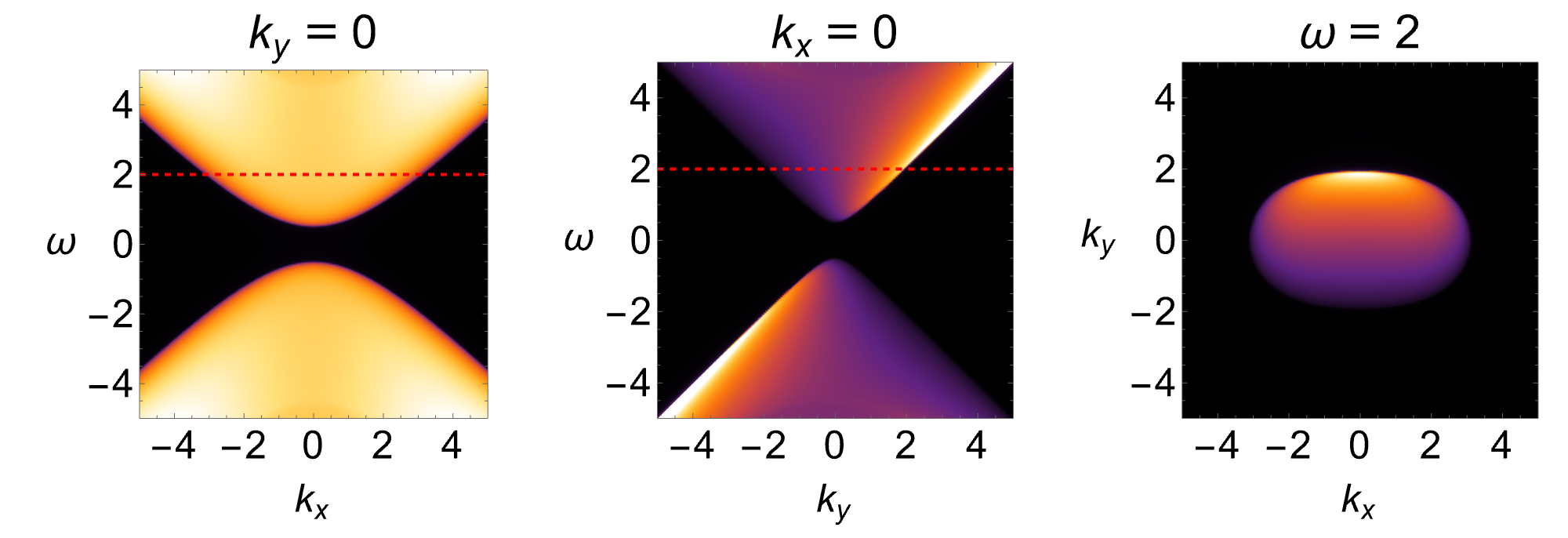}
				\end{minipage} & \begin{minipage}{0.25\textwidth}
					\centering
					\footnotesize
					$\mathcal{B}_{5x} = 2$ and $\mathcal{B}_{5} = 2.5$\\
					\vspace{5pt}
					\includegraphics[width =1.0\textwidth]{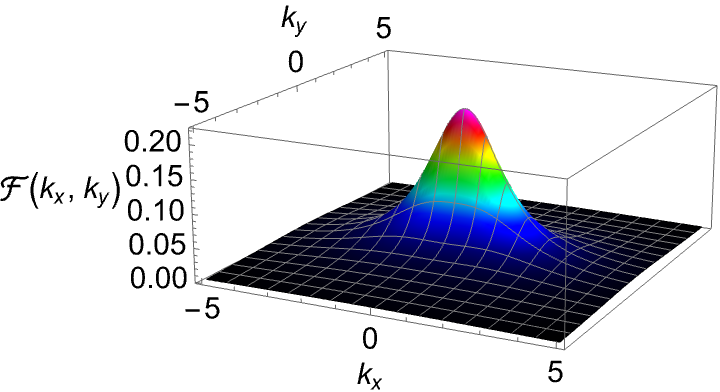}
				\end{minipage} & \begin{minipage}{0.25\textwidth}
					\centering
					\footnotesize
					$\mathcal{B}_{5x} = 2$ and $\mathcal{B}_{5} = -2.5$\\
					\vspace{5pt}
					\includegraphics[width =1.0\textwidth]{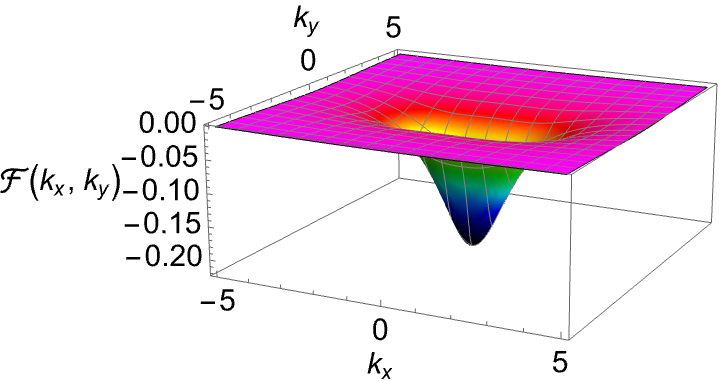}
				\end{minipage} \\
				\cline{2-5}
				& $B_{5z}$ & \begin{minipage}{0.45\textwidth}
					\vspace{10pt}
					\centering
					\footnotesize
					$\mathcal{B}_{5z} = 1$ and $\mathcal{B}_{5} = 1.5$\\
					\vspace{3pt}
					\includegraphics[width =1.0\textwidth]{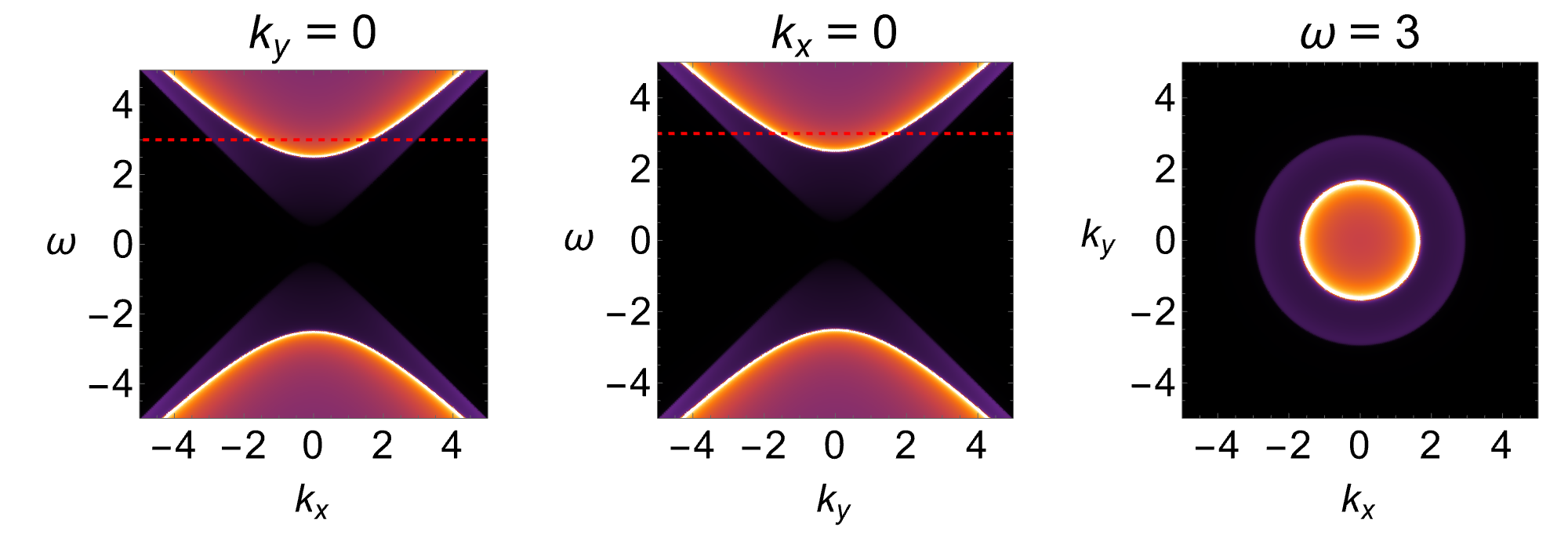}
				\end{minipage} & \begin{minipage}{0.25\textwidth}
					\centering
					\footnotesize
					$\mathcal{B}_{5z} = 1$ and $\mathcal{B}_{5} = 1.5$\\
					\vspace{5pt}
					\includegraphics[width =1.0\textwidth]{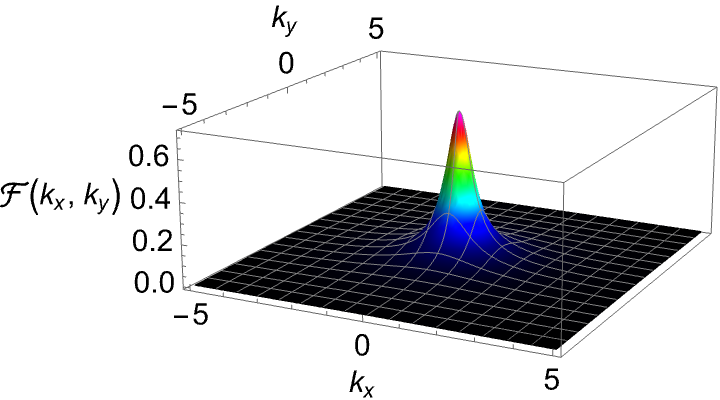}
				\end{minipage} & \begin{minipage}{0.25\textwidth}
					\centering
					\footnotesize
					$\mathcal{B}_{5z} = 1$ and $\mathcal{B}_{5} = -1.5$\\
					\vspace{5pt}
					\includegraphics[width =1.0\textwidth]{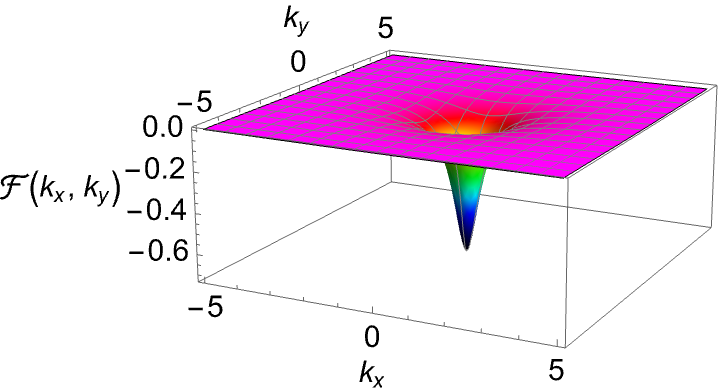}
				\end{minipage} \\
				\hline
				\multirow{4}{*}{\begin{minipage}{0.25\textwidth}
						\centering
						{\large\textbf{Antisymmetric 2-tensor}}\\[5pt]
						$\mathcal{L}_{\mathrm{int}} = \dfrac{1}{2}\bar{\psi}B_{MN}\Gamma^{MN}\psi$
				\end{minipage}}  & $B_{ti}$ &  \begin{minipage}{0.45\textwidth}
					\vspace{10pt}
					\centering
					\footnotesize
					$B_{tx} = 2$ and $\mathcal{B}_{5} = 2.5$\\
					\vspace{3pt}
					\includegraphics[width =1.0\textwidth]{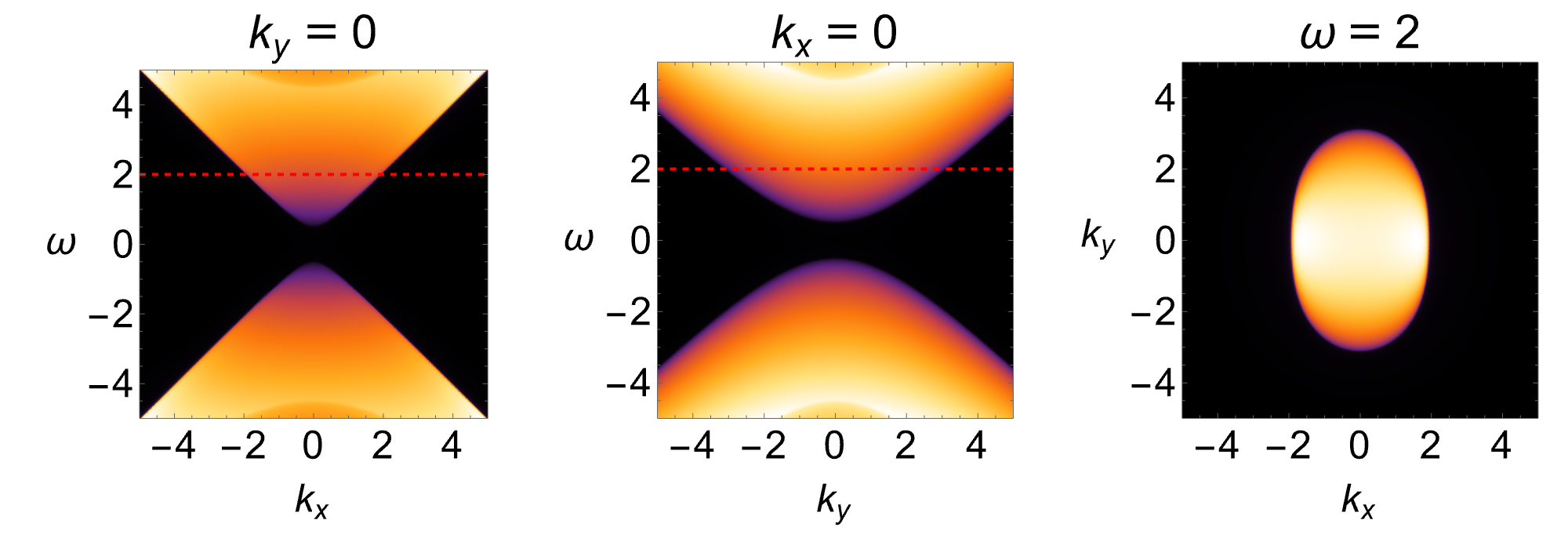}
				\end{minipage} & \begin{minipage}{0.25\textwidth}
					\vspace{10pt}
					\centering
					\footnotesize
					$B_{tx} = 2$ and $\mathcal{B}_{5} = 2.5$\\
					\vspace{5pt}
					\includegraphics[width =1.0\textwidth]{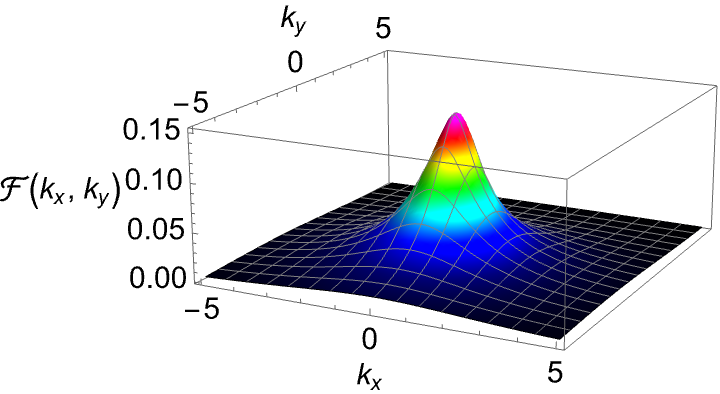}
				\end{minipage} & \begin{minipage}{0.25\textwidth}
					\vspace{10pt}
					\centering
					\footnotesize
					$B_{tx} = 2$ and $\mathcal{B}_{5} = -2.5$\\
					\vspace{5pt}
					\includegraphics[width =1.0\textwidth]{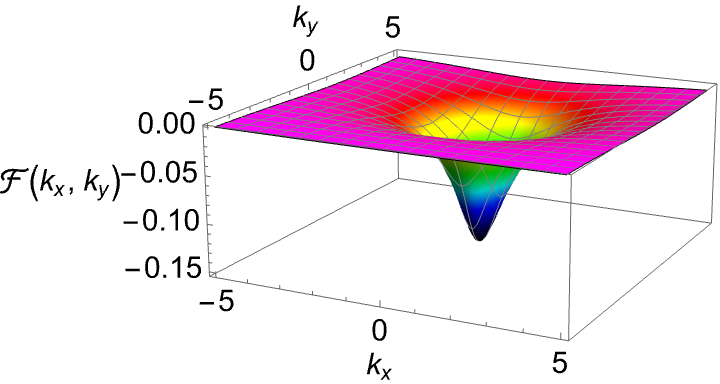}
				\end{minipage} \\
				\cline{2-5}
				& $B_{tz}$ &  \begin{minipage}{0.45\textwidth}
					\vspace{10pt}
					\centering
					\footnotesize
					$B_{tz} = 2$ and $\mathcal{B}_{5} = 2.5$\\
					\vspace{3pt}
					\includegraphics[width =1.0\textwidth]{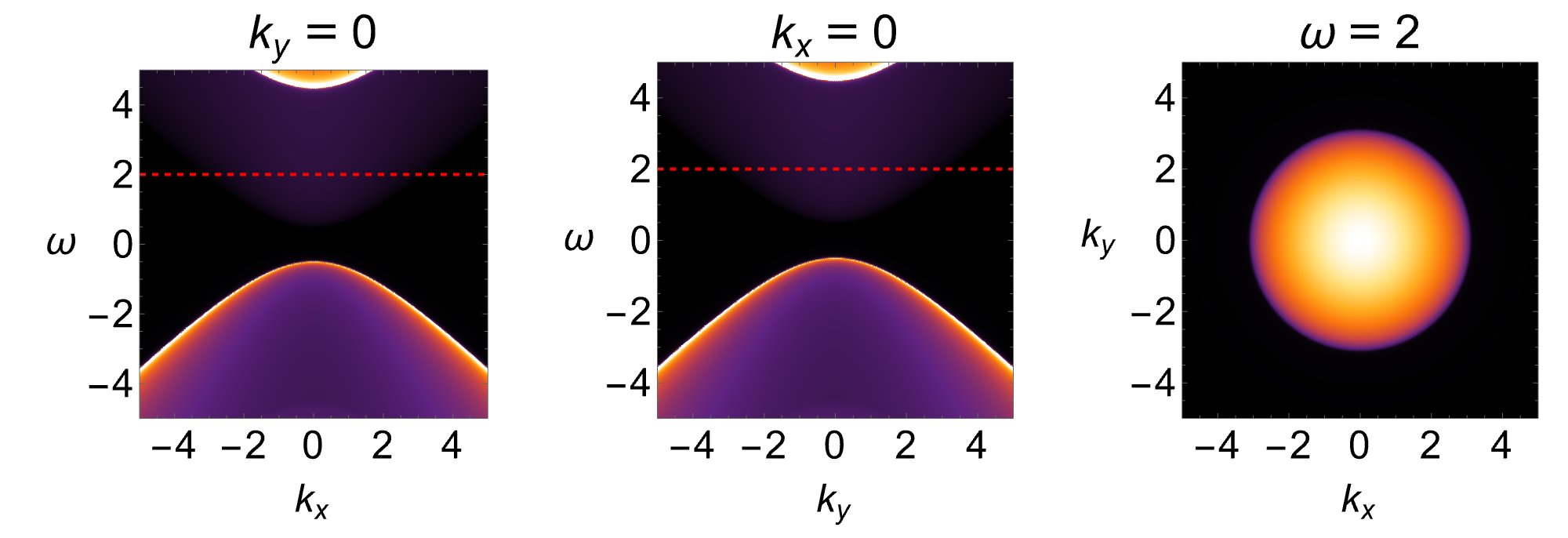}
				\end{minipage} & \multicolumn{2}{c|}{\begin{minipage}{0.5\textwidth}
						\centering
						\large Identical to pseudoscalar case.
				\end{minipage}} \\
				\cline{2-5}
				& $B_{ij}$ &  \begin{minipage}{0.45\textwidth}
					\vspace{10pt}
					\centering
					\footnotesize
					$B_{xy} = 2$ and $\mathcal{B}_{5} = 0.5$\\
					\vspace{3pt}
					\includegraphics[width =1.0\textwidth]{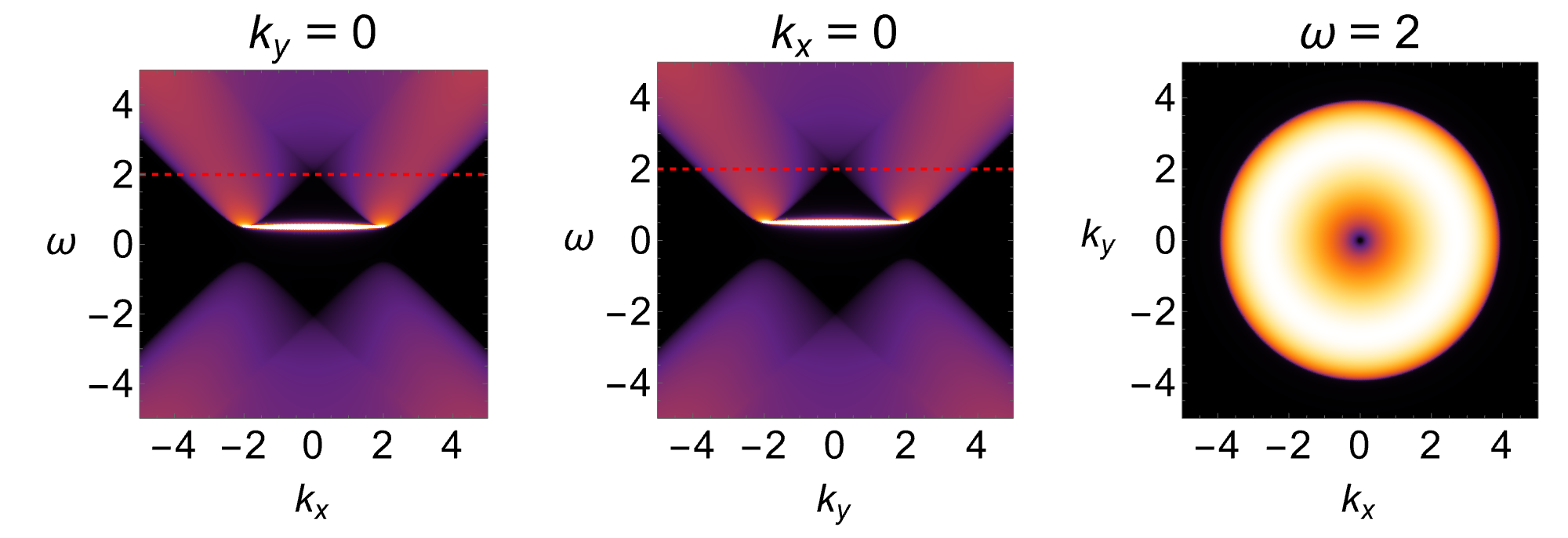}
				\end{minipage} & \begin{minipage}{0.25\textwidth}
					\vspace{10pt}
					\centering
					\footnotesize
					$B_{xy} = 2$ and $\mathcal{B}_{5} = 0.5$\\
					\vspace{5pt}
					\includegraphics[width =1.0\textwidth]{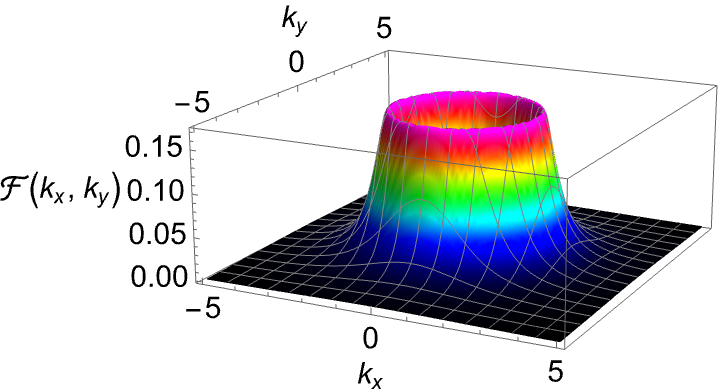}
				\end{minipage} & \begin{minipage}{0.25\textwidth}
					\vspace{10pt}
					\centering
					\footnotesize
					$B_{xy} = 2$ and $\mathcal{B}_{5} = -0.5$\\
					\vspace{5pt}
					\includegraphics[width =1.0\textwidth]{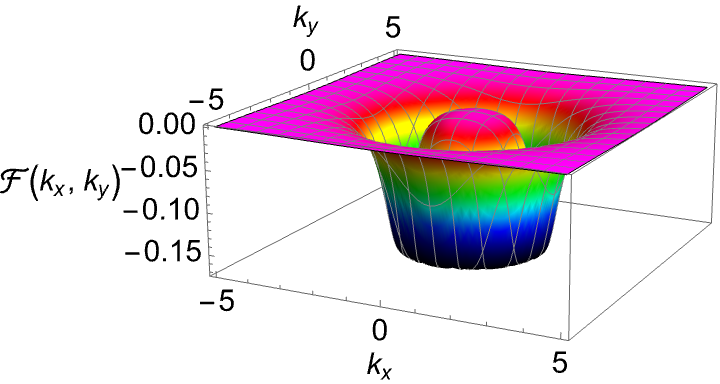}
				\end{minipage} \\
				\cline{2-5}
				& $B_{iz}$ & \begin{minipage}{0.45\textwidth}
					\vspace{10pt}
					\centering
					\footnotesize
					$B_{xz} = 2$ and $\mathcal{B}_{5} = 1$\\
					\vspace{3pt}
					\includegraphics[width =1.0\textwidth]{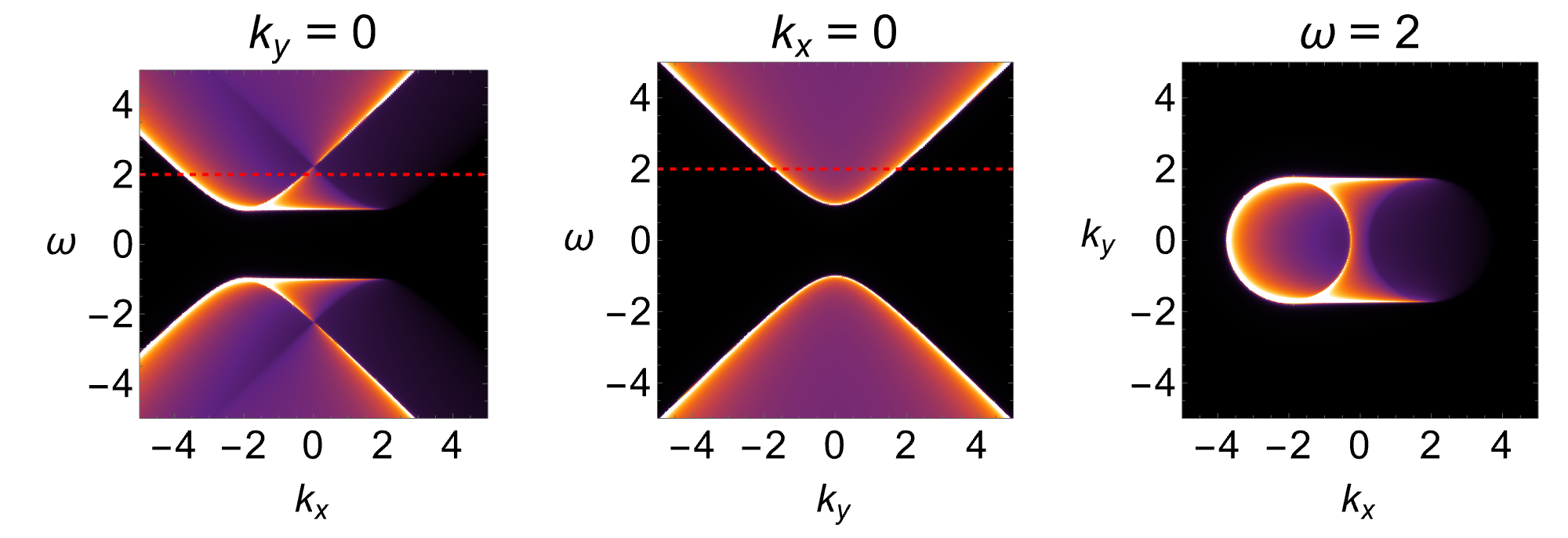}
				\end{minipage} & \begin{minipage}{0.25\textwidth}
					\vspace{10pt}
					\centering
					\footnotesize
					$B_{xz} = 2$ and $\mathcal{B}_{5} = 1$\\
					\vspace{5pt}
					\includegraphics[width =1.0\textwidth]{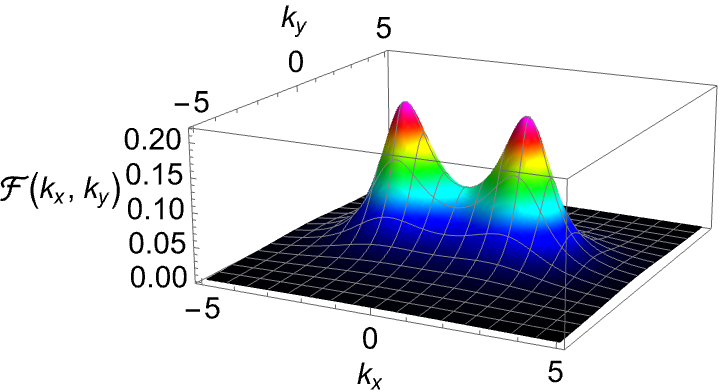}
				\end{minipage} & \begin{minipage}{0.25\textwidth}
					\vspace{10pt}
					\centering
					\footnotesize
					$B_{xz} = 2$ and $\mathcal{B}_{5} = -1$\\
					\vspace{5pt}
					\includegraphics[width =1.0\textwidth]{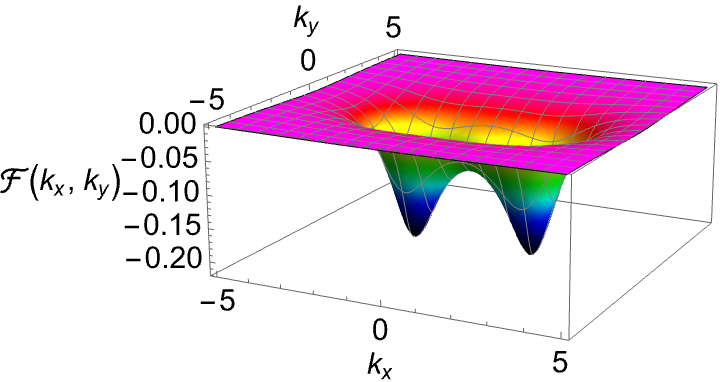}
				\end{minipage} \\
				\hline
			\end{tabular}
		}
		\caption{\label{table6}
			Constant-momentum and constant-frequency slices (red dotted lines) of spectral densities within the range $A(\omega, \mathbf{k}) \in [0, 10]$, along with the corresponding Berry curvature, are shown for one-flavor spinors under each interaction type in $\mathrm{AdS}_{4}$. All results are obtained from gapped spectral densities by introducing a pseudoscalar coupling $B_{5}$.}
	\end{table}
	Compared to the Berry curvature for pseudoscalar coupling, the Berry curvature shapes for other interactions can be classified into three types:
	\begin{itemize}
		\item Identical to pseudoscalar case: $B$, $B_{t}$, $B_{z}$, $B_{tz}$.
		\item Shift of Berry curvature: $B_{i}$.
		\item Deformation of Berry curvature: $B_{5t}$, $B_{5i}$, $B_{5z}$, $B_{ti}$, $B_{ij}$, $B_{iz}$.
	\end{itemize}
	Detailed analysis of their structure are provided in Appendix~\ref{appendix:C}.
	
	From the Berry curvature profiles, we compute the Chern numbers using \eqref{eq3.4}.
	Although the shape of the Berry curvature varies across interaction types and strengths, we find that the Chern number in the gapped regime is universally given by
	\begin{equation}
		\label{eq3.19}
		C = \dfrac{1}{2}\mathrm{sgn}(\mathcal{B}_{5}) \quad \forall B_{I},
	\end{equation}
	identical to the pseudoscalar case.
	This highlights the fundamental role of the pseudoscalar coupling in defining quantized topological invariants in the one-flavor case.
	
	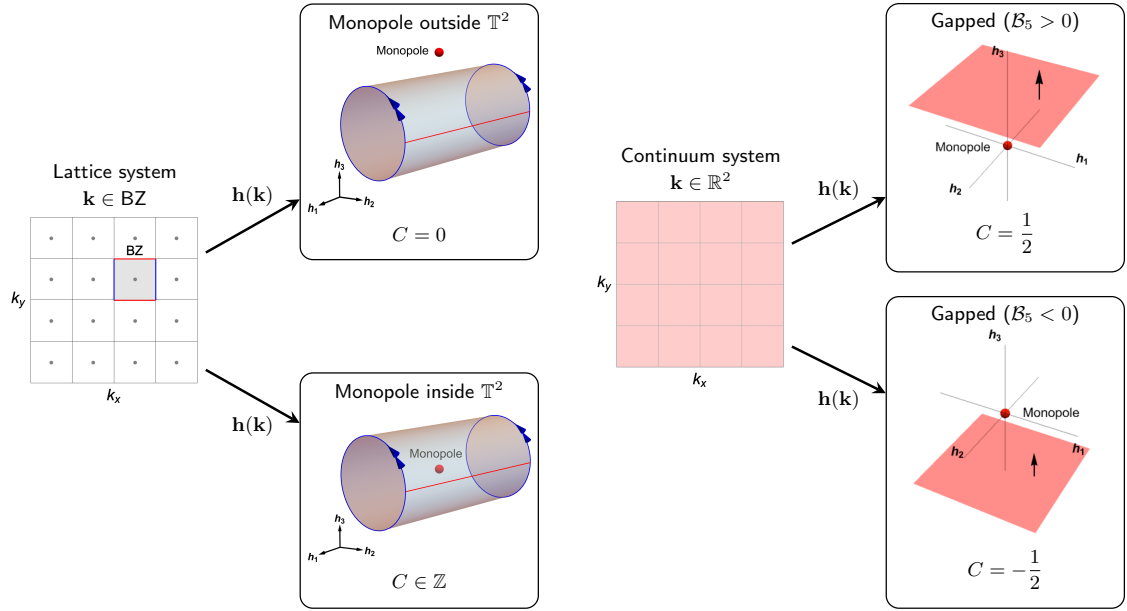
\begin{figure}[t]
		\centering\sffamily
		\begin{subfigure}[t]{0.49\textwidth}
			\centering
			\resizebox{\textwidth}{!}{
				\begin{tikzpicture}[node distance=5.0cm, scale=0.7, transform shape]
					\node (temp0) [none] {
						\begin{minipage}{3.0cm}
							\includegraphics[width=\textwidth]{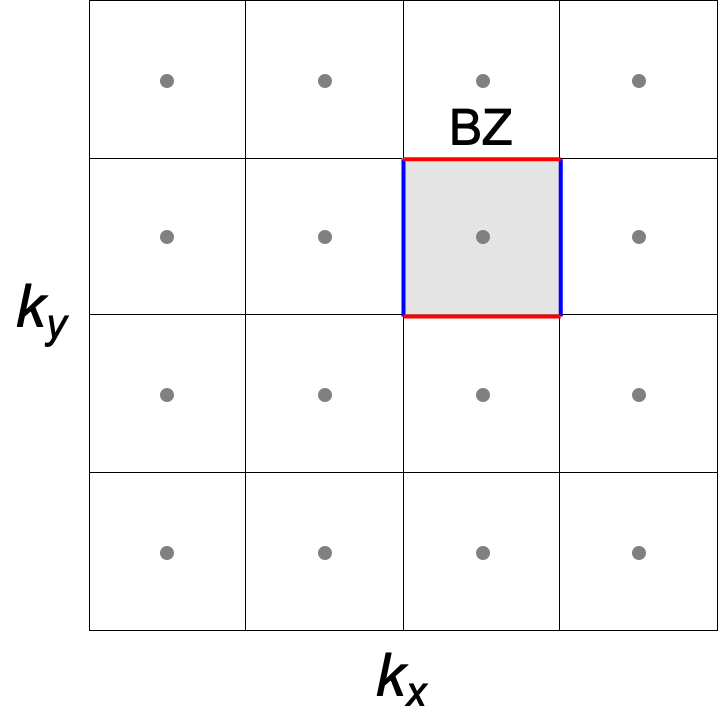}
						\end{minipage}
					};
					\node (temp1) [startstop1, right of=temp0, yshift=2.85cm] {
						\begin{minipage}{3.5cm}
							\centering\footnotesize
							Monopole outside $\mathbb{T}^{2}$\vspace{5pt}\\
							\includegraphics[width=\textwidth]{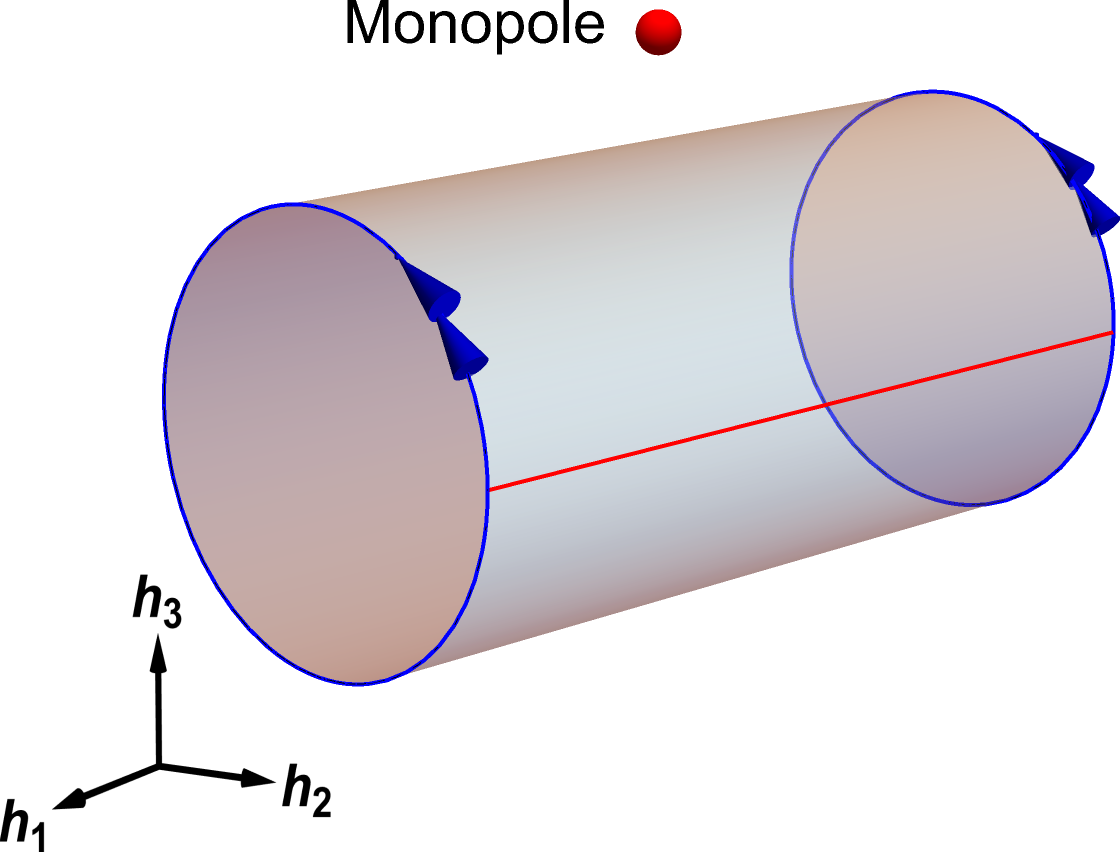}\\
							$C = 0$\vspace{5pt}
						\end{minipage}
					};
					\node (temp2) [startstop1, right of=temp0, yshift=-2.85cm] {
						\begin{minipage}{3.5cm}
							\centering\footnotesize
							Monopole inside $\mathbb{T}^{2}$\vspace{5pt}\\
							\includegraphics[width=\textwidth]{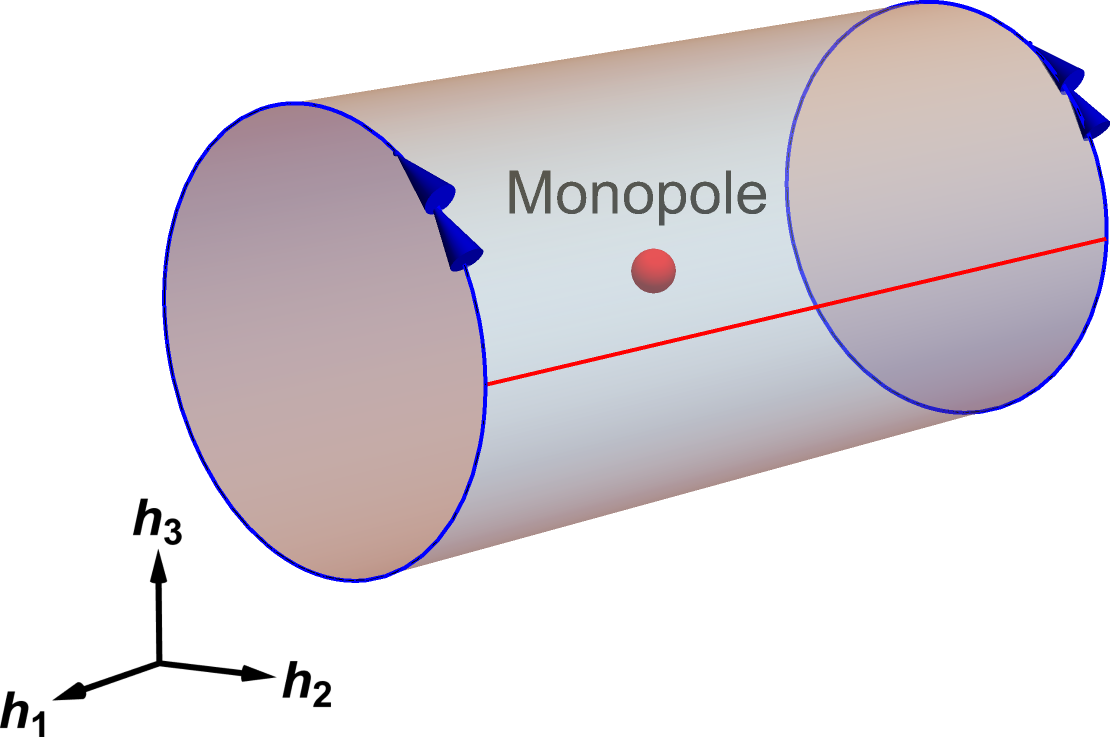}\\
							$C \in \mathbb{Z}$\vspace{5pt}
						\end{minipage}
					};
					\node (temp0p) [none, above of=temp0, xshift=0.2cm, yshift=-3.0cm] {
						\begin{minipage}{3.0cm}
							\centering\footnotesize
							Lattice system\\
							$\mathbf{k}\in\text{BZ}$
						\end{minipage}
					};
					\draw[arrow] (temp0) -- node[anchor=south] {\begin{minipage}{1.5cm}
							\centering\footnotesize
							$\mathbf{h}(\mathbf{k})$\vspace{5pt}
					\end{minipage}} (temp1);
					\draw[arrow] (temp0) -- node[anchor=north] {\begin{minipage}{1.5cm}
							\centering\vspace{5pt}\footnotesize
							$\mathbf{h}(\mathbf{k})$
					\end{minipage}} (temp2);
				\end{tikzpicture}
			}
			\caption{Integer topological invariant in lattice systems.}
			\label{fig:intBZ}
		\end{subfigure}
		\hfill
		\begin{subfigure}[t]{0.49\textwidth}
			\centering
			\resizebox{\textwidth}{!}{
				\begin{tikzpicture}[node distance=5.0cm, scale=0.7, transform shape]
					\node (temp0) [none] {
						\begin{minipage}{3.0cm}
							\includegraphics[width=\textwidth]{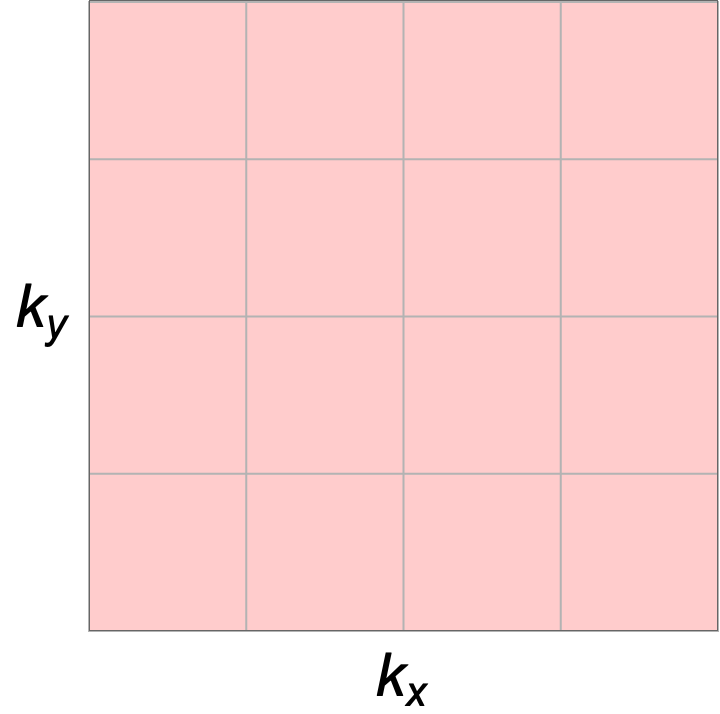}
						\end{minipage}
					};
					\node (temp2) [startstop1, right of=temp0, yshift=2.5cm] {
						\begin{minipage}{3.5cm}
							\centering\footnotesize
							Gapped ($\mathcal{B}_{5} > 0$)\\
							\includegraphics[width=\textwidth]{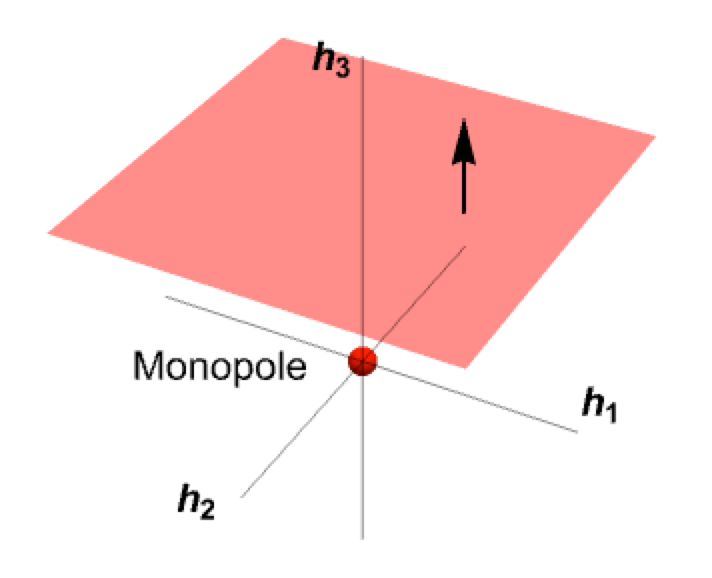}\\
							$C = \dfrac{1}{2}$\vspace{5pt}
						\end{minipage}
					};
					\node (temp3) [startstop1, right of=temp0, yshift=-2.5cm] {
						\begin{minipage}{3.5cm}
							\centering\footnotesize
							Gapped ($\mathcal{B}_{5} < 0$)\\
							\includegraphics[width=\textwidth]{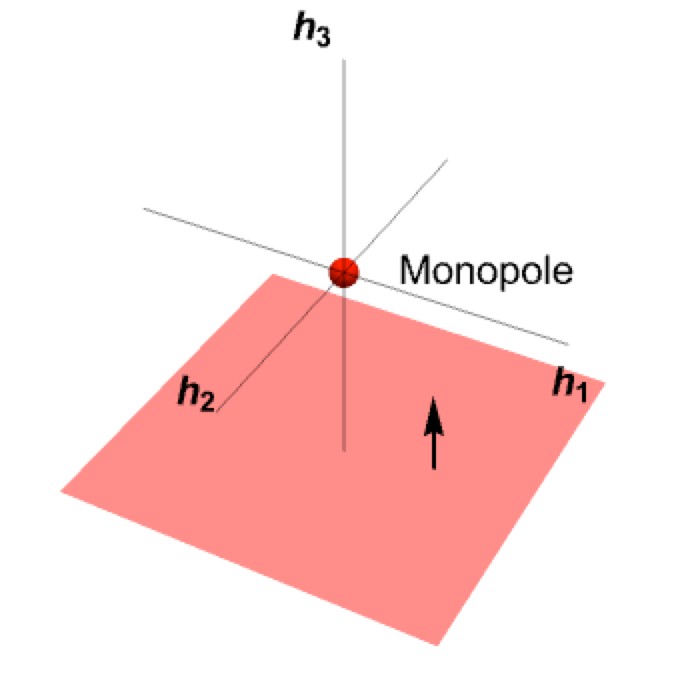}\\
							$C = -\dfrac{1}{2}$\vspace{5pt}
						\end{minipage}
					};
					\node (temp0p) [none, above of=temp0, xshift=0.2cm, yshift=-3.0cm] {
						\begin{minipage}{2.6cm}
							\centering\footnotesize
							Continuum system\\
							$\mathbf{k}\in\mathbb{R}^{2}$
						\end{minipage}
					};
					\draw[arrow] (temp0) -- node[anchor=south] {\begin{minipage}{1.5cm}
							\centering\footnotesize
							$\mathbf{h}(\mathbf{k})$\vspace{5pt}
					\end{minipage}} (temp2);
					\draw[arrow] (temp0) -- node[anchor=north] {\begin{minipage}{1.5cm}
							\centering\vspace{5pt}\footnotesize
							$\mathbf{h}(\mathbf{k})$
					\end{minipage}} (temp3);
				\end{tikzpicture}
			}
			\caption{Fractional topological invariant in continuum systems.}
			\label{fig:fracR2}
		\end{subfigure}
		\caption{\label{Figure3}
			Conceptual comparison between topological invariants in lattice and continuum systems via the mapping $\mathbf{h}(\mathbf{k})$. 
			(a) In lattice systems, the Brillouin zone (BZ) is compact, and the map defines a closed torus in $\mathbf{h}$-space, yielding an integer-valued Chern number.
			(b) In continuum systems, $\mathbf{k}$-space maps to an open surface in $\mathbf{h}$-space; for gapped phases, this covers either upper or lower surface, resulting in a fractional Chern number.}
	\end{figure}
	The fractional Chern number in \eqref{eq3.19} arises since the continuum momentum space in our theory lacks the periodic identification of lattice models.
	In a two-dimensional lattice system, the first Brillouin zone (BZ) is considered, and the map $\mathbf{k} \in \text{BZ} \mapsto \mathbf{h}(\mathbf{k}) = (h_{1}(\mathbf{k}), h_{2}(\mathbf{k}), h_{3}(\mathbf{k}))$ defines a closed surface in $\mathbf{h}$-space, topologically a torus via compactification.
	When the monopole lies outside the torus, the Chern number is zero; when it lies inside, it becomes a non-zero integer.
	In contrast, in our continuum model, $\mathcal{B}_{5}$ lifts the $\sigma_{3}$ component of $H_{\mathrm{topo}}(\mathbf{k})$, opening a gap and shifting the surface traced by $\mathbf{h}(\mathbf{k})$ to lie entirely on either the upper or lower surface.\footnote{When $\mathbf{h}(\mathbf{k})$ is normalized, the image occupies only the northern or southern hemisphere of the Bloch sphere.}
	As a result, only half of the monopole flux contributes to the invariant so that fractional Chern number arises.
	The sign of this invariant is determined by the direction of the lift: it is positive for $\mathcal{B}_{5} > 0$ and negative for $\mathcal{B}_{5} < 0$.
	We illustrate this comparison in the figure~\ref{Figure3}.

	\section{\label{section4}Two-flavor spinors at zero temperature}
	In condensed matter systems, multiple degrees of freedom—such as spin or orbital angular momentum—are frequently present.
	Motivated by this, we extend the topological framework developed for one-flavor spinors to the case of two-flavor systems.
	The analysis is performed for both standard–standard (SS) and standard–alternative (SA) quantization schemes, whose precise definitions are provided in the subsequent section.
	
	\subsection{Dirac equation and source identification}
	\label{section4.1}
	For simplicity, we assume the masses of the flavors are identical.
	We use the superscript $(i)$, with $i = 1, 2$, to label individual flavors in the two-flavor analysis.
	The actions for the two-flavor spinors are given by   
	\begin{align}
		\label{eq4.1}
		S_{\mathrm{bulk}} &= i\sum_{i = 1}^{2}\int_{\mathcal{M}}d^{4}x\sqrt{-g}\,\bar{\psi}^{(i)}(\slashed{D} - m)\psi^{(i)},\\
		\label{eq4.2}
		S_{\mathrm{bdy}}^{(\mathrm{SS})} &= -i\int_{\partial\mathcal{M}}d^{3}x\sqrt{-gg^{zz}}\left(\bar{\psi}^{(1)}\psi^{(1)} + \bar{\psi}^{(2)}\psi^{(2)}\right),\\
		\label{eq4.3}
		S_{\mathrm{bdy}}^{(\mathrm{SA})} &= -i\int_{\partial\mathcal{M}}d^{3}x\sqrt{-gg^{zz}}\left(\bar{\psi}^{(1)}\psi^{(1)} - \bar{\psi}^{(2)}\psi^{(2)}\right).
	\end{align}
	As we will show later, nontrivial topology arises when two distinct interactions with different internal structures are included in the interaction action.
	To capture this, we analyze the two-flavor case by considering a general interaction action of the form
	\begin{equation}
		\label{eq4.4}
		S_{\mathrm{int}} = i\int_{\mathcal{M}}d^{4}x\sqrt{-g}
		\begin{pmatrix}
			\bar{\psi}^{(1)} & \bar{\psi}^{(2)}
		\end{pmatrix}
		\left[\tau_{a}\otimes\left(B_{I}\Gamma^{I}\right) + \tau_{b}\otimes\left(\tilde{B}_{J}\Gamma^{J}\right)\right]
		\begin{pmatrix}
			\psi^{(1)}\\
			\psi^{(2)}
		\end{pmatrix},
	\end{equation}
	where $B_{I}\Gamma^{I}$ are defined as in \eqref{eq2.7}, and $\tilde{B}_{J}\Gamma^{J}$ represents an independent interaction channel of the same form, with $\tilde{B}_{I} = \tilde{\mathcal{B}}_{I}z^{p_{I}}$ as in \eqref{eq2.9}.
	The matrices $(\tau_{a}, \tau_{b})$ denote arbitrary Pauli matrices specifying the interaction structure.
	The remaining actions $S_{g}$ and $S_{B}$ are the same as in the one-flavor case, and the action for $\tilde{B}_{I}$ is defined analogously to $S_{B}$.
	For clarity, we assign the gapping parameters to the $\tau_{b}$ sector, i.e., to the $\tilde{B}_{J}\Gamma^{J}$ component.
	Unlike the one-flavor case, not only the pseudoscalar but scalar order parameter also can serve to open a gap in two-flavor case as follows: 
	\begin{itemize}
		\item In SS quantization one takes a pseudoscalar coupling $\tilde{B}_{I}\Gamma^{I} = -i \tilde{B}_{5}\Gamma^5$ to gap the spectrum.
		\item In several SA cases, the gap instead arises by a scalar coupling $\tilde{B}_{I}\Gamma^{I} = -\tilde{B}\mathbbm{1}_{4\times4}$. 
		All other SA cases still require the pseudoscalar coupling to produce a gap.
	\end{itemize}
	Therefore, $\tilde{B}_{I}$ can include either scalar or pseudoscalar order parameter field.
	From the total action, the bulk equations of motion for $\psi^{(i)}$ are given by
	\begin{equation}
		\label{eq4.5}
		\begin{pmatrix}
			\slashed{D} - m & 0\\
			0 & \slashed{D} - m
		\end{pmatrix}\begin{pmatrix}
			\psi^{(1)}\\
			\psi^{(2)}
		\end{pmatrix} + \left[\tau_{a}\otimes(B_{I}\Gamma^{I}) + \tau_{b}\otimes(\tilde{B}_{J}\Gamma^{J})\right]
		\begin{pmatrix}
			\psi^{(1)}\\
			\psi^{(2)}
		\end{pmatrix} = 0.
	\end{equation}
	As in the one-flavor case, we adopt the following ansatz for the two-spinor field:
	\begin{equation}
		\label{eq4.6}
		\psi^{(i)}
		= (-gg^{zz})^{-1/4}e^{-i\omega t + ik_{x}x + ik_{y}y} 
		\zeta^{(i)} .
	\end{equation}
	
	To define the source and condensation for two-flavor spinors, we decompose each flavor field as
	\begin{equation}
		\label{eq4.7}
		\psi^{(1)} = \begin{pmatrix}
			\psi^{(1)}_{+}\\
			\psi^{(1)}_{-}
		\end{pmatrix}  \quad\text{and}\quad
		\psi^{(2)} = \begin{pmatrix}
			\psi^{(2)}_{+}\\
			\psi^{(2)}_{-}
		\end{pmatrix} ,
	\end{equation}
	where each $\psi_{\pm}^{(i)}$ ($i = 1, 2$) is a two-component spinor field. 
	From this field decomposition, we decompose $\zeta^{(i)}$ $(i = 1, 2)$ in \eqref{eq4.6} in the same manner as
	\begin{equation}
		\label{eq4.8}
		\zeta^{(1)} = \begin{pmatrix}
			\zeta^{(1)}_{+}\\
			\zeta^{(1)}_{-}
		\end{pmatrix}  
		\quad\text{and}\quad
		\zeta^{(2)}= \begin{pmatrix}
			\zeta^{(2)}_{+}\\
			\zeta^{(2)}_{-}
		\end{pmatrix} .
	\end{equation}
	Under this decomposition, we identify the source and condensation to construct the Green’s function for each quantization choice, as detailed below.
	
	\noindent
	\textbf{SS case.}
	For the SS quantization method, we choose the boundary action as $S_{\mathrm{bdy}}^{(\mathrm{SS})}$, which can be rewritten as
	\begin{equation}
		\label{eq4.9}
		\begin{gathered}
			S_{\mathrm{bdy}}^{(\mathrm{SS})} = -i\sum_{i = 1}^{2}\int_{\partial\mathcal{M}}d^{3}x \, \bar{\zeta}^{(i)}\zeta^{(i)} =  -\sum_{i = 1}^{2}\int_{\partial\mathcal{M}}d^{3}x \, {\zeta^{(i)}_{+}}^{\dagger}\zeta_{-}^{(i)} + \text{h.c.}
		\end{gathered}
	\end{equation}
	When we vary the bulk action with respect to $\psi^{(i)}$ and add the boundary action variation, one can find that the total action variation can be expressed only in terms of the variation of $\zeta^{(1)}_{+}$ and $\zeta^{(2)}_{+}$ if the equations of motion are satisfied. 
	Therefore, we identify the source and condensation as the boundary quantities of $(\zeta^{(1)}_{+}, \zeta^{(2)}_{+})^{\mathrm{T}}$ and $(\zeta^{(1)}_{-}, \zeta^{(2)}_{-})^\text{T}$, respectively.
	For notational clarity, we denote these bulk quantities as
	\begin{equation}
		\label{eq4.10}
		\xi^{(\mathrm{S})}_{\mathrm{(SS)}} \equiv \begin{pmatrix}
			\zeta^{(1)}_{+}\\
			\zeta^{(2)}_{+}
		\end{pmatrix}\quad\text{and}\quad\xi^{(\mathrm{C})}_{\mathrm{(SS)}} \equiv \begin{pmatrix}
			\zeta^{(1)}_{-}\\
			\zeta^{(2)}_{-}
		\end{pmatrix}.
	\end{equation}
	
	\noindent
	\textbf{SA case.}
	In the SA quantization, we choose the boundary action as $S_{\mathrm{bdy}}^{(\mathrm{SA})}$, which can be rewritten as
	\begin{equation}
		\label{eq4.11}
		S_{\mathrm{bdy}}^{(\mathrm{SA})} = -i\int_{\partial\mathcal{M}}d^{3}x\left(\bar{\zeta}^{(1)}\zeta^{(1)} - \bar{\zeta}^{(2)}\zeta^{(2)}\right) = -\int_{\partial\mathcal{M}}d^{3}x\left({\zeta_{+}^{(1)}}^{\dagger}\zeta_{-}^{(1)} - {\zeta_{+}^{(2)}}^{\dagger}\zeta_{-}^{(2)}\right) + \text{h.c.}
	\end{equation}
	We do the same variation on the total action as we did in the SS case.
	In this case, the total action variation depends only on $\delta\zeta_{+}^{(1)}$ and $\delta\zeta_{-}^{(2)}$. 
	Accordingly, we identify the source and condensation in SA quantization as the boundary quantities of
	\begin{equation}
		\label{eq4.12}
		\xi^{(\mathrm{S})}_{\mathrm{(SA)}} \equiv \begin{pmatrix}
			\zeta^{(1)}_{+}\\
			\zeta^{(2)}_{-}
		\end{pmatrix}\quad\text{and}\quad\xi^{(\mathrm{C})}_{\mathrm{(SA)}} \equiv \begin{pmatrix}
			\zeta^{(1)}_{+}\\
			\zeta^{(2)}_{-}
		\end{pmatrix},
	\end{equation}
	respectively.
	
	To extract the source and condensation from $\xi^{(\mathrm{S})}_{\mathrm{(SS, SA)}}$ and $\xi^{(\mathrm{C})}_{\mathrm{(SS, SA)}}$, we examine the boundary behavior of them by solving the equations of motion in \eqref{eq4.5}.
	We observe that the leading terms of $\xi^{(\mathrm{S})}_{\mathrm{(SS, SA)}}$ and $\xi^{(\mathrm{C})}_{\mathrm{(SS, SA)}}$ are identical to those of the one-flavor case, as expressed in \eqref{eq2.18}.
	Thus, we use the same notation, $\mathcal{J}$ and $\mathcal{C}$, as in \eqref{eq2.18} for the source and condensation of two-flavor spinors, respectively, under each quantization choice.
	Note, however, that these quantities are now four-component objects.
	
	\subsection{\label{section4.2}Green's function and spectral densities}
	From the source and condensation identification, we define the retarded Green’s function for two-flavor spinors under SS and SA quantization.
	
	\noindent
	\textbf{SS case.}
	Under the chosen $\mathcal{J}$ and $\mathcal{C}$ as the boundary quantities of \eqref{eq4.10}, we extend the calculation and notations in the section~\ref{section2.2} and Appendix~\ref{appendix:A} from two components to four components using the same procedure.
	The resultant Green’s function is expressed as
	\begin{equation}
		\label{eq4.13}
		G_{R}^{(\mathrm{SS})} = \lim_{z \to 0} z^{2m} \mathbb{C}^{(\mathrm{SS})}(\mathbb{S}^{(\mathrm{SS})})^{-1},
	\end{equation}
	where $\mathbb{S}^{(\mathrm{SS})}$ and $\mathbb{C}^{(\mathrm{SS})}$ are now $4\times4$ matrices.
	
	\noindent
	\textbf{SA case.}
	We employ the same calculation we did in SS case for the Green's function derivation in SA quantization, and the result is given by
	\begin{equation}
		\label{eq4.14}
		G_{R}^{(\mathrm{SA})} = \lim_{z \to 0} z^{2m} (\sigma_{0}\oplus-\sigma_{0})\mathbb{C}^{(\mathrm{SA})}(\mathbb{S}^{(\mathrm{SA})})^{-1}.
	\end{equation}
	Notably, this Green's function can be expressed analytically in terms of the SS one, as shown in Appendix~\ref{appendix:D}. 
	Defining the SS Green's function in \eqref{eq4.13} as
	\begin{equation}
		\label{eq4.15}
		G_{R}^{(\mathrm{SS})} = \begin{pmatrix}
			\mathbb{G}_{11} & \mathbb{G}_{12}\\
			\mathbb{G}_{21} & \mathbb{G}_{22}
		\end{pmatrix},
	\end{equation}
	where $\mathbb{G}_{ij}$ are $2\times2$ matrices, \eqref{eq4.14} can be expressed as
	\begin{equation}
		\label{eq4.16}
		G_{R}^{(\mathrm{SA})} = \begin{pmatrix}
			\mathbb{G}_{11} - \mathbb{G}_{12}\mathbb{G}_{22}^{-1}\mathbb{G}_{21} & \quad \mathbb{G}_{12}\mathbb{G}_{22}^{-1}\\
			\mathbb{G}_{22}^{-1}\mathbb{G}_{21} & -\mathbb{G}_{22}^{-1}
		\end{pmatrix}.
	\end{equation}
	Therefore, once the Green's function in SS quantization is determined, SA one follows straightforwardly.
	
	The spectral density for each quantization method is obtained in the same manner as in the one-flavor case, following \eqref{eq2.21}.
	In the two-flavor case, it is convenient to compute the Green’s function numerically for each interaction type.
	To this end, we employ the flow equation formalism, which takes the form of a matrix-valued Riccati equation, as introduced in \cite{PhysRevD.79.025023}.
	The derivation of the flow equations is provided in Appendix~\ref{appendix:A}.
	
	\subsection{Non-Abelian Berry phase}
	From the defined two-flavor Green's function in \eqref{eq4.13} and \eqref{eq4.14}, the definition of topological Hamiltonian in two-flavor case is identically given by \eqref{eq3.1}.
	As we did in one-flavor case, we consider the gapped phase.
	
	The topological structure falls into two distinct cases: non-degenerate and degenerate for lowest two eigenvalues of Hamiltonian. 
	When calculating the Berry curvature in a degenerate system using the Abelian formalism, there is a gauge ambiguity due to the freedom in choosing a basis within the degenerate subspace. 
	This is naturally resolved in the non-Abelian formalism, where the traced Berry curvature over the occupied-spectra indices is gauge invariant and thus independent of the specific choice of eigenvectors. 
	Also, when there is no degeneracy, the non-Abelian formalism naturally reduces to Abelian case as well.
	The non-Abelian Berry connection is defined as \cite{Vanderbilt_2018, PhysRevLett.126.246801}
	\begin{equation}
		\label{eq4.17}
		\boldsymbol{\mathcal{A}}_{\mu}(\mathbf{k}) \equiv (\mathcal{A}_{ab})_{\mu}(\mathbf{k}) = i\bra{n_{a}(\mathbf{k})}\partial_{\mu}\ket{n_{b}(\mathbf{k})},
	\end{equation}
	where $\ket{n_{a,b}(\mathbf{k})}$ ($a, b = 1, 2$) denote orthonormal eigenstates spanning the degenerate occupied spectra, and $\mu = x, y$ labels momentum components $(k_{x}, k_{y})$.
	The corresponding non-Abelian Berry curvature tensor is given by
	\begin{equation}
		\label{eq4.18}
		\boldsymbol{\mathcal{F}}_{\mu\nu}(\mathbf{k}) = \partial_{\mu}\boldsymbol{\mathcal{A}}_{\nu}(\mathbf{k}) - \partial_{j}\boldsymbol{\mathcal{A}}_{\mu}(\mathbf{k}) - i\left[\boldsymbol{\mathcal{A}}_{\mu}(\mathbf{k}), \boldsymbol{\mathcal{A}}_{\nu}(\mathbf{k})\right].
	\end{equation}
	
	One can show that the Chern number can be calculated by choosing an eigenvector set for $H_{\mathrm{topo}}(\mathbf{k})$ and the result is invariant under the different eigenvector choices.
	To see this, we consider a local $\mathrm{U}(2)$ transformation with a matrix $U(\mathbf{k})$ acting on the chosen degenerate orthonormal eigenvector set $\ket{n_{i}(\mathbf{k})}$ ($i = 1, 2$) and the corresponding gauge transformation of Berry connection, such as
	\begin{equation}
		\label{eq4.19}
		\begin{gathered}
			\ket{n_{i}(\mathbf{k})} \mapsto \sum_{j = 1}^{2}U_{ji}(\mathbf{k})\ket{n_{j}(\mathbf{k})}
			\quad\Longrightarrow\quad
			\boldsymbol{\mathcal{A}}_{\mu}(\mathbf{k}) \mapsto U^{\dagger}(\mathbf{k})\boldsymbol{\mathcal{A}}_{\mu}(\mathbf{k})U(\mathbf{k}) + i U^{\dagger}(\mathbf{k})\partial_{\mu}U(\mathbf{k}).
		\end{gathered}
	\end{equation}
	The corresponding curvature transforms in a gauge covariant way, such that
	\begin{equation}
		\label{eq4.20}
		\boldsymbol{\mathcal{F}}_{\mu\nu}(\mathbf{k}) \mapsto U^{\dagger}(\mathbf{k})\boldsymbol{\mathcal{F}}_{\mu\nu}(\mathbf{k})U(\mathbf{k}).
	\end{equation}
	The trace of it,
	\begin{equation}
		\label{eq4.21}
		\mathcal{F}(\mathbf{k}) \equiv \mathrm{Tr}\boldsymbol{\mathcal{F}}_{xy}(\mathbf{k}),
	\end{equation}
	is gauge invariant, so we adopt this as the definition of Berry curvature for computing the Chern number in two-flavor spinors.
	
	Similar to \eqref{eq3.7}, we evaluate the Chern number in a numerical way: the Berry phase accumulation over a small plaquette in $\mathbf{k}$-space can be calculated via a discretized Wilson loop as illustrated in the figure~\ref{Figure2}. 
	For two-flavor case, the overlap function becomes matrix-valued, such as\footnote{To ensure numerical stability, we apply singular value decomposition to each overlap matrix $\boldsymbol{U}_{\mathbf{k}\mathbf{k}’}$: for $\boldsymbol{U}_{\mathbf{k}\mathbf{k}’} = V \Sigma W^{\dagger}$, we replace it by the unitary matrix $VW^{\dagger}$ in the Wilson loop computation.} \cite{Vanderbilt_2018, 10.21468/SciPostPhysLectNotes.51}
	\begin{equation}
		\label{eq4.22}
		\begin{gathered}
			\phi(\mathbf{k}_{0}) = \text{Im}\log(\det(\boldsymbol{U}_{\mathbf{k}_{1}\mathbf{k}_{4}}\boldsymbol{U}_{\mathbf{k}_{4}\mathbf{k}_{3}}\boldsymbol{U}_{\mathbf{k}_{3}\mathbf{k}_{2}}\boldsymbol{U}_{\mathbf{k}_{2}\mathbf{k}_{1}}))\\
			\text{where}\quad \boldsymbol{U}_{\mathbf{k}\mathbf{k}^{\prime}} \equiv (U_{ab})_{\mathbf{k}\mathbf{k}^{\prime}} = \bra{n_{a}(\mathbf{k})}\ket{n_{b}(\mathbf{k}^{\prime})}.
		\end{gathered}
	\end{equation}
	The Chern number is then obtained using the same formula as in \eqref{eq3.8}.
	
	\subsection{Topological invariant}
	Using non-Abelian Berry phase formalism, we obtain Chern numbers for two-flavor spinors with the chosen pair $(\tau_{a}, \tau_{b})$ in \eqref{eq4.4} under SS and SA quantization.
	We distinguish two cases: (1) only one of the Pauli-matrix sectors is gapped, and (2) both sectors are gapped by independent gapping parameters.
	
	\noindent
	\textbf{Vector, axial vector, and tensor case.}
	We first consider the case in which $B_{I}$ in the $\tau_{a}$ sector of \eqref{eq4.4} correspond to vector, axial vector, and tensor components.
	For the SA quantization, the gapping parameter corresponds to a scalar when $\tau_{b} = \tau_{1}$ or $\tau_{2}$, and to a pseudoscalar otherwise.
	We summarize all Chern numbers for chosen interaction and quantization in the table~\ref{table7}.
	\begin{table}\small
		\resizebox{1.0\textwidth}{!}
		{\begin{tabular}{| c | p{4.0cm} | p{3.0cm} | p{3.0cm} | p{4.0cm} |}
				\hline
				\multicolumn{5}{|c|}{\begin{minipage}{1.0\textwidth}
						\centering\vspace{10pt}
						\textbf{Vector, axial vector, and tensor case: }
						$
						\begin{pmatrix}
							\bar{\psi}^{(1)} & \bar{\psi}^{(2)}
						\end{pmatrix}
						\left[\tau_{a}\otimes(B_{I}\Gamma^{I}) + \tau_{b}\otimes(\tilde{B}_{J}\Gamma^{J})\right]
						\begin{pmatrix}
							\psi^{(1)}\\
							\psi^{(2)}
						\end{pmatrix}
						$
						\vspace{5pt}
				\end{minipage}} \\
				\hline
				\multirow{2}{*}{\begin{minipage}{0.25\textwidth}
						\centering\vspace{5pt}
						Order parameter: \\
						$\tau_{a}\otimes(B_{I}\Gamma^{I})$
						\vspace{5pt}
				\end{minipage}} & \multicolumn{4}{c|}{\begin{minipage}{0.7\textwidth}
						\centering\vspace{5pt}
						Gapping parameter: $\tau_{b}\otimes(\tilde{B}_{J}\Gamma^{J})$
						\vspace{5pt}
				\end{minipage}} \\
				\cline{2-5}
				& \begin{minipage}{1.0\linewidth}\vspace{5pt}
					\centering
					$\tau_{0}$\vspace{5pt}
				\end{minipage} & \begin{minipage}{1.0\linewidth}\vspace{5pt}
					\centering
					$\tau_{1}$\vspace{5pt}
				\end{minipage} & \begin{minipage}{1.0\linewidth}\vspace{5pt}
					\centering
					$\tau_{2}$\vspace{5pt}
				\end{minipage} & \begin{minipage}{1.0\linewidth}\vspace{5pt}
					\centering
					$\tau_{3}$\vspace{5pt}
				\end{minipage} \\
				\hline
				\begin{minipage}{0.05\linewidth}\vspace{5pt}
					\centering
					$\tau_{0}$\vspace{5pt}
				\end{minipage} & \multirow{4}{*}{\begin{minipage}{1.0\linewidth}
						\centering
						$\begin{cases}
							\text{SS: }C = \mathrm{sgn}(\mathcal{B}_{5}^{(2)})\\
							\text{SA: }C = 0
						\end{cases}$
				\end{minipage}} & \centering\multirow{4}{*}{(SS, $\mathrm{SA}^{\ast}$): $C = 0$} & \centering\multirow{4}{*}{(SS, $\mathrm{SA}^{\ast}$): $C = 0$} &\multirow{4}{*}{\begin{minipage}{1.0\linewidth}
						\centering
						$\begin{cases}
							\text{SS: }C = 0\\
							\text{SA: }C = \mathrm{sgn}(\mathcal{B}_{5}^{(2)})
						\end{cases}$
				\end{minipage}} \\
				\cline{1-1}
				\begin{minipage}{0.05\linewidth}\vspace{5pt}
					\centering
					$\tau_{1}$\vspace{5pt}
				\end{minipage} & & & &  \\
				\cline{1-1}
				\begin{minipage}{0.05\linewidth}\vspace{5pt}
					\centering
					$\tau_{2}$\vspace{5pt}
				\end{minipage} & & && \\
				\cline{1-1}
				\begin{minipage}{0.05\linewidth}\vspace{5pt}
					\centering
					$\tau_{3}$\vspace{5pt}
				\end{minipage} & & & & \\
				\hline
			\end{tabular}
		}\\[5pt]
		$\ast$: gapping parameter in $\tau_{b}$ sector corresponds to scalar coupling, otherwise pseudoscalar.
		\caption{\label{table7}
			Summary of Chern numbers of gapped spectra for two-flavor case under different interaction structures with vector, axial vector, and tensor components and quantization methods, i.e., standard-standard (SS) and standard-alternative (SA) quantization. 
		}
	\end{table}
	From this, all possible Chern number is
	\begin{equation}
		\label{eq4.23}
		C = \mathrm{sgn}(\tilde{\mathcal{B}}_{5})\quad\text{or}\quad 0,
	\end{equation}
	which now becomes an integer, effectively doubling or subtracting the values obtained in the one-flavor case.
	
	\noindent
	\textbf{Scalar and pseudoscalar case. }We now consider the case where $B_{I}$ is either a scalar or a pseudoscalar. 
	This setup introduces two independent gapping parameters in the interaction action, leading to three distinct combinations for $B_{I}$/$\tilde{B}_{J}$ choice: pseudoscalar/pseudoscalar, pseudoscalar/scalar, and scalar/scalar interaction.
	Scalar/pseudoscalar case is symmetrically equivalent to the pseudoscalar/scalar case by exchanging $\tau_{a}\leftrightarrow\tau_{b}$.
	\begin{itemize}
		\item \textbf{Pseudoscalar/pseudoscalar.}
		All possible Chern numbers are listed in the table~\ref{table8} when the two parameters are both given by pseudoscalar, $B_{I}\Gamma^{I} = -i B_{5}\Gamma^{5}$ and $\tilde{B}_{J}\Gamma^{J} = -i \tilde{B}_{5}\Gamma^{5}$.
		\begin{table}[t]
			\centering
			\resizebox{\textwidth}{!}{
				\begin{tabular}[c]{|c|c|c|c|c|}
					\hline
					\multicolumn{5}{|c|}{
						\begin{minipage}{1.5\textwidth}\vspace{10pt}
							\centering\Large
							\textbf{SS pseudoscalar/pseudoscalar: }$\mathcal{L}_{\mathrm{int}} = 
							\begin{pmatrix}
								\bar{\psi}^{(1)} & \bar{\psi}^{(2)}
							\end{pmatrix}
							\left[\tau_{a}\otimes(B_{5}\Gamma^{5}) + \tau_{b}\otimes(\tilde{B}_{5}\Gamma^{5})\right]
							\begin{pmatrix}
								\psi^{(1)}\\
								\psi^{(2)}
							\end{pmatrix}
							$
							\vspace{10pt}
						\end{minipage}
					} \\
					\hline
					\diagbox[width=1.6cm]{\Large$\tau_{a}$}{\Large$\tau_{b}$}  & \multicolumn{1}{c|}{\Large$\tau_{0}$} & \multicolumn{1}{c|}{\Large$\quad\quad \tau_{1} \quad\quad$} & \multicolumn{1}{c|}{\Large$\quad\quad \tau_{2} \quad\quad$} & \multicolumn{1}{c|}{\Large$\tau_{3}$} \\
					\hline
					\centering \Large$\tau_{0}$ & \begin{minipage}{0.4\textwidth}
						\vspace{5pt}
						\centering\Large
						$C = \mathrm{sgn}(\mathcal{B}_{5} + \tilde{\mathcal{B}}_{5})$
						\vspace{5pt}
					\end{minipage} & 
					\multicolumn{3}{c|}{
						\begin{minipage}{1.0\textwidth}
							\vspace{10pt}
							\centering\Large
							$C = \dfrac{1}{2}[\mathrm{sgn}(\mathcal{B}_{5} + \tilde{\mathcal{B}}_{5}) + \mathrm{sgn}(\mathcal{B}_{5} - \tilde{\mathcal{B}}_{5})]$
							\vspace{10pt}
						\end{minipage}
					} \\
					\hline
					\centering \Large$\tau_{1}$ & 
					\centering \multirow{3}{*}{
						\begin{minipage}{0.6\textwidth}
							\vspace{10pt}
							\centering\Large
							$C = \dfrac{1}{2}[\mathrm{sgn}(\mathcal{B}_{5} + \tilde{\mathcal{B}}_{5}) + \mathrm{sgn}(\tilde{\mathcal{B}}_{5} - \mathcal{B}_{5})]$
							\vspace{10pt}
						\end{minipage}
					} &  \multicolumn{3}{c|}{}  \\
					\cline{1-1}
					\centering \Large$\tau_{2}$ & & \multicolumn{3}{c|}{\centering\Large
						$C = 0$}  \\
					\cline{1-1}
					\centering \Large$\tau_{3}$ & & \multicolumn{3}{c|}{}  \\
					\hline
					\hline
					\multicolumn{5}{|c|}{
						\begin{minipage}{1.5\textwidth}\vspace{10pt}
							\centering\Large
							\textbf{SA pseudoscalar/pseudoscalar: }$\mathcal{L}_{\mathrm{int}} = 
							\begin{pmatrix}
								\bar{\psi}^{(1)} & \bar{\psi}^{(2)}
							\end{pmatrix}
							\left[\tau_{a}\otimes\left(B_{5}\Gamma^{5}\right) + \tau_{b}\otimes(\tilde{B}_{5}\Gamma^{5})\right]
							\begin{pmatrix}
								\psi^{(1)}\\
								\psi^{(2)}
							\end{pmatrix}
							$
							\vspace{10pt}
						\end{minipage}
					} \\
					\hline
					\diagbox[width=1.6cm]{\Large$\tau_{a}$}{\Large$\tau_{b}$} & \Large$\tau_{0}$ & \multicolumn{1}{c|}{\Large$\quad\tau_{1}\quad$} & \multicolumn{1}{c|}{\Large$\quad\tau_{2}\quad$} & \Large$\tau_{3}$ \\
					\hline
					\Large$\tau_{0}$ & 
					\multicolumn{3}{c|}{\multirow{3}{*}{\Large$C = 0$}} & 
					\begin{minipage}{0.6\textwidth}\vspace{10pt}
						\centering\Large
						$C = \dfrac{1}{2}[\mathrm{sgn}(\mathcal{B}_{5} + \tilde{\mathcal{B}}_{5}) + \mathrm{sgn}(\tilde{\mathcal{B}}_{5} - \mathcal{B}_{5})]$
						\vspace{10pt}
					\end{minipage} \\
					\cline{1-1}\cline{5-5}
					\Large$\tau_{1}$ & 
					\multicolumn{3}{c|}{} & \multirow{2}{*}{
						\begin{minipage}{0.6\textwidth}\vspace{5pt}
							\centering\Large
							$C = \mathrm{sgn}(\tilde{\mathcal{B}}_{5})$
							\vspace{5pt}
						\end{minipage}
					} \\
					\cline{1-1}
					\Large$\tau_{2}$ & 
					\multicolumn{3}{c|}{} &  \\
					\hline
					\Large$\tau_{3}$ & 
					\begin{minipage}{0.6\textwidth}\vspace{10pt}
						\centering\Large
						$C = \dfrac{1}{2}[\mathrm{sgn}(\mathcal{B}_{5} + \tilde{\mathcal{B}}_{5}) + \mathrm{sgn}(\mathcal{B}_{5} - \tilde{\mathcal{B}}_{5})]$
						\vspace{10pt}
					\end{minipage} & 
					\multicolumn{2}{c|}{
						\begin{minipage}{0.3\textwidth}
							\centering\Large
							$C = \mathrm{sgn}(\mathcal{B}_{5})$
						\end{minipage}
					} & \Large$C = \mathrm{sgn}(\mathcal{B}_{5} + \tilde{\mathcal{B}}_{5})$ \\
					\hline
				\end{tabular}
			}
			\caption{\label{table8}Classification of Chern numbers for two-flavor spinors under pseudoscalar/pseudoscalar couplings in SS and SA quantization, organized by flavor structure $(\tau_{a}, \tau_{b})$ of the interaction action.
				The tables are symmetric in $\tau_{a}\leftrightarrow\tau_{b}$ due to the action form.}
		\end{table}
		The classification of the Chern number in this case is given by
		\begin{equation}
			\label{eq4.24}
			C = \begin{cases}
				\mathrm{sgn}(\mathcal{B}_{5}), \\[5pt]
				\mathrm{sgn}(\mathcal{B}_{5} + \tilde{\mathcal{B}}_{5}),\\[5pt]
				\dfrac{1}{2}[\mathrm{sgn}(\mathcal{B}_{5} + \tilde{\mathcal{B}}_{5}) + \mathrm{sgn}(\mathcal{B}_{5} - \tilde{\mathcal{B}}_{5})],\\[5pt]
				0,
			\end{cases}
		\end{equation}
		depending on the interaction structure.
		Other entries in the table~\ref{table8} are symmetrically obtained from these cases by exchanging $\tau_{a} \leftrightarrow \tau_{b}$, i.e., $\mathcal{B}_{5} \leftrightarrow \tilde{\mathcal{B}}_{5}$.
		In the third case of \eqref{eq4.24}, two gapping orders compete each other and a gap closing occurs when the strengths of them become identical, i.e., $|\mathcal{B}_{5}| = |\tilde{\mathcal{B}}_{5}|$.

		\item \textbf{Pseudoscalar/scalar.}
		We consider the case where $B_{I}$ is the pseudoscalar order while $\tilde{B}_{J}$ is the scalar one.
		We list the possible Chern numbers in this case in the table~\ref{table9}.
		\begin{table}[t]
			\centering
			\resizebox{\textwidth}{!}{
				\scriptsize
				\begin{tabular}[c]{|c|c|c|c|c|}
					\hline
					\multicolumn{5}{|c|}{
						\begin{minipage}{1.0\textwidth}\vspace{10pt}
							\centering
							\textbf{SS pseudoscalar/scalar: }$\mathcal{L}_{\mathrm{int}} = 
							\begin{pmatrix}
								\bar{\psi}^{(1)} & \bar{\psi}^{(2)}
							\end{pmatrix}
							\left[\tau_{a}\otimes(B_{5}\Gamma^{5}) + \tau_{b}\otimes(-i\tilde{B}\mathbbm{1}_{4\times4})\right]
							\begin{pmatrix}
								\psi^{(1)}\\
								\psi^{(2)}
							\end{pmatrix}
							$
							\vspace{10pt}
						\end{minipage}
					} \\
					\hline
					\diagbox[width=1.6cm]{$\tau_{a}$}{$\tau_{b}$}  & \multicolumn{1}{c|}{$\qquad\qquad\quad\tau_{0}\quad\qquad\qquad$} & \multicolumn{1}{c|}{$\qquad\qquad\quad\tau_{1}\quad\qquad\qquad$} & \multicolumn{1}{c|}{$\qquad\qquad \quad\tau_{2}\quad\qquad\qquad$} & \multicolumn{1}{c|}{$\qquad\qquad\quad\tau_{3}\quad\qquad\qquad$} \\
					\hline
					\centering $\tau_{0}$ & 
					\multicolumn{4}{c|}{\begin{minipage}{0.5\textwidth}\vspace{5pt}
							\centering
							$C = \mathrm{sgn}(\mathcal{B}_{5})$\vspace{5pt}
					\end{minipage}} \\
					\hline
					\centering $\tau_{1}$ & \multicolumn{4}{c|}{}\\
					\cline{1-1}
					\centering $\tau_{2}$ & \multicolumn{4}{c|}{$C = 0$}\\
					\cline{1-1}
					\centering $\tau_{3}$ & \multicolumn{4}{c|}{}\\
					\hline
					\hline
					\multicolumn{5}{|c|}{
						\begin{minipage}{1.0\textwidth}\vspace{10pt}
							\centering
							\textbf{SA pseudoscalar/scalar: }$\mathcal{L}_{\mathrm{int}} = 
							\begin{pmatrix}
								\bar{\psi}^{(1)} & \bar{\psi}^{(2)}
							\end{pmatrix}
							\left[\tau_{a}\otimes(B_{5}\Gamma^{5}) + \tau_{b}\otimes(-i\tilde{B}\mathbbm{1}_{4\times4})\right]
							\begin{pmatrix}
								\psi^{(1)}\\
								\psi^{(2)}
							\end{pmatrix}
							$
							\vspace{10pt}
						\end{minipage}
					} \\
					\hline
					\diagbox[width=1.6cm]{$\tau_{a}$}{$\tau_{b}$}  & \multicolumn{1}{c|}{$\qquad\qquad\tau_{0}\qquad\qquad$} & \multicolumn{1}{c|}{$\qquad\qquad\tau_{1}\qquad\qquad$} & \multicolumn{1}{c|}{$\qquad\qquad\tau_{2}\qquad\qquad$} & \multicolumn{1}{c|}{$\qquad\qquad\tau_{3}\qquad\qquad$} \\
					\hline
					\centering $\tau_{0}$ & \multicolumn{4}{c|}{}\\
					\cline{1-1}
					\centering $\tau_{1}$ & \multicolumn{4}{c|}{$C = 0$}\\
					\cline{1-1}
					\centering $\tau_{2}$ & \multicolumn{4}{c|}{}\\
					\hline
					\centering $\tau_{3}$ & $C = \mathrm{sgn}(\mathcal{B}_{5})$ & 
					\multicolumn{2}{c|}{\begin{minipage}{0.4\textwidth}\vspace{5pt}
							\centering
							$C = \dfrac{1}{2}[\mathrm{sgn}(\mathcal{B}_{5} + \tilde{\mathcal{B}}) + \mathrm{sgn}(\mathcal{B}_{5} - \tilde{\mathcal{B}})]$\vspace{5pt}
					\end{minipage}} & $C = \mathrm{sgn}(\mathcal{B}_{5})$ \\
					\hline
				\end{tabular}
			}
			\caption{\label{table9}
				Classification of Chern numbers for two-flavor spinors under pseudoscalar/scalar couplings in SS and SA quantization, organized by flavor structure $(\tau_{a}, \tau_{b})$ of the interaction action.
			}
		\end{table}
		The all possible classification of Chern numbers in pseudoscalar/scalar case is then
		\begin{equation}
			\label{eq4.25}
			C = \begin{cases}
				\mathrm{sgn}(\mathcal{B}_{5}),\\[5pt]
				\dfrac{1}{2}[\mathrm{sgn}(\mathcal{B}_{5} + \tilde{\mathcal{B}}) + \mathrm{sgn}(\mathcal{B}_{5} - \tilde{\mathcal{B}})],\\[5pt]
				0.
			\end{cases}
		\end{equation}
		For the second case in this formula, gap closing happens when $|\mathcal{B}_{5}| = |\tilde{\mathcal{B}}|$.
		
		\item \textbf{Scalar/scalar.}
		We consider the case in which both $B_{I}$ and $\tilde{B}_{J}$ are scalar order parameter fields.
		In this case, the gapped phases yield $C = 0$, independent of the values of the order parameters and the choice of $(\tau_{a}, \tau_{b})$.
	\end{itemize}
	
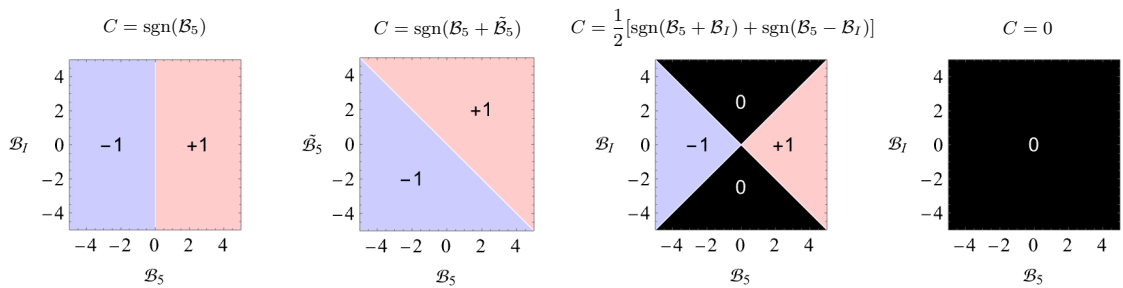
\begin{figure}[tbp]
		\centering\sffamily
		\resizebox{1.0\textwidth}{!}{
			\begin{tikzpicture}[node distance=5cm, scale=0.7, transform shape]
				\node (temp1) [none] {
					\begin{minipage}{4cm}
						\includegraphics[width=1.0\textwidth]{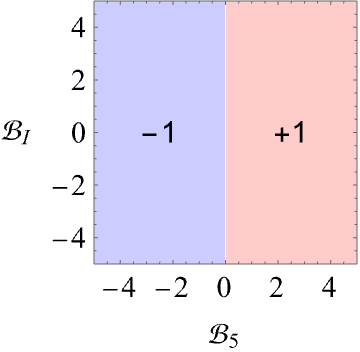}
					\end{minipage}
				};
				\node (temp2) [none, right of=temp1] {
					\begin{minipage}{4cm}
						\includegraphics[width=1.0\textwidth]{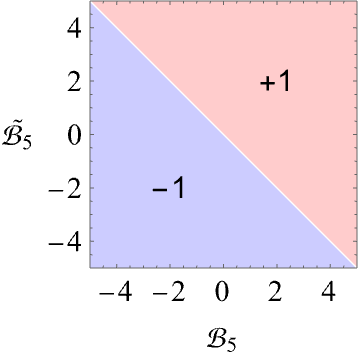}
					\end{minipage}
				};
				\node (temp3) [none, right of=temp2] {
					\begin{minipage}{4cm}
						\includegraphics[width=1.0\textwidth]{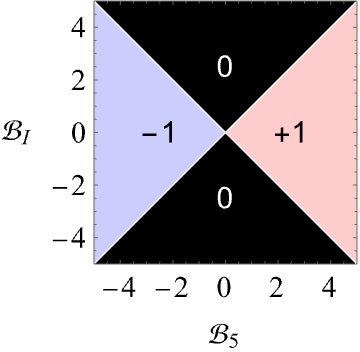}
					\end{minipage}
				};
				\node (temp4) [none, right of=temp3] {
					\begin{minipage}{4cm}
						\includegraphics[width=1.0\textwidth]{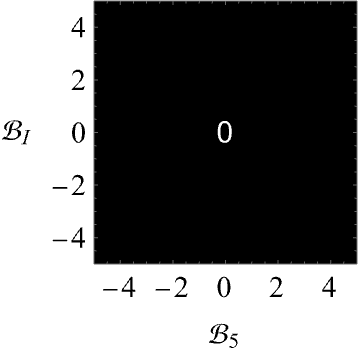}
					\end{minipage}
				};
				
				\node (temp1p) [none, above of=temp1, xshift=0.5cm, yshift=-2.5cm]{
					\footnotesize
					$C = \mathrm{sgn}(\mathcal{B}_{5})$
				};
				\node (temp2p) [none, above of=temp2, xshift=0.5cm, yshift=-2.5cm]{
					\footnotesize
					$C = \mathrm{sgn}(\mathcal{B}_{5} + \tilde{\mathcal{B}}_{5})$
				};
				\node (temp3p) [none, above of=temp3, xshift=0.2cm, yshift=-2.5cm]{
					\footnotesize
					$C = \dfrac{1}{2}[\mathrm{sgn}(\mathcal{B}_{5} + \mathcal{B}_{I}) + \mathrm{sgn}(\mathcal{B}_{5} - \mathcal{B}_{I})]$
				};
				\node (temp4p) [none, above of=temp4, xshift=0.4cm, yshift=-2.5cm]{
					\footnotesize
					$C = 0$
				};
			\end{tikzpicture}
		}
		\caption{\label{Figure4}
			Classification of four possible topological phase diagram in Chern number for gapped spectra in two-flavor spinor case. Phase boundaries correspond to gap-closing points between regions with $C = \pm1$. In the third diagram, the order parameter $\mathcal{B}_{I}$ can be either a pseudoscalar $\tilde{\mathcal{B}}_{5}$ or a scalar $\mathcal{B}$.
		}
	\end{figure}

	\noindent
	In summary, the Chern number in gapped two-flavor case is classified according to the gapping term structure in it as follow:
	\begin{itemize}
		\item $C = \mathrm{sgn}(\mathcal{B}_{5})$: single pseudoscalar order.
		\item $C = \mathrm{sgn}(\mathcal{B}_{5} + \tilde{\mathcal{B}}_{5})$: sum of two independent pseudoscalar order parameters.
		\item $C = \dfrac{1}{2}[\mathrm{sgn}(\mathcal{B}_{5} + \mathcal{B}_{I}) + \mathrm{sgn}(\mathcal{B}_{5} - \mathcal{B}_{I})]\,\,\,\text{with}\,\,\, \mathcal{B}_{I} \in \{\mathcal{B}, \tilde{\mathcal{B}}_{5}\}$: there exists two competing gapping orders.
		\item $C = 0$: otherwise.
	\end{itemize}
	The topological phase diagram for each case is illustrated in the figure~\ref{Figure4}.
	We comment on the first diagram: 
	for certain axial vector and tensor component of $B_{I}$, a gap does not open when $0 < \abs{\mathcal{B}_{5}} < \abs{\mathcal{B}_{I}}$, so the topology is ill-defined in this intermediate regime, as discussed in Appendix~\ref{appendix:B}.

	As shown in the table~\ref{table7}–\ref{table9}, a nonzero Chern number arises in the SS quantization when at least one gapping term appears in the $\tau_{0}$ sector.
	In contrast, for the SA quantization, a nonzero Chern number appears when the gapping term is in the $\tau_{3}$ sector.
	Remarkably, this can be understood from a duality between the SS and SA quantization via boundary Green's function.
	A detailed analysis of this duality is provided in Appendix~\ref{appendix:D}.

	\section{\label{section5}Topology at finite temperature}
	We have identified distinct classes of Chern numbers associated with gapped fermionic spectra at the boundary. 
	In this section, we examine their stability  under physical deformations including the temperature and clarify the regime in which our H-MFT   approach can be applied for obtaining well-defined topological invariants in many-body system.
	
	\subsection{Stability of the Chern number}
	To see that our Chern number behaves as a genuine topological invariant, we test its stability under two key deformations: the first is the bulk fermion mass $\abs{m} < 1/2$, which determines the anomalous dimension of the dual boundary operator and effectively controls the strength of electron-electron interactions \cite{PhysRevD.110.106017}. The second is  the temperature $T$.
	
	To see the effects of these parameters, we numerically solve the Green's function using the flow equation formalism.
	Then, we evaluate the Chern number  using \eqref{eq3.7} and \eqref{eq4.22}, respectively.
	We find that the Chern number remains invariant under deformations of $m$ and $T$ as long as the spectral gap remains open,  although the Berry curvature  are generically   deformed.
	
	The invariance of Chern number under the deformation of the temperature, $T$,  is supported by the scaling behavior of the Green’s function:
	\begin{equation}
		\label{eq5.1}
		G_{R}(\omega, \mathbf{k}, \mathcal{B}_{I}, m, T) = T^{-2m}\mathcal{G}\left(\omega, \mathbf{k}, \mathcal{B}_{I}, m, T\right)\mathbf{G}\left(\frac{\omega}{T}, \frac{\mathbf{k}}{T}, \frac{\mathcal{B}_{I}}{T^{\Delta_{I}}}, m\right), 
	\end{equation}
	where $\mathcal{G}$ is the overall function encoding singularity structures,   $\mathbf{G}$   is the remaining part of matrix   Green’s function and $\Delta_{I}$ is a scaling dimension.
	This scaling behavior originates from the conformal symmetry of the boundary theory, inherited from the asymptotically AdS geometry.
	It turns out that such symmetry  protect the matrix structure of the Green’s function, and thereby the associated topological Hamiltonian so that the topological number  remains invariant under deformation of $T$.
	
	Under the deformation of $m$, the Green’s function's matrix structure $\mathbf{G}$ can be modified for the cases where the Berry curvature shape is deformed relative  to that of the pseudoscalar, but  it turns out that the Chern number itself remains invariant.

	\subsection{Physical regime of topological invariants}
	Our Chern number results indicate that a topological invariant can be defined via the  Berry potential formalism in strongly interacting many-body systems; here, we clarify the physical regime in which this construction is applicable.
	
	The Berry phase is defined for single-particle states, but its generalization to interacting many-body systems is not completely clear. 	
	One generalization is the Uhlmann phase $\phi_{\mathrm{U}}$, defined via the adiabatic process of density matrices along the parameter $t$ \cite{UHLMANN1986229, UHLMANN1993253, Uhlmann1991}:
	\begin{equation}
		\label{eq5.2}
		\phi_{\mathrm{U}} = \arg\mathrm{Tr}(\rho_{\mathbf{k}_{0}}\mathcal{P}e^{\oint A_{\mathrm{U}}(t) dt}),
	\end{equation}
	where $\rho_{\mathbf{k}_{0}}$ is a reference density matrix at a point $\mathbf{k}_{0}$, $\mathcal{P}$ is the path ordering operator, and $A_{\mathrm{U}}(t)$ is the Uhlmann connection.
	Heuristically, the resultant topological invariant $C_{\mathrm{U}}$ can be approximated as a weighted sum of the Chern numbers $C_{i}$ associated with single-particle states $\ket{\psi_{i}}$,
	\begin{equation}
		C_{\mathrm{U}} \approx \sum_{i} w_{i} C_{i},
	\end{equation}
	with a weighting function $w_{i}$.
	Due to the factor $w_{i}$ in front of each integer-valued $C_{i}$, the Uhlmann phase often lacks integer-valued topological invariant and is sensitive to control parameters such as temperature or disorder \cite{10.21468/SciPostPhysCore.6.1.024, PhysRevB.106.024310}. 
	Moreover, its construction relies on an ensemble average over microscopically well-defined single-particle states—an assumption valid only in non-interacting or weakly coupled systems.
	
	Our well-defined integer Chern number is  challenges the conventional understanding of topological invariants in many-body systems.
	This is made possible because in  our H-MFT approach, a finite temperature configuration is a state described by a well defined geometry (see the figure~\ref{Figure1}): there is a well defined infalling boundary condition that turns out to give a consistent description of retarded Green function at the finite temperature.
	Such a system exhibits hydrodynamic behavior \cite{PhysRevB.76.144502, PhysRevB.93.075426, Seo:2016vks}, acting as a macroscopic fluid.

	On the other hand, in the non-interacting or perturbative field theory picture, what temperature does is to change the distribution of the particles over the single particle state, which certainly is not a proper picture to describe a strongly coupled many-body system near a quantum critical point. 
	
	From these discussion, our treatment of topology at finite temperature is definitely different from the conventional approach and the origin of the difference is very fundamental.
	Summarizing, we extends the applicability of Berry potential formalism to strongly coupled regimes where conventional Uhlmann-phase approaches fail to give well-defined topological invariants. 
	
	\section{\label{sec:conclusion}Conclusion}
	In this work, we have completed a systematic classification of all gapped fermionic phases in strongly correlated many-body systems using the holographic mean-field theory (H-MFT) applied to all possible bulk fermion bilinear interactions in $\mathrm{AdS}_{4}$.
	Our central results are (i) the fully analytic derivation of topological Hamiltonians for all bilinear interactions, followed by exact computation of the associated Berry curvature and Chern number, and (ii) the establishment of new approach for defining topological invariants in strongly interacting many-body systems which gives well defined integer with stability independent of the deformation parameters especially even at finite temperature.
	
	For a single flavor, we demonstrated that once a pseudoscalar coupling $\mathcal{B}_{5}$ is introduced to open a gap, the Berry curvature becomes smooth across momentum space and integrates to the universal fractional value $C = \tfrac{1}{2} \, \mathrm{sgn}(\mathcal{B}_{5})$, regardless of the presence of other scalar, vector, axial vector, or antisymmetric tensor backgrounds.
	The sign of $\mathcal{B}_{5}$ thus acts as the only true topological control parameter, while all other order parameters merely deform the spectral density or displace the curvature in momentum space.
	The fractional Chern number arises from the continuum nature of momentum space, which encloses only half of the Berry curvature flux due to the absence of lattice compactification.
	
	We extended this analysis to two-flavor spinors, constructing the Green’s functions for both SS and SA quantization schemes.
	These lead to two-fold degeneracy in the occupied spectra, necessitating the use of the non-Abelian Berry phase to define a gauge-invariant topological quantity.
	The resulting Chern numbers are integers, whose values depend on the structure of the gapping terms and the choice of quantization, and can be classified into four distinct types.
	
	Importantly, we found that the Chern number remains invariant under deformations of interaction strength, bulk fermion mass $m$, and temperature $T$.
	This robustness confirms that the topological number is a genuine invariant applicable in strongly interacting many-body systems, in contrast to weakly interacting cases where topological invariants such as the Uhlmann phase are not applicable in strongly correlated regime.
	The key reason is that the bulk geometry is fixed as an $\mathrm{AdS}_{4}$ black hole background, the dual boundary state can be effectively treated as a macroscopic single-particle state.
	Consequently, the  Berry potential formalism remains valid and extends naturally to the strongly interacting regime where H-MFT applies.
	
	Our classification provides the first complete ``periodic table’’ of holographic topological phases in four bulk dimensions.  
	Natural extensions include topology in higher dimensions and correspondence to tenfold classification of topological insulators and superconductors.
	Moreover, it would be valuable to generalize the H-MFT framework to capture topological features in experimentally relevant strongly correlated systems, such the interacting surfaces of three-dimensional topological insulators.
	Our work is applicable to recent holographic approaches that encode lattice structure directly into the bulk Dirac equation to make a realistic model for strongly correlated electronic systems \cite{PhysRevD.110.106017}. 
	Now that the topological features can be accessed holographically, we expect that there is a correspondence between lattice parameters and holographic order parameters.

	\section*{Acknowledgment}
	The authors would like to thank  Moon-Jip Park  for the helpful discussion. This work is supported by NRF of Korea with grant No. NRF-2021R1A2B5B02002603, RS-2023-00218998. 
	
	\appendix
	
	\section{Derivation of retarded Green's function}
	\label{appendix:A}
	In this appendix, we present methods to obtain the Green’s function from both the Dirac equation and the flow equation formalism.
	
	\noindent
	\textbf{Analytic Green's function.}
	To analytically express the retarded Green’s function from the chosen source and condensation, we largely follow the formalism developed in \cite{Yuk:2022lof, byun2025symmetric}.
	
	Because $\xi^{(\mathrm{S})}$ and $\xi^{(\mathrm{C})}$ in \eqref{eq2.17} are two-component spinors, each has two independent solutions. 
	Therefore, the general solution of the spinors can be represented as a linear combination of these two solutions with constant coefficients.
	It is convenient to express $\xi^{(\mathrm{S})}$ and $\xi^{(\mathrm{C})}$ in matrix form, separating their basis solutions from the corresponding coefficients.
	For instance, if we denote the two basis solutions for $\xi^{(\mathrm{S})}$ as $(\xi^{(\mathrm{S}, 1)}_{1}, \xi^{(\mathrm{S}, 1)}_{2})^\text{T}$ and $(\xi^{(\mathrm{S}, 2)}_{1}, \xi^{(\mathrm{S}, 2)}_{2})^\text{T}$, with the corresponding coefficients $c_{1}$ and $c_{2}$, then $\xi^{(\mathrm{S})}$ can be expressed as
	\begin{equation}
		\begin{gathered}
			\label{eqA.1}
			\xi^{(\mathrm{S})} = c_{1}
			\begin{pmatrix}
				\xi^{(\mathrm{S}, 1)}_{1}\\
				\xi^{(\mathrm{S}, 1)}_{2}
			\end{pmatrix} + c_{2}
			\begin{pmatrix}
				\xi^{(\mathrm{S}, 2)}_{1}\\
				\xi^{(\mathrm{S}, 2)}_{2}
			\end{pmatrix} = 
			\begin{pmatrix}
				\xi^{(\mathrm{S}, 1)}_{1} & \xi^{(\mathrm{S}, 2)}_{1}\\
				\xi^{(\mathrm{S}, 1)}_{2} & \xi^{(\mathrm{S}, 2)}_{2}
			\end{pmatrix}
			\begin{pmatrix}
				c_{1}\\
				c_{2}
			\end{pmatrix} = \mathbb{S}(z){\bf c},\\
			\text{where}\quad\mathbb{S}(z) = 
			\begin{pmatrix}
				\xi^{(\mathrm{S}, 1)}_{1} & \xi^{(\mathrm{S}, 2)}_{1}\\
				\xi^{(\mathrm{S}, 1)}_{2} & \xi^{(\mathrm{S}, 2)}_{2}
			\end{pmatrix}\quad\text{and}\quad{\bf c} = 
			\begin{pmatrix}
				c_{1}\\
				c_{2}
			\end{pmatrix}.
		\end{gathered}
	\end{equation}
	Likewise, we can represent the $\xi^{(\mathrm{C})}$ in a similar way. 
	On the other hand, the Dirac equation relates $\xi^{(\mathrm{C})}$ to $\xi^{(\mathrm{S})}$, implying that they share the same coefficient vector ${\bf c}$, so that
	\begin{equation}
		\label{eqA.2}
		\xi^{(\mathrm{C})} = \mathbb{C}(z){\bf c},
	\end{equation}
	where $\mathbb{C}(z)$ is a $2\times2$ matrix-valued function. 
	From the boundary behavior of $\xi^{(\mathrm{S})}$ and $\xi^{(\mathrm{C})}$ in \eqref{eq2.18}, we observe that the leading terms $z^{\pm m}$ arise from those of $\mathbb{S}(z)$ and $\mathbb{C}(z)$ in \eqref{eqA.1} and \eqref{eqA.2}, respectively, since ${\bf c}$ is a constant vector.
	Therefore, the boundary behavior of $\mathbb{S}(z)$ and $\mathbb{C}(z)$ is given by 
	\begin{equation}
		\label{eqA.3}
		\mathbb{S}(z) \approx z^{m}\mathbb{S}_{0}\quad\text{and}\quad\mathbb{C}(z) \approx z^{-m}\mathbb{C}_{0},
	\end{equation}
	with constant $2\times2$ matrices $\mathbb{S}_{0}$ and $\mathbb{C}_{0}$.
	It then follows that the boundary behavior of the spinors $\xi^{(\mathrm{S})}$ and $\xi^{(\mathrm{C})}$ is given by
	\begin{equation}
		\label{eqA.4}
		\xi^{(\mathrm{S})} \approx z^{m}\mathbb{S}_{0}{\bf c}\quad\text{and}\quad\xi^{(\mathrm{C})} \approx z^{-m}\mathbb{C}_{0}{\bf c}. 
	\end{equation}
	Comparing \eqref{eq2.18} with \eqref{eqA.4}, the source and condensation can be rewritten as
	\begin{equation}
		\label{eqA.5}
		\mathcal{J} = \mathbb{S}_{0}{\bf c}\quad\text{and}\quad\mathcal{C} = \mathbb{C}_{0}{\bf c} \quad \hbox{ so that }  \mathcal{C} = \mathbb{C}_{0}\mathbb{S}^{-1}_{0}\mathcal{J}. 
	\end{equation}
	
	We can now obtain the retarded Green’s function in terms of $\mathbb{S}_{0}$ and $\mathbb{C}_{0}$ from the total action.
	Because the bulk action does not contribute to the total action due to the equation of motion, the Green’s function is solely determined by the boundary action in \eqref{eq2.16}.
	To achieve this, we rewrite the boundary action, denoted as the effective action $S_{\mathrm{eff}}$, in terms of $\mathcal{J}$ and $\mathcal{C}$ as
	\begin{equation}
		\label{eqA.6}
		S_{\mathrm{eff}} = -\int_{\partial\mathcal{M}}d^{3}x\,{\xi^{(\mathrm{S})}}^{\dagger}\xi^{(\mathrm{C})} + \text{h.c.} = -\int_{\partial\mathcal{M}}d^{3}x\,\mathcal{J}^{\dagger}\mathcal{C} + \text{h.c.}
	\end{equation}
	From the equation \eqref{eqA.5}, the effective action can be rewritten  in terms of the source only,   
	\begin{equation}
		\label{eqA.7}
		S_{\mathrm{eff}} = -\int_{\partial\mathcal{M}}d^{3}x\,\mathcal{J}^{\dagger}(\mathbb{C}_{0}\mathbb{S}^{-1}_{0})\mathcal{J} + \text{h.c.}
	\end{equation}
	According to the linear response theory, this effective action can be expressed as
	\begin{equation}
		\label{eqA.8}
		S_{\mathrm{eff}} = -\int_{\partial\mathcal{M}}d^{3}x\,\mathcal{J}^{\dagger}G_{R}\mathcal{J} + \text{h.c.}
	\end{equation}
	where $G_{R}$ is the retarded Green's function for the source $\mathcal{J}$. 
	By comparing \eqref{eqA.7} and \eqref{eqA.8}, we identify the retarded Green's function as 
	\begin{equation}
		\label{eqA.9}
		G_{R} = \mathbb{C}_{0}\mathbb{S}^{-1}_{0}.
	\end{equation}
	Meanwhile, the Dirac equation or the flow equation provides the solutions of bulk quantities $\mathbb{G}(z)$ defined by
	\begin{equation}
		\label{eqA.10}
		\mathbb{G}(z) \equiv \mathbb{C}(z)\mathbb{S}^{-1}(z),
	\end{equation}
	whose boundary behavior determines the retarded Green’s function $G_{R}$. 
	Thus, to determine $G_{R}$ from $\mathbb{G}(z)$, we should look for  the boundary behavior of $\mathbb{G}(z)$.
	From \eqref{eqA.3} and \eqref{eqA.9},   
	\begin{equation}
		\label{eqA.11}
		\mathbb{G}(z) = \mathbb{C}(z)\mathbb{S}^{-1}(z) \approx z^{-2m}\mathbb{C}_{0}\mathbb{S}^{-1}_{0} = z^{-2m}G_{R}.
	\end{equation}
	Then, we can find the retarded Green's function from the bulk quantity $\mathbb{G}(z)$ as
	\begin{equation}
		\label{eqA.12}
		G_{R} = \lim_{z\to0}z^{2m}\mathbb{G}(z),
	\end{equation}
	which gives \eqref{eq2.20}.
	Therefore, once we know $\mathbb{G}(z)$, we can obtain $G_{R}$. 
	
	\noindent
	\textbf{Flow equation.}
	For numerical calculation of Green's function, it is convenient to employ flow equation \cite{PhysRevD.79.025023, Yuk:2022lof}, to take the advantage that we can get the Green's function without solving Dirac equation directly.
	
	To set up the flow equation, we begin by rewriting the Dirac equation \eqref{eq2.13} in terms of $\xi^{(\mathrm{S})}$ and $\xi^{(\mathrm{C})}$ by using \eqref{eq2.15} and \eqref{eq2.17}, such that
	\begin{align}
		\partial_{z}\xi^{(\mathrm{S})} + \mathbb{M}_{1}\xi^{(\mathrm{S})} + \mathbb{M}_{2}\xi^{(\mathrm{C})} &= 0,\\
		\partial_{z}\xi^{(\mathrm{C})} + \mathbb{M}_{3}\xi^{(\mathrm{C})} + \mathbb{M}_{4}\xi^{(\mathrm{S})} &= 0,
	\end{align}
	where $\mathbb{M}_{1}$, $\mathbb{M}_{2}$, $\mathbb{M}_{3}$, and $\mathbb{M}_{4}$ are $2\times2$ matrix-valued functions.
	Then, we substitute the expression in \eqref{eqA.1} and \eqref{eqA.2} into these equations of motion, which gives
	\begin{gather}
		\label{eqA.15}
		\partial_{z}\mathbb{S}(z) + \mathbb{M}_{1}\mathbb{S}(z) + \mathbb{M}_{2}\mathbb{C}(z) = 0,\\
		\label{eqA.16}
		\partial_{z}\mathbb{C}(z) + \mathbb{M}_{3}\mathbb{C}(z) + \mathbb{M}_{4}\mathbb{S}(z) = 0,
	\end{gather}
	because ${\bf c}$ is an arbitrary vector in the solution space. 
	When we take a derivative with respect to $z$ on the bulk quantity $\mathbb{G}(z)$ in \eqref{eqA.10}, since $\partial_{z}\mathbb{S}^{-1} = \mathbb{S}(z)^{-1}(\partial_{z}\mathbb{S}(z))\mathbb{S}^{-1}$, we have
	\begin{equation}
		\label{eqA.17}
		\begin{aligned}
			\partial_{z}\mathbb{G} &= \partial_{z}(\mathbb{C}(z)\mathbb{S}(z)^{-1})\\
			&= (\partial_{z}\mathbb{C}(z))\mathbb{S}(z)^{-1} + \mathbb{C}(z)\partial_{z}\mathbb{S}(z)^{-1}\\
			&= (\partial_{z}\mathbb{C}(z))\mathbb{S}(z)^{-1} - \mathbb{C}(z)\mathbb{S}(z)^{-1}(\partial_{z}\mathbb{S}(z))\mathbb{S}(z)^{-1}.
		\end{aligned}
	\end{equation}
	If we substitute the expressions for $\partial_{z}\mathbb{S}(z)$ and $\partial_{z}\mathbb{C}(z)$ from \eqref{eqA.15} and \eqref{eqA.16} into above equation, we obtain
	\begin{equation}
		\label{eqA.18}
		\partial_{z}\mathbb{G} - \mathbb{G}(z)\mathbb{M}_{2}\mathbb{G}(z) - \mathbb{G}(z)\mathbb{M}_{1} + \mathbb{M}_{3}\mathbb{G}(z) + \mathbb{M}_{4} = 0,
	\end{equation}
	which we refer to as the flow equation.
	To solve this differential equation, we impose the near-horizon boundary condition for $\mathbb{G}(z)$, which is given by \cite{Yuk:2022lof}
	\begin{equation}
		\label{eqA.19}
		\mathbb{G}(z_{\mathrm{H}}) = i\mathbbm{1}_{2\times 2}.
	\end{equation}
	Applying this boundary condition, we get $\mathbb{G}(z)$ and we also solve the Green's function from \eqref{eqA.12}.
	The flow equation for two-flavor spinors is derived in the same manner, which results in the same form of \eqref{eqA.18}.
	In this case, all matrices reduce to $4 \times 4$ form, and the boundary condition in \eqref{eqA.19} becomes $i\mathbbm{1}_{4\times4}$.

	\section{\label{appendix:B}Topological invariant in gappless phase}
	As discussed in the section~\ref{section3.2}, the critical case exhibits a degeneracy at the gap-closing point, where the Berry curvature becomes ill-defined due to a singularity.
	Consequently, the definition of the Chern number in \eqref{eq3.4} cannot be rigorously applied. 
	This issue persists across other interaction types in the gapless regime, except in the presence of a pseudoscalar term.
	This breakdown can be explicitly seen within the another definition of Berry curvature: the expression in perturbation theory for the Berry curvature associated with the upper band,
	\begin{equation}
		\label{eqB.1}
		\begin{aligned}
			\mathcal{F}(\mathbf{k}) &= \dfrac{i}{(E_{+} - E_{-})^{2}}\left(\bra{n_{+}}\partial_{k_{x}}H(\mathbf{k})\ket{n_{-}}\bra{n_{-}}\partial_{k_{y}}H(\mathbf{k})\ket{n_{+}}\right. \\
			&\qquad\qquad\qquad\qquad\qquad\qquad\qquad\left.- \bra{n_{+}}\partial_{k_{y}}H(\mathbf{k})\ket{n_{-}}\bra{n_{-}}\partial_{k_{x}}H(\mathbf{k})\ket{n_{+}}\right)\\
			&= -\dfrac{2}{(E_{+} - E_{-})^{2}}\,\text{Im}\left(\bra{n_{+}}\partial_{k_{x}}H(\mathbf{k})\ket{n_{-}}\bra{n_{-}}\partial_{k_{y}}H(\mathbf{k})\ket{n_{+}}\right),
		\end{aligned}
	\end{equation}
	where $E_{\pm}$ and $\ket{n_{\pm}}$ are the eigenvalues and eigenvectors of $H(\mathbf{k})$, clearly diverges at the gap-closing point where $E_{+} = E_{-}$.
	
	To define a meaningful topological invariant in such gapless cases, we introduce a prescription based on the Cauchy principal value of the Berry curvature integral:
	\begin{equation}
		\label{eqB.2}
		C \equiv \dfrac{1}{2\pi}PV\int_{\mathbb{R}^{2}}dk_{x}dk_{y}\mathcal{F}(\mathbf{k}).
	\end{equation}
	For example, in the critical (non-interacting) case where the Berry curvature is ill-defined at $\mathbf{k} = (0,0)$, the integral is performed by excluding a small region around this point. 
	The result of the principal value integration is
	\begin{equation}
		\label{eqB.3}
		C = 0.
	\end{equation}
	This vanishing Chern number persists for all interacting cases in the absence of pseudoscalar coupling. 
	In these cases, the Berry curvature either vanishes throughout the $\mathbf{k}$-space or integrates to zero due to its odd functional structure in $\mathbf{k}$-space, leading to cancellation, e.g., $B_{ti}$ coupling.
	
	For axial-vector couplings $B_{5i}$ ($i = x, y, z$) and tensor couplings $B_{ti}$ ($i = x, y, z$), the spectral density remains gapless in the intermediate regime $0 < \abs{\mathcal{B}_{5}} < \abs{\mathcal{B}_{I}}$, where degeneracies are not fully resolved.
	Because of the divergence in \eqref{eqB.1}, no well-defined topological invariant can be assigned, so we consider the topology of such systems only in the regime $\abs{\mathcal{B}_{5}} > \abs{\mathcal{B}_{I}}$.

	\section{\label{appendix:C}Summary of spectral and topological features}
	We provide detailed comments on the Green’s function and associated topological features for each interaction type, as obtained in the section~\ref{section3.3}.
	As discussed therein, the Chern number universally takes the value $C = \frac{1}{2}\mathrm{sgn}(\mathcal{B}_{5})$, regardless of the interaction type, as long as a spectral gap is present.
	The interaction types below are classified according to the boundary point of view, following the classification introduced in \cite{Sukrakarn:2023ncp}.
	
	\subsection*{Scalar: $\mathcal{L}_{\mathrm{int}} = -i\bar{\psi}B\psi$}
	We consider scalar interaction of the form $B(z) = \mathcal{B} z$. 
	For this interaction, the spectral density remains gapless unless a pseudoscalar term is introduced. 
	The addition of $\mathcal{B}_{5} \neq 0$ opens a spectral gap immediately, so we consider $\mathcal{B}_{5} \neq 0$.
	Despite the presence of a scalar order parameter, the resultant Berry curvature is identical to that of the pure pseudoscalar case given in \eqref{eq3.15}.
	
	\subsection*{Vector: $\mathcal{L}_{\mathrm{int}} = \bar{\psi}B_{M}\Gamma^{M}\psi$}
	We consider vector-type interactions of the form $B_{M}(z) = \mathcal{B}_{M}$. 
	These vector couplings effectively shift the momentum along the direction of the coupling.
	\begin{itemize}
		\item \textbf{Temporal vector.}
		For a temporal coupling $B_{t}$, the analytic Green’s function corresponds to that of the critical case with a frequency shift $\omega \to \omega + \mathcal{B}_{t}$.
		This also holds when there is additional pseudoscalar term for gap opening.
		A true spectral gap around $\omega = 0$ appears only when $|\mathcal{B}_{5}| > |\mathcal{B}_{t}|$; otherwise, the upper spectral branch crosses into $\omega < 0$, and the system is no longer gapped at the Fermi level.
		The corresponding topological Hamiltonian yields the same Berry curvature as the pure pseudoscalar case.
		\item \textbf{Spatial vector.}
		For spatial components $B_{i}$ ($i = x, y$), a nonzero $\mathcal{B}_{5}$ opens a gap. 
		These couplings result in a momentum shift $k_{i} \to k_{i} - \mathcal{B}_{i}$ in Green's function, topological Hamiltonian, and Berry curvature.
		Although the Berry curvature is spatially shifted, the Chern number remains invariant.
		\item \textbf{Radial scalar.}
		For radial coupling $B_{z}$, the Green’s function in the absence of a pseudoscalar term is identical to the critical case.
		Hence, when a pseudoscalar coupling is introduced, the resulting gapped phase becomes identical to the pure pseudoscalar case, with the same spectral and topological structure.
	\end{itemize}

	\subsection*{Axial vector: $\mathcal{L}_{\mathrm{int}} = \bar{\psi}B_{5M}\Gamma^{M}\Gamma^{5}\psi$}
	The axial vector-type couplings $B_{5M}(z) = \mathcal{B}_{5M}$ yield a gapped spectral density either when $\mathcal{B}_{5} \neq 0$ or when the pseudoscalar coupling dominates the axial vector component, i.e., $|\mathcal{B}_{5}| > |\mathcal{B}_{5M}|$, depending on the component of $B_{5M}$.
	\begin{itemize}
		\item \textbf{Temporal axial vector.}
		For the temporal component $B_{5t}$, the spectral density remains gapless unless a pseudoscalar coupling $\mathcal{B}_{5} \neq 0$ is introduced, which opens a gap—similar to the scalar and vector cases.
		The Berry curvature is symmetrically deformed but the Chern number is invariant.
		\item \textbf{Spatial axial vector.}
		For spatial components $B_{5i}$ $(i = x, y)$, the introduction of a pseudoscalar coupling $\mathcal{B}_{5}$ does not immediately resolve the singularity in the Berry curvature.
		The parameters $\mathcal{B}_{5i}$ and $\mathcal{B}_{5}$ compete to determine whether a spectral gap forms.
		When $0 < |\mathcal{B}_{5}| < |\mathcal{B}_{5i}|$, the spectral density remains gapless, and the Berry curvature exhibits persistent singularities located at $(k_{i}, k_{\perp}) = \left( \pm \sqrt{{\mathcal{B}_{5i}}^{2} - {\mathcal{B}_{5}}^{2}},\, 0 \right)$.
		A gapped spectrum appears only when $|\mathcal{B}_{5}| > |\mathcal{B}_{5i}|$, in which case the Berry curvature becomes smooth and is deformed along the $k_{\perp}$ direction compared to the pure pseudoscalar case with the same $\mathcal{B}_{5}$.
		The Chern number in this regime coincides with that of the pure pseudoscalar case.
		\item \textbf{Radial pseudoscalar.}
		The coupling $B_{5z}$ corresponds to the case where the gap opens when $\abs{\mathcal{B}_{5}} > \abs{\mathcal{B}_{5z}}$.
		Notably, the spectral densities differ qualitatively between $\mathcal{B}_{5} > 0$ and $\mathcal{B}_{5} < 0$: the gap corresponds to zeros of the spectral density, while in the latter, it becomes poles.
		The Berry curvature in this case is quantitatively deformed relative to the pure pseudoscalar case.
	\end{itemize}

	\subsection*{Antisymmetric 2-tensor: $\mathcal{L}_{\mathrm{int}} = \dfrac{1}{2}\bar{\psi}B_{MN}\Gamma^{MN}\psi$}
	We consider antisymmetric tensor interactions of the form $B_{MN}(z) = \mathcal{B}_{MN} z^{-1}$.
	As in the axial vector case, a gapped spectral density arises either when $\mathcal{B}_{5} \neq 0$, or when the pseudoscalar term dominates, i.e., $|\mathcal{B}_{5}| > |\mathcal{B}_{MN}|$, depending on the component of $B_{MN}$.
	\begin{itemize}
		\item \textbf{Time-space-like tensor.}
		Similar to the $B_{5i}$ case, the introduction of a pseudoscalar term with $0 < |\mathcal{B}_{5}| < |\mathcal{B}_{ti}|$ fails to open a spectral gap and does not regularize the singularities in the Berry curvature.
		A gapped phase arises only when $|\mathcal{B}_{5}| > |\mathcal{B}_{ti}|$, in which case the Berry curvature becomes smooth and is deformed along the $k_{i}$ direction compared to the pure pseudoscalar case with the same $\mathcal{B}_{5}$. 
		The Chern number in this regime coincides with that of the pure pseudoscalar case.
		\item \textbf{Radial temporal vector.}
		When $0 < |\mathcal{B}_{5}| < |\mathcal{B}_{tz}|$, the spectrum remains gapless and the Berry curvature retains singular features.
		A gap opens for $|\mathcal{B}_{5}| > |\mathcal{B}_{tz}|$, and in this regime, the Berry curvature becomes identical to the pure pseudoscalar case.
		\item \textbf{Space-space-like tensor.}
		For the $B_{ij}$ couplings, the spectrum is gapped for any nonzero $\mathcal{B}_{5}$, which removes the singularities in the Berry curvature and this is deformed symmetrically, but invariant Chern number.
		\item \textbf{Radial spatial vector.}
		For the $B_{iz}$ couplings, a nonzero pseudoscalar coupling $\mathcal{B}_{5}$ immediately opens a spectral gap, and the Berry curvature becomes a smooth function with two peaks along the $k_{i}$ axis.
		The Chern number remains invariant.
	\end{itemize}

	\section{\label{appendix:D}Duality between SS and SA quantization}
	In this section, we demonstrate that the SS and SA quantization schemes are related by a spinor transformation that maps both Green’s functions and topological invariants, largely helping to understand the results under different quantization choices.
	
	\noindent
	\textbf{Green's function.}
	Once we know the Green's function in SS quantization, we can figure out that in SA quantization as well.
	This follows from the transformation of spinors, which maps the source and condensation between the two schemes.
	
	We begin by defining two projection operators, $\mathbb{P}_{+}$ and $\mathbb{P}_{-}$, and introduce their combination into an $8 \times 8$ involution matrix $\mathbb{P}$ as follows:
	\begin{equation}
		\label{eqC.1}
		\mathbb{P} = \begin{pmatrix}
			\mathbb{P}_{+} & \mathbb{P}_{-}\\
			\mathbb{P}_{-} & \mathbb{P}_{+}
		\end{pmatrix} \quad\text{where}\quad
		\begin{cases}
			\mathbb{P}_{+} = \text{diag}(1, 1, 0, 0),\\[5pt]
			\mathbb{P}_{-} = \text{diag}(0, 0, 1, 1).
		\end{cases}
	\end{equation}
	Using \eqref{eq4.10} and \eqref{eq4.12}, this operator transforms the pairs of bulk-extended quantities of source and condensation between the SS and SA quantizations, such that
	\begin{equation}
		\label{eqC.2}
		\mathbb{P}\begin{pmatrix}
			\xi^{(\mathrm{S})}_{(\mathrm{SS})}\\[5pt]
			\xi^{(\mathrm{C})}_{(\mathrm{SS})}
		\end{pmatrix} = \begin{pmatrix}
			\xi^{(\mathrm{S})}_{(\mathrm{SA})}\\[5pt]
			\xi^{(\mathrm{C})}_{(\mathrm{SA})}
		\end{pmatrix}\quad\text{and}\quad\mathbb{P}\begin{pmatrix}
			\xi^{(\mathrm{S})}_{(\mathrm{SA})}\\[5pt]
			\xi^{(\mathrm{C})}_{(\mathrm{SA})}
		\end{pmatrix} = \begin{pmatrix}
			\xi^{(\mathrm{S})}_{(\mathrm{SS})}\\[5pt]
			\xi^{(\mathrm{C})}_{(\mathrm{SS})}
		\end{pmatrix}.
	\end{equation}
	On the other hand, we can express the spinors as
	\begin{equation}
		\begin{cases}
			\xi^{(\mathrm{S})}_{(\mathrm{SS})} = \mathbb{S}^{(\mathrm{SS})}\mathbf{c},\\[5pt]
			\xi^{(\mathrm{C})}_{(\mathrm{SS})} = \mathbb{C}^{(\mathrm{SS})}\mathbf{c},
		\end{cases}
		\quad\text{and}\quad
		\begin{cases}
			\xi^{(\mathrm{S})}_{(\mathrm{SA})} = \mathbb{S}^{(\mathrm{SA})}\mathbf{c},\\[5pt]
			\xi^{(\mathrm{C})}_{(\mathrm{SA})} = \mathbb{C}^{(\mathrm{SA})}\mathbf{c},
		\end{cases}
	\end{equation}
	similar to \eqref{eqA.1} and \eqref{eqA.2}.
	By using this, \eqref{eqC.2} can be expressed as
	\begin{equation}
		\label{eqC.3}
		\mathbb{S}^{(\mathrm{SA})} = \mathbb{P}_{+}\mathbb{S}^{(\mathrm{SS})} + \mathbb{P}_{-}\mathbb{C}^{(\mathrm{SS})}
		\quad\text{and}\quad
		\mathbb{C}^{(\mathrm{SA})} = \mathbb{P}_{+}\mathbb{C}^{(\mathrm{SS})} + \mathbb{P}_{-}\mathbb{S}^{(\mathrm{SS})}.
	\end{equation}
	By putting these expressions into the Green's function definition in SA quantization \eqref{eq4.14}, this becomes
	\begin{equation}
		\label{eqC.4}
		\begin{aligned}
			G_{R}^{(\mathrm{SA})} &= \lim_{z \to 0} z^{2m} (\sigma_{0}\oplus-\sigma_{0})\mathbb{C}^{(\mathrm{SA})}(\mathbb{S}^{(\mathrm{SA})})^{-1}\\
			&= \lim_{z \to 0} z^{2m} (\sigma_{0}\oplus-\sigma_{0})(\mathbb{P}_{+}\mathbb{C}^{(\mathrm{SS})} + \mathbb{P}_{-}\mathbb{S}^{(\mathrm{SS})})(\mathbb{P}_{+}\mathbb{S}^{(\mathrm{SS})} + \mathbb{P}_{-}\mathbb{C}^{(\mathrm{SS})})^{-1}.
		\end{aligned}
	\end{equation}
	By rearranging this and using \eqref{eq4.13}, we obtain
	\begin{equation}
		\label{eqC.5}
		G_{R}^{(\mathrm{SA})} = (\sigma_{0}\oplus-\sigma_{0})(\mathbb{P}_{+}G_{R}^{(\mathrm{SS})} + \mathbb{P}_{-})(\mathbb{P}_{+} + \mathbb{P}_{-}G_{R}^{(\mathrm{SS})})^{-1}.
	\end{equation}
	Therefore, we can represent the retarded Green's function in SA quantization in terms of that in SS quantization as \eqref{eq4.16}.
	
	\noindent
	\textbf{Chern number.}
	In two-flavor spinor case, the Chern number associated with a certain $(\tau_{a}, \tau_{b})$ interaction structure in SA quantization is dual to that of a different sector in SS quantization. 
	This duality arises because the boundary action in SA quantization can be obtained from that in SS case via the field redefinition:
	\begin{equation}
		\label{eqE.1}
		\psi^{(2)} \,\to\, \psi^{(2)}\Gamma^{5} \quad \Longrightarrow \quad S_{\mathrm{bdy}}^{(\mathrm{SS})} \to S_{\mathrm{bdy}}^{(\mathrm{SA})}
	\end{equation}
	Since the definition of the Green’s function originates from the boundary action, boundary quantities—such as the spectral density and topological invariants—are related by this transformation.
	We emphasize that this duality between the boundary quantities holds specifically in the $\mathrm{AdS}_{4}$ geometry.
	
	For example, consider the $(\tau_{0}, \tau_{0})$ case in SS quantization, where the interaction action is given by
	\begin{equation}
		\label{eqE.2}
		S_{\mathrm{int}} = i\int_{\mathcal{M}}d^{4}x\sqrt{-g}
		\begin{pmatrix}
			\bar{\psi}^{(1)} & \bar{\psi}^{(2)}
		\end{pmatrix}
		\begin{pmatrix}
			(B_{I} + \tilde{B}_{I})\Gamma^{I} & 0\\
			0 & (B_{I} + \tilde{B}_{I})\Gamma^{I}
		\end{pmatrix}
		\begin{pmatrix}
			\psi^{(1)}\\
			\psi^{(2)}
		\end{pmatrix}.
	\end{equation}
	Applying the transformation $\psi^{(2)} \to \psi^{(2)} \Gamma^{5}$ leads to
	\begin{equation}
		\label{eqE.3}
		S_{\mathrm{int}} = i\int_{\mathcal{M}}d^{4}x\sqrt{-g}
		\begin{pmatrix}
			\bar{\psi}^{(1)} & \bar{\psi}^{(2)}
		\end{pmatrix}
		\begin{pmatrix}
			(B_{I} + \tilde{B}_{I})\Gamma^{I} & 0\\
			0 & -(B_{I} + \tilde{B}_{I})\Gamma^{I}
		\end{pmatrix}
		\begin{pmatrix}
			\psi^{(1)}\\
			\psi^{(2)}
		\end{pmatrix},
	\end{equation}
	which corresponds to the $(\tau_{3}, \tau_{3})$ case in SA quantization. 
	Therefore, the interaction structures $(\tau_{0}, \tau_{0})$ in SS and $(\tau_{3}, \tau_{3})$ in SA are dual to each other under this $\Gamma^{5}$ rotation.
	This duality extends to their topological properties: the Chern number for the $(\tau_{0}, \tau_{0})$ sector in SS quantization exactly matches that of the $(\tau_{3}, \tau_{3})$ sector in SA, as verified in the table~\ref{table7}–\ref{table9}.

	\bibliographystyle{jhep}
	\bibliography{TBH_ref.bib}
	
\end{document}